\documentclass[preprint]{aastex61}
\usepackage{tabularx}
\usepackage{tabularht}
\usepackage{rotating}
\usepackage{hyperref}
\hypersetup{linkcolor=red,citecolor=blue,filecolor=cyan,urlcolor=magenta}
\usepackage[hyphenbreaks]{breakurl}
\shorttitle{SEP events observed by the PAMELA mission}
\shortauthors{Bruno et al.}

\begin{document}

\title{Solar energetic particle events observed by the PAMELA mission}

\correspondingauthor{Alessandro Bruno}
\email{alessandro.bruno-1@nasa.gov}
\author{A.~Bruno}\affil{Heliophysics Division, NASA Goddard Space Flight Center, Greenbelt, MD, USA.}\affil{INFN, Sezione di Bari, I-70126 Bari, Italy.}
\author{G.~A.~Bazilevskaya}\affil{Lebedev Physical Institute, RU-119991 Moscow, Russia.}
\author{M.~Boezio}\affil{INFN, Sezione di Trieste, I-34149 Trieste, Italy.}
\author{E.~R.~Christian}\affil{Heliophysics Division, NASA Goddard Space Flight Center, Greenbelt, MD, USA.}
\author{G.~A.~de~Nolfo}\affil{Heliophysics Division, NASA Goddard Space Flight Center, Greenbelt, MD, USA.}
\author{M.~Martucci}\affil{Department of Physics, University of Rome ``Tor Vergata'', I-00133 Rome, Italy.}\affil{INFN, Laboratori Nazionali di Frascati, Via Enrico Fermi 40, I-00044 Frascati, Italy.}
\author{M.~Merge'}\affil{INFN, Sezione di Rome ``Tor Vergata'', I-00133 Rome, Italy.}\affil{Department of Physics, University of Rome ``Tor Vergata'', I-00133 Rome, Italy.}
\author{V.~V.~Mikhailov}\affil{National Research Nuclear University MEPhI, RU-115409 Moscow, Russia.}
\author{R.~Munini}\affil{INFN, Sezione di Trieste, I-34149 Trieste, Italy.}
\author{I.~G.~Richardson}\affil{Heliophysics Division, NASA Goddard Space Flight Center, Greenbelt, MD, USA.}\affil{Department of Astronomy, University of Maryland, College Park, MD, USA.}
\author{J.~M.~Ryan}\affil{Space Science Center, University of New Hampshire, Durham, NH, USA.}
\author{S.~Stochaj}\affil{Electrical and Computer Engineering, New Mexico State University, Las Cruces, NM, USA.}
\author{O.~Adriani}\affil{Department of Physics and Astronomy, University of Florence, I-50019 Sesto Fiorentino, Florence, Italy.}\affil{INFN, Sezione di Florence, I-50019 Sesto Fiorentino, Florence, Italy.}
\author{G.~C.~Barbarino}\affil{Department of Physics, University of Naples ``Federico II'', I-80126 Naples, Italy.}\affil{INFN, Sezione di Naples, I-80126 Naples, Italy.}
\author{R.~Bellotti}\affil{INFN, Sezione di Bari, I-70126 Bari, Italy.}\affil{Department of Physics, University of Bari, I-70126 Bari, Italy.}
\author{E.~A.~Bogomolov}\affil{Ioffe Physical Technical Institute, RU-194021 St. Petersburg, Russia.}
\author{M.~Bongi}\affil{Department of Physics and Astronomy, University of Florence, I-50019 Sesto Fiorentino, Florence, Italy.}\affil{INFN, Sezione di Florence, I-50019 Sesto Fiorentino, Florence, Italy.}
\author{V.~Bonvicini}\affil{INFN, Sezione di Trieste, I-34149 Trieste, Italy.}
\author{S.~Bottai}\affil{INFN, Sezione di Florence, I-50019 Sesto Fiorentino, Florence, Italy.}
\author{F.~Cafagna}\affil{INFN, Sezione di Bari, I-70126 Bari, Italy.}
\author{D.~Campana}\affil{INFN, Sezione di Naples, I-80126 Naples, Italy.}
\author{P.~Carlson}\affil{KTH, Department of Physics, and the Oskar Klein Centre for Cosmoparticle.}
\author{M.~Casolino}\affil{INFN, Sezione di Rome ``Tor Vergata'', I-00133 Rome, Italy.}\affil{RIKEN, Advanced Science Institute, Wako-shi, Saitama, Japan.}
\author{G.~Castellini}\affil{IFAC, I-50019 Sesto Fiorentino, Florence, Italy.}
\author{C.~De~Santis}\affil{INFN, Sezione di Rome ``Tor Vergata'', I-00133 Rome, Italy.}\affil{Department of Physics, University of Rome ``Tor Vergata'', I-00133 Rome, Italy.}
\author{V.~Di~Felice}\affil{INFN, Sezione di Rome ``Tor Vergata'', I-00133 Rome, Italy.}\affil{Agenzia Spaziale Italiana (ASI) Science Data Center, Via del Politecnico snc, I-00133 Rome, Italy.}
\author{A.~M.~Galper}\affil{National Research Nuclear University MEPhI, RU-115409 Moscow, Russia.}
\author{A.~V.~Karelin}\affil{National Research Nuclear University MEPhI, RU-115409 Moscow, Russia.}
\author{S.~V.~Koldashov}\affil{National Research Nuclear University MEPhI, RU-115409 Moscow, Russia.}
\author{S.~Koldobskiy}\affil{National Research Nuclear University MEPhI, RU-115409 Moscow, Russia.}
\author{S.~Y.~Krutkov}\affil{Ioffe Physical Technical Institute, RU-194021 St. Petersburg, Russia.}
\author{A.~N.~Kvashnin}\affil{Lebedev Physical Institute, RU-119991 Moscow, Russia.}
\author{A.~Leonov}\affil{National Research Nuclear University MEPhI, RU-115409 Moscow, Russia.}
\author{V.~Malakhov}\affil{National Research Nuclear University MEPhI, RU-115409 Moscow, Russia.}
\author{L.~Marcelli}\affil{INFN, Sezione di Rome ``Tor Vergata'', I-00133 Rome, Italy.}\affil{Department of Physics, University of Rome ``Tor Vergata'', I-00133 Rome, Italy.}
\author{A.~G.~Mayorov}\affil{National Research Nuclear University MEPhI, RU-115409 Moscow, Russia.}
\author{W.~Menn}\affil{Department of Physics, Universit\"{a}t Siegen, D-57068 Siegen, Germany.}
\author{E.~Mocchiutti}\affil{INFN, Sezione di Trieste, I-34149 Trieste, Italy.},
\author{A.~Monaco}\affil{Department of Physics, University of Bari, I-70126 Bari, Italy.}\affil{INFN, Sezione di Bari, I-70126 Bari, Italy.}
\author{N.~Mori}\affil{INFN, Sezione di Florence, I-50019 Sesto Fiorentino, Florence, Italy.}
\author{G.~Osteria}\affil{INFN, Sezione di Naples, I-80126 Naples, Italy.}
\author{B.~Panico}\affil{INFN, Sezione di Naples, I-80126 Naples, Italy.}
\author{P.~Papini}\affil{INFN, Sezione di Florence, I-50019 Sesto Fiorentino, Florence, Italy.}
\author{M.~Pearce}\affil{KTH, Department of Physics, and the Oskar Klein Centre for Cosmoparticle Physics, AlbaNova University Centre, SE-10691 Stockholm, Sweden.}
\author{P.~Picozza}\affil{INFN, Sezione di Rome ``Tor Vergata'', I-00133 Rome, Italy.}\affil{Department of Physics, University of Rome ``Tor Vergata'', I-00133 Rome, Italy.}
\author{M.~Ricci}\affil{INFN, Laboratori Nazionali di Frascati, Via Enrico Fermi 40, I-00044 Frascati, Italy.}
\author{S.~B.~Ricciarini}\affil{INFN, Sezione di Florence, I-50019 Sesto Fiorentino, Florence, Italy.}\affil{IFAC, I-50019 Sesto Fiorentino, Florence, Italy.}
\author{M.~Simon}\affil{Department of Physics, Universit\"{a}t Siegen, D-57068 Siegen, Germany.}
\author{R.~Sparvoli}\affil{INFN, Sezione di Rome ``Tor Vergata'', I-00133 Rome, Italy.}\affil{Department of Physics, University of Rome ``Tor Vergata'', I-00133 Rome, Italy.}
\author{P.~Spillantini}\affil{National Research Nuclear University MEPhI, RU-115409 Moscow, Russia.}\affil{IAPS/INAF, I-00133 Rome, Italy.}
\author{Y.~I.~Stozhkov}\affil{Lebedev Physical Institute, RU-119991 Moscow, Russia.}
\author{A.~Vacchi}\affil{INFN, Sezione di Trieste, I-34149 Trieste, Italy.}\affil{Department of Mathematics, Computer Science and Physics, University of Udine, I-33100 Udine, Italy.}
\author{E.~Vannuccini}\affil{INFN, Sezione di Florence, I-50019 Sesto Fiorentino, Florence, Italy.}
\author{G.~I.~Vasilyev}\affil{Ioffe Physical Technical Institute, RU-194021 St. Petersburg, Russia.}
\author{S.~A.~Voronov}\affil{National Research Nuclear University MEPhI, RU-115409 Moscow, Russia.}
\author{Y.~T.~Yurkin}\affil{National Research Nuclear University MEPhI, RU-115409 Moscow, Russia.}
\author{G.~Zampa}\affil{INFN, Sezione di Trieste, I-34149 Trieste, Italy.}
\author{N.~Zampa}\affil{INFN, Sezione di Trieste, I-34149 Trieste, Italy.}



\begin{abstract}
Despite the significant progress achieved in recent years, the physical mechanisms underlying the origin of solar
energetic particles (SEPs) are still a matter of debate. The complex nature of both particle acceleration and
transport poses challenges to developing a universal picture of SEP events that encompasses both the low-energy
(from tens of keV to a few hundreds of MeV) observations made by space-based instruments and the GeV particles
detected by the worldwide network of neutron monitors in ground-level enhancements (GLEs). The high-precision
data collected by the Payload for Antimatter Matter Exploration and Light-nuclei Astrophysics (PAMELA)
satellite experiment offer a unique opportunity to study the SEP fluxes between $\sim$80 MeV and a few GeV,
significantly improving the characterization of the most energetic events. In particular, PAMELA can measure for
the first time with good accuracy the spectral features at moderate and high energies, providing important
constraints for current SEP models. In addition, the PAMELA observations allow the relationship between low and
high-energy particles to be investigated, enabling a clearer view of the SEP origin. No qualitative distinction
between the spectral shapes of GLE, sub-GLE and non-GLE events is observed, suggesting that GLEs are not a
separate class, but are the subset of a continuous distribution of SEP events that are more intense at high energies.
While the spectral forms found are to be consistent with diffusive shock acceleration theory, which predicts
spectral rollovers at high energies that are attributed to particles escaping the shock region during acceleration,
further work is required to explore the relative influences of acceleration and transport processes on SEP spectra.
\end{abstract}


\section{Introduction}\label{Introduction}
Solar energetic particle (SEP) events pose a recognized hazard to spacecraft and high-altitude aircraft, and are a health risk for
astronauts and flight crews, making them an important constituent of what we call space weather. Historically the origin of SEPs
has been inferred by observations of their morphology and
composition. According to the classification scheme discussed,
for example by \citet{ref:KAHLER1978,ref:KAHLER1984,ref:CLIVER1982,ref:MASON1984,ref:CANE1986,ref:REAMES1988}, 
SEP events can be subdivided into two distinct
categories, gradual and impulsive events, which are related to
different acceleration mechanisms. The gradual events, associated
with type II radio emission, are believed to be accelerated high in
the corona by shocks driven by coronal mass ejections (CMEs; \citet{ref:REAMES1999,ref:ROUILLARD2011}). They are characterized by
elemental abundances, charge states, and temperatures typical of
the ambient corona, and they produce by far the highest SEP
intensities near Earth. The impulsive events, generally much less
intense, are linked to short-duration soft X-ray flare emission from
low altitudes \citep{ref:PALLAVICINI1977} and fast-drift type III radio
emission reflecting electron escape into the interplanetary medium \citep{ref:WILD1963}. They are thought to be accelerated at flare sites mostly by processes in association with magnetic reconnection \citep{ref:SHIMOJO-SHIBATA2000,ref:DRAKE2013} or wave-particle interactions \citep{ref:FISK1978,ref:TEMERINROTH1992,ref:MILLERROBERTS1995,ref:MILLERREAMES1996,ref:ROTHTEMERIN1997}
and are characterized by enrichments in $^{3}$He, electrons and heavy ions such as Fe \citep{ref:REAMES1999,ref:TYLKA2005,ref:MEWALDT2012}. The two-class scenario was subsequently revised to include so-called ``hybrid'' events \citep{ref:CLIVER1996}, exhibiting some characteristics of both classes.
Recent studies have shown that SEP events are in general originated by a mixture of impulsive and gradual processes, and the event evolution depends on their relative importance and on the magnetic connection to Earth, albeit there is still no consensus about the details of the individual mechanisms \citep{ref:CANE2003,ref:CANE2010,ref:TYLKA-LEE2006,ref:GOPALSWAMY2012,ref:KAHLER2012,ref:MEWALDT2012,ref:MEWALDT2015,ref:REAMES2013,ref:CLIVER2016,ref:BAZILEVSKAYA2017}.

The most energetic SEP events induce atmospheric showers
whose secondary products can be detected by ground-based
detectors such as neutron monitors (NMs), muon hodoscopes,
and ionization chambers. How the particles in such rare events,
known as ground-level enhancements (GLEs), are accelerated
remains controversial, and in part due to the relatively few
observations above a few hundred MeV, they have often been
treated as a special, distinct category of SEP events compared
to those observed at lower energies. In particular, while GLEs
are considered gradual events, a direct flare contribution has
become a matter of debate \citep{ref:GRECHNEV2008,ref:ASCHWANDEN2012,ref:MCCRACKEN2012,ref:KAHLER2017}. For example, the double-pulse time profile registered in a number of cases has suggested two components: a rapid onset related to an impulsive injection of flare particles, followed by a gradual phase attributed to shock-accelerated particles \citep{ref:VASHENYUK2006,ref:VASHENYUK2011,ref:MCCRACKEN2008}.

Aside from the relevant space weather implications \citep{ref:SHEASMART2012}, GLEs are of particular interest because they represent SEP acceleration at its most efficient \citep{ref:MEWALDT2012}. In addition, the high-energy protons of GLE events can reach 1 au with minimal interplanetary scattering \citep{ref:CLIVER1982}. Thus, their spectra provide important constraints on SEP origin. For example, in the scenario of diffusive shock acceleration,
high-energy cutoffs (or ``rollovers'') may reflect changes in the acceleration efficiency, resulting from either the three-dimensionality of the shock front (curvature), limited acceleration time scales, and/or vanishing power in the magnetic field wave spectrum (causing the diffusion coefficient to increase rapidly 
with the heliocentric distance), each contributing to releasing particles from the shock and terminating acceleration \citep{ref:ELLISON_RAMATY1985,ref:LEERYAN86,ref:LEE2005,ref:TYLKA-LEE2006}. 
Also, since both the shock speed and the magnetic field strength decrease with increasing heliocentric distance, the maximum acceleration energy decreases as the shock propagates out into the interplanetary space, thus more energetic ions are typically accelerated at earlier times when the shock is closer to the Sun, although some particles can remain trapped behind the shock, only escaping and propagating upstream at later times \citep{ref:ZANK2000}.
Some studies have argued that SEP spectral breaks occurring at $\sim$30 MeV/n are indicative of the limits of shock acceleration (see e.g. \citet{ref:DESAI2016} and references therein),
although interplanetary transport may also play a relevant role, producing distinctive features in the SEP spectra measured at 1 au \citep{ref:LILEE2015,ref:ZHAO2016}. In particular, the spectra of GLE events above the breaks were found to be typically harder with respect to non-GLE events \citep{ref:MEWALDT2012}; in addition, a further steepening at higher energies ($\gtrsim$500 MeV) has been suggested through a comparison of spacecraft and NM data (e.g. \citet{ref:DEBRUNNER1988,ref:TYLKADIETRICH2009}). However, until recently, SEP measurements were relegated to $\lesssim$600 MeV with the exception of ground-based instruments, whose spectral shapes must be modeled -- based on a number of assumptions -- to account for the effects of cosmic ray (CR) interactions within the terrestrial magnetosphere and atmosphere, and for which there is no compositional information.

Thanks to its unique observational capabilities, the Payload for Antimatter Matter Exploration and Light-nuclei Astrophysics (PAMELA) mission provides accurate and detailed SEP measurements in a wide energy range, bridging the gap between the low-energy observations of in-situ space-based instruments and GLE data from the worldwide network of NMs. In particular, PAMELA can detect, for the first time, with good sensitivity, the rollover in the high-energy spectra predicted by diffusive shock acceleration theory, enabling a more complete and clearer view of the SEP origin and transport.

\section{PAMELA observations}
PAMELA is a space-borne experiment designed for the precise measurement of charged CRs -- protons, electrons, their antiparticles, and light nuclei -- in the kinetic energy interval from several tens of MeV up to several hundreds of GeV \citep{ref:PHYSICSREPORTS,ref:NUOVOCIMENTO}.
The instrument consists of a magnetic spectrometer equipped with a silicon tracking system, a time-of-flight system shielded by an anticoincidence system, an electromagnetic calorimeter and a neutron detector. The Resurs-DK1 satellite, which hosts the apparatus, was launched into a semi-polar (70 deg inclination) and elliptical (350--610 km altitude) orbit on 2006 June 15; in 2010 it was moved to an approximately circular orbit at an altitude of $\sim$580 km.
It operated up until the loss of contact in 2016 January. 
PAMELA made a comprehensive survey of the interplanetary and magnetospheric radiation in the near-Earth environment (see e.g. \citet{ref:SOLARMOD,ref:TRAPPED,ref:ALBEDO,ref:GSTORM}).
In particular, PAMELA made measurements of SEP events in solar cycles 23 and 24, including spectral, compositional, and angular observations \citep{ref:SEP2006,ref:MAY17PAPER}.

\subsection{Data analysis}\label{Data analysis}
Proton intensities are evaluated with a 48-minute time resolution,
corresponding to spacecraft semi-orbits. However, due to the shielding effect of Earth's magnetosphere, low-rigidity (momentum / charge) interplanetary CRs can be registered only when the satellite passes through relatively high magnetic latitude regions, 
so the effective ``duty cycle'' is higher for higher rigidity particles.
To discard trapped/albedo particles and avoid magnetospheric effects \citep{ref:BRUNO_HAWAII}, interplanetary CR fluxes are conservatively estimated by selecting protons with a rigidity 1.3 times higher than the local St\"ormer vertical cutoff, based on the International Geomagnetic Reference Field (IGRF) model \citep{ref:IGRF11}, thereby avoiding the variable and high gradient penumbral region. Details about apparatus performance, proton selection, detector efficiencies and experimental uncertainties can be found in \citet{ref:SOLARMOD,ref:PHYSICSREPORTS,ref:MARTUCCI2018}.

The removal of the background due to galactic CRs (hereafter GCRs) is a delicate aspect of the SEP spectra assessment. To account for short-time variations in the GCR intensities related to solar activity, including Forbush decrease effects \citep{ref:USOSKINPAMELA},
the time dependent GCR component is computed for each semi-orbit, by extrapolating to lower energies the fit of the measured spectrum performed above the maximum SEP energy up to 100 GeV, based on the force-field model \citep{ref:FORCE-FIELD}:
\begin{equation}\label{eq:force-field}
F_{gcr}(E) = F_{LIS}(E + \phi)\times\left[\frac{E\times(E + 2m_{p})}{(E + \phi)\times(E + \phi + 2m_{p})}\right],
\end{equation}
where $m_{p}$ is the proton mass. This function describes the shape of the GCR spectrum with a single (time-dependent) parameter: the modulation potential $\phi$. The parameterization by \citet{ref:POTGIETER2014}, normalized to PAMELA data, is used for the local interstellar spectrum $F_{LIS}$.

Pitch-angle anisotropies with respect to the local interplanetary
magnetic field (IMF) direction are accounted for by estimating the instrument ``asymptotic'' exposure along the satellite orbit, based on an accurate trajectory tracing analysis implementing a realistic description of the Earth's magnetosphere; details about the developed methodology can be found in \citet{ref:BRUNO_JPCS}. A comprehensive investigation of the angular distributions extended to all the SEP events observed by PAMELA will be the object of forthcoming publications.

SEP energy spectra are evaluated in 22 logarithmic bins spanning the energy range from $\sim$80 MeV to $\sim$3 GeV. The mean energies are computed according to \citet{ref:LAFFERTYWYATT}, assuming a power law spectrum:
\begin{equation}\label{LAFFERTYWYATT_Eq}
E_{mean}=\left[\frac{E_{max}^{1-\gamma}-E_{min}^{1-\gamma}}{(E_{max}-E_{min})\times(1-\gamma)}\right]^{-\frac{1}{\gamma}}
\end{equation}
where $\gamma$ is the spectral index, and $E_{max}$ and $E_{min}$ are the upper and the lower energy limits of the considered bin. A $\gamma$ = 3 spectrum is used, although $E_{mean}$ is insensitive to $\gamma$ ($\Delta E_{mean}\lesssim$0.3\% for 1$<\gamma<$6) due to the relatively small bin widths.

The statistical uncertainties on measured SEP spectra are calculated by accounting for the GCR background subtraction, by using 68.27\% confidence level intervals for a Poisson signal $F_{tot}$ in presence of a background $F_{gcr}$ \citep{ref:FELDMANCOUSINS}. Total systematic errors, accounting for uncertainties on selection efficiencies, background subtraction and other corrections, are estimated to be $\sim$ 20\%.

Event fluences are evaluated using the flux intensities $F_{sep,i}(E)$ from the various semi-orbits that register a signal during the SEP event duration interval $T$:
\begin{equation}\label{eq:fluences}
\Phi_{sep}(E) = \int_{T}F_{sep}(E)dt \simeq \sum_{i=1}^{n} \left[F_{sep,i}(E)\times\Delta t_{i}\right] = \frac{T}{n}\times\sum_{i=1}^{n} F_{sep,i}(E),
\end{equation}
where $n$ is the number of time
bins with width $\Delta t_{i}$=$T/n$ determined  by the 48-minute data time resolution.
The integration interval is computed by identifying the event start/stop bins in the flux temporal profiles. When a new event commences while a preceding one was still in progress, the onset time of the second event is set as the end time of the first event. Consequently, the spectrum for the second event will include a contribution from the decay of the previous event.

As mentioned, the PAMELA duty cycle relative to the orbital period increases with growing particle rigidity due to geomagnetic effects; it also varies with the geographic longitude as a consequence of the asymmetries between the terrestrial rotational and magnetic axes \citep{ref:ALBEDO}. In particular, for semi-orbits far away from the Earth's magnetic poles, the minimum effective geomagnetic cutoff can be higher than the PAMELA threshold, so the SEP intensity information can be missing for the lowest energy bins. Data gaps are corrected by means of interpolation algorithms; results are cross-checked with the comparison with the GOES proton fluxes calibrated using the PAMELA SEP data according to \citet{ref:BRUNO_GOES}.

Accounting for a possible rollover in the high-energy SEP spectra, event-integrated fluences are fitted by using a functional form based on \citet{ref:ELLISON_RAMATY1985} (hereafter referred as E-R), consisting of a power law spectrum modulated by an exponential: 
\begin{equation}\label{eq:E-R_function}
\Phi_{sep}(E) = A \times \left(E/E_{s}\right)^{-\gamma} \times e^{-E/E_{0}},
\end{equation}
where $A$ is the normalization, $\gamma$ is the spectral index, and $E_{0}$ is the cutoff or rollover energy; the scaling energy $E_{s}$ is fixed to the PAMELA energy threshold (80 MeV). For well-connected high-energy events, the release point can be computed to be a few solar radii \citep{ref:KAHLER1994}. If we separate the shock acceleration at these distances from the transport of those particles to Earth, we can interpret the power law in terms of the compression ratio of the shock, while the cutoff energy can be interpreted in terms of the limits of the acceleration process, with the intervening transport to Earth giving rise to the late phase isotropy and the extended duration. Contrary to power law functions, Equation \ref{eq:E-R_function} does not extend the spectrum to infinite energies and is consistent with the idea that shock acceleration is limited in time and space (see Section \ref{Introduction}).
Aside from theoretical motivations, it describes the SEP spectra with a reduced set of parameters when compared to other functional forms (e.g. the double power law Band function \citep{ref:BAND1993,ref:TYLKADIETRICH2009}), thus minimizing parameter cross-correlations.

For a comparison, the spectra are also fitted by using a simple power law function, and an $F$-test is performed to compare the two fitting models providing a quantitative estimate of the model that best fits the data. The $F$-statistic is given by the ratio between the corresponding reduced chi-squared: $F = \tilde{\chi}^{2}_{PL} / \tilde{\chi}^2_{ER}$. The associated $p$-value is used to support or reject the null hypothesis (the power law model):
a small numerical value ($p$ $\ll$ 1) implies a very significant rejection.

The same fitting procedure is applied to the assessment of the peak spectra.
For each energy bin, the peak intensity is evaluated by taking the most intense flux value registered during the SEP event.
In general, with respect to fluences, peak spectra are characterized by much larger uncertainties since they rely on single semi-orbit (48-minute) data.

\begin{table}
\centering\footnotesize
\begin{tabular}{c|c|c|c|c|c|c|c}
& SEP Event & \multicolumn{3}{c}{Flare} & \multicolumn{3}{|c}{CME}\\
\hline
\# & Onset time & Onset time & Class & Location & 1$^{st}$-app. time & V$_{sky}$ & Width\\
\hline
1 & 2006 12/05, 15:00 & 12/05, 10:19 & X9.0 & S06E79 & \nodata & \nodata & \nodata \\
2 & 2006 12/06, 23:15 & 12/06, 18:29 & X6.5 & S05E64 & 12/06, 20:12 & \nodata & H\\
3 & 2006 12/13, 02:55 & 12/13, 02:14 & X3.4 & S06W23 & 12/13, 02:54 & 1774 & H\\
4 & 2006 12/14, 22:55 & 12/14, 21:07 & X1.5 & S06W46 & 12/14, 22:30 & 1042 & H\\
5 & 2011 03/21, 03:30 & 03/21, 02:11\tablenotemark{a} & $\lesssim$X1.3\tablenotemark{a} & N23W129\tablenotemark{a} & 03/21, 02:24 & 1341 & H\\
6 & 2011 06/07, 07:00 & 06/07, 06:16 & M2.5 & S21W54 & 06/07, 06:49 & 1255 & H\\
7 & 2011 09/06, 02:30 & 09/06, 01:35 & M5.3 & N14W07 & 09/06, 02:24 & 782 & H\\
8 & 2011 09/06, 23:35 & 09/06, 22:12 & X2.1 & N14W18 & 09/06, 23:05 & 575 & H\\
9 & 2011 11/04, 00:15 & 11/03, 22:45\tablenotemark{b} & $\lesssim$X1.4\tablenotemark{b} & N09E154\tablenotemark{b} & 11/03, 23:30 & 991 & H\\
10 & 2012 01/23, 04:20 & 01/23, 03:38 & M8.7 & N28W21 & 01/23, 04:00 & 2175 & H\\
11 & 2012 01/27, 18:40 & 01/27, 17:37 & X1.7 & N27W71 & 01/27, 18:27 & 2508 & H\\
12 & 2012 03/07, 01:40 & 03/07, 00:02 & X5.4 & N17E27 & 03/07, 00:24 & 2684 & H\\
13 & 2012 03/13, 17:50 & 03/13, 17:12 & M7.9 & N17W66 & 03/13, 17:36 & 1884 & H\\
14 & 2012 05/17, 01:50 & 05/17, 01:25 & M5.1 & N11W76 & 05/17, 01:48 & 1582 & H\\
15 & 2012 07/07, 00:05 & 07/06, 23:01 & X1.1 & S13W59 & 07/06, 23:24 & 1828 & H\\
16 & 2012 07/08, 17:45 & 07/08, 16:23 & M6.9 & S17W74 & 07/08, 16:54 & 1497 & 157\\
17 & 2012 07/12, 17:15 & 07/12, 15:37 & X1.4 & S15W01 & 07/12, 16:48 & 885 & H\\
18 & 2012 07/19, 06:25 & 07/19, 04:17 & M7.7 & S13W88 & 07/19, 05:24 & 1631 & H\\
19 & 2012 07/23, 06:30? & 07/23, 02:31\tablenotemark{c} & $\lesssim$X2.5\tablenotemark{c} & S17W132\tablenotemark{c} & 07/23, 02:36 & 2003 & H\\
20 & 2013 04/11, 08:00 & 04/11, 06:55 & M6.5 & N09E12 & 04/11, 07:24 & 861 & H\\
21 & 2013 05/22, 13:50 & 05/22, 13:08 & M5.0 & N15W70 & 05/22, 13:25 & 1466 & H\\
22 & 2013 09/30, 02:15 & 09/29, 21:43 & C1.3 & N17W29 & 09/29, 22:12 & 1179 & H\\
23 & 2013 10/28, 17:55 & 10/28, 15:07 & M4.4 & S06E28 & 10/28, 15:36 & 812 & H\\
24 & 2013 11/02, 07:25 & 11/02, 04:00 & \nodata & N03W139 & 11/02, 04:48 & 828 & H\\
25 & 2014 01/06, 08:05 & 01/06, 07:30\tablenotemark{d} & $\lesssim$X3.5\tablenotemark{e} & S15W112\tablenotemark{e} & 01/06, 08:00 & 1402 &  H\\
26 & 2014 01/07, 19:20 & 01/07, 18:04 & X1.2 & S15W11 & 01/07, 18:24 & 1830 & H\\
27 & 2014 02/25, 03:00 & 02/25, 00:39 & X4.9 & S12E82 & 02/25, 01:25 & 2147 & H\\
28 & 2014 04/18, 13:30 & 04/18, 12:31 & M7.3 & S20W34 & 04/18, 13:25 & 1203 & H\\
29 & 2014 09/01, 17:00 & 09/01, 10:54\tablenotemark{f} & $\lesssim$X2.4\tablenotemark{e} & N14E127\tablenotemark{e} & 09/01, 11:12 & 1901 & H\\
30 & 2014 09/10, 19:45 & 09/10, 17:21 & X1.6 & N14E02 & 09/10, 18:00 & 1267 & H\\
\hline
\end{tabular}
\tablerefs{
(\textit{a}) \citet{ref:ROUILLARD2012}, 
(\textit{b}) \citet{ref:MEWALDT2013}, 
(\textit{c}) \citet{ref:NITTA2013},\\
(\textit{d}) \citet{ref:THAKUR2014},
(\textit{e}) \citet{ref:ACKERMANN2017},
(\textit{f}) \citet{ref:PLOTNIKOV2017}.} 
\caption{List of the major SEP events observed by PAMELA between 2006 July and 2014 September. For each event, the onset time (UT) and the associated flare onset time/class/location information is reported, along with the parent CME first appearance time, sky-plane velocity (km s$^{-1}$) and angular width (deg, or ``H'' in case of full halo CMEs). See the text for details.}
\label{tab:used_events}
\end{table}

\subsection{Data set}\label{Data set}
Table \ref{tab:used_events} lists the 30 major SEP events detected by PAMELA between 2006 July and 2014 September. 
The first column reports the event number. 
The second column gives the SEP event onset times based on the Geostationary Operational Environmental Satellite (GOES) 5-minute resolution proton fluxes for energies $>$100 MeV (\url{https://umbra.nascom.nasa.gov/sdb/goes/particle/}). 
Columns 3, 4 and 5 display the associated flare onset/class/location data from the GOES X-ray archive (\url{ftp://ftp.ngdc.noaa.gov/STP/space-weather/solar-data/solar-features/solar-flares/x-rays/goes/}).
In the case of events originating on the far side of the Sun, the flare size is estimated from observations made by the Extreme Ultraviolet Imager (EUVI) on board the Solar Terrestrial Relations Observatory (STEREO) spacecraft \citep{ref:ROUILLARD2012,ref:MEWALDT2013,ref:NITTA2013,ref:ACKERMANN2017};
no estimate is available
for the 2013 November 2 event.
Finally, the last three columns report the parent CME first appearance times, sky-plane
speeds and angular widths from the CDAW catalog of the Large Angle and Spectrometric Coronagraph (LASCO) on board the Solar and Heliospheric Observatory (SOHO) (\url{https://cdaw.gsfc.nasa.gov/CME_list/}). 

\begin{figure}[!t]
\centering
\begin{tabular}{cc}
\includegraphics[width=3.4in]{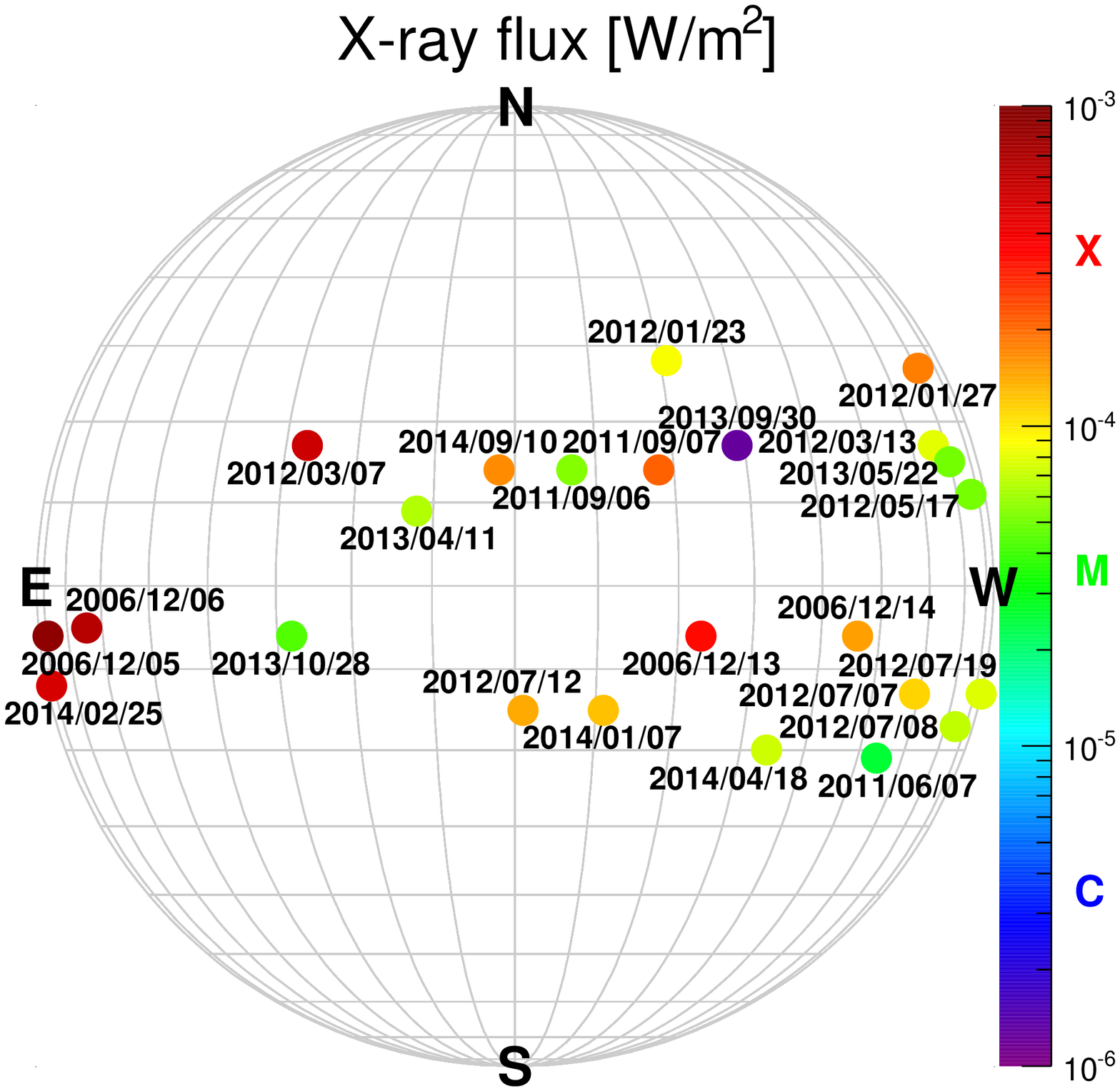} & \includegraphics[width=3.4in]{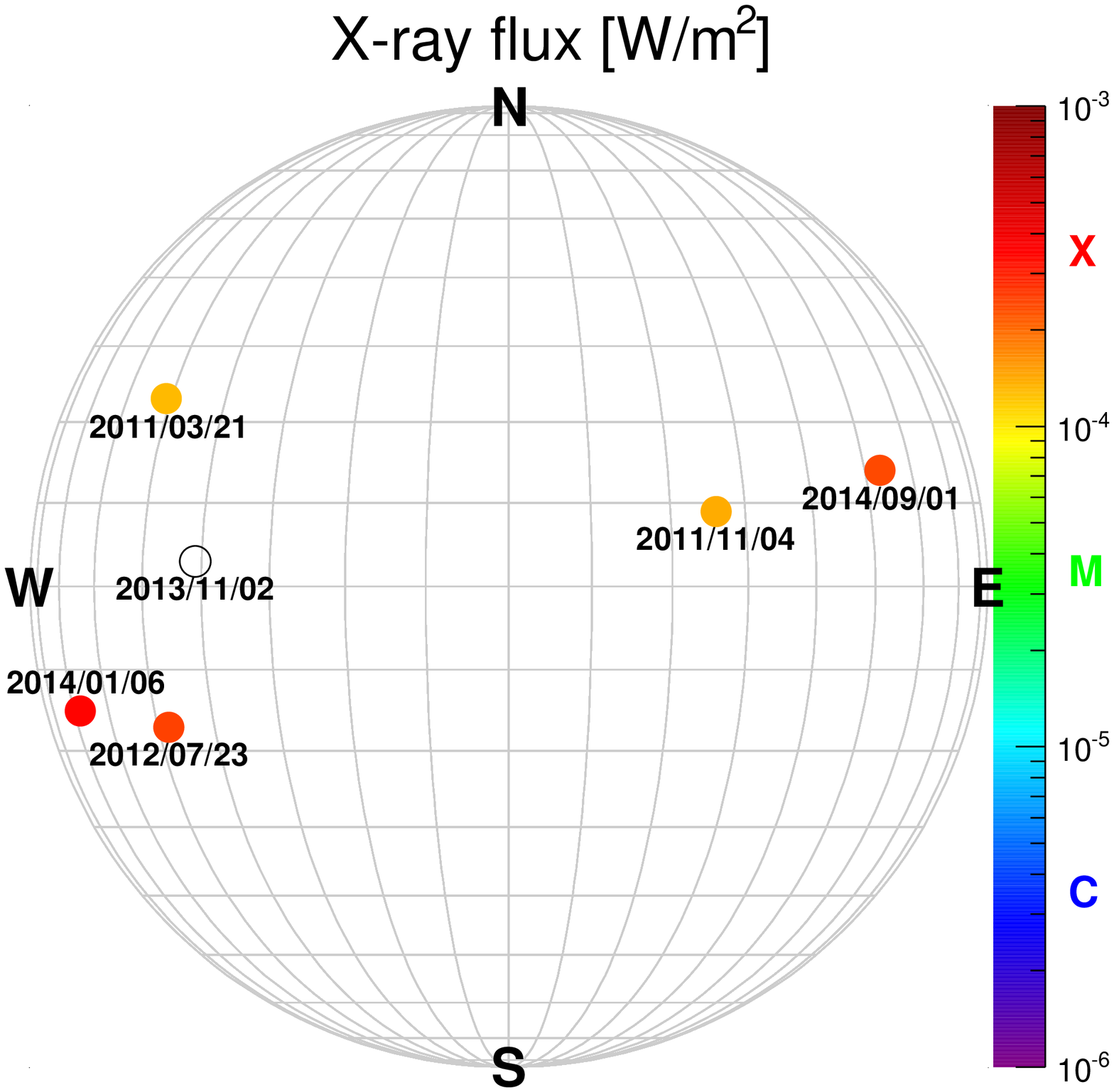} \\
\end{tabular}
\caption{Soft X-ray peak fluxes (color codes) as a function of flare heliographic locations, for the SEP events registered by PAMELA (see Table \ref{tab:used_events}). Front and back side eruptions are reported in the left and the right panels, respectively. Upper limits are provided in case of far side events; the flare size information is missing for the 2013 November 2 event.}
\label{fig:flares_vs_location}
\end{figure}

The heliographic distribution of the associated flares is illustrated in Figure \ref{fig:flares_vs_location}; the color code denotes the soft X-ray peak flux.
Six back side events are included (right panel): 2011 March 21, 2011 November 3, 2012 July 23, 2013 November 2, 2014 January 6 and 2014 September 1.
The front side sample (left panel) consists of 17 events occurring in the western hemisphere and 7 events in the eastern hemisphere.
Apart from the 2013 September 30 event, linked to a C-class flare and a quiescent filament eruption, the PAMELA SEP data set is associated with $\ge$M-class flares, including 17 X-class eruptions, and with full halo CMEs except for the 2012 July 8 event (partial halo CME).
All registered events were generated within $\sim$30 deg from the solar equator, with the largest latitudes for the 2012 January 23 and 27 eruptions (N28 and N27, respectively).

It should be noted that 8 events in the PAMELA list (2006 December 5, 2011 September 6 and 7, 2011 November 4, 2012 July 19, 2013 October 28, 2013 November 2, 2014 September 1) do not meet the NOAA criterion for a SEP event that is based on the 10 sr$^{-1}$s$^{-1}$cm$^{-2}$ flux threshold for protons with kinetic energies above 10 MeV, so they are not reported in the NOAA ``Solar Proton Events Affecting the Earth Environment''
catalog (\url{ftp://ftp.swpc.noaa.gov/pub/indices/SPE.txt}). 

According to convention, a SEP event is counted as a GLE if at least two independent NMs -- including a near sea level station -- have registered a simultaneous statistically significant increase related to the SEP arrival (\url{http://www.nmdb.eu/}). 
Analogously, in this work we classify as ``sub-GLEs'' the events unambiguously detected 
by only one\footnote{A more recent Antarctic NM station, called ``Dome C'' (3233 m altitude), started operation in early 2015 after the events discussed in this study.} NM: the South Pole station (hereafter SPNM) at a 2820 m altitude.
The two GLEs (numbered 70 and 71) that occurred during the operation of PAMELA, on 2006 December 13, event late in cycle 23 and 2012 May 17, near the peak of cycle 24, were both detected by PAMELA. Unfortunately, a large gap in the PAMELA data, related to an on board system reset of the satellite, occurred during the 2006 December 13 event \citep{ref:SEP2006}, so this event is excluded from this analysis.
Similarly, the measurement of a sub-GLE on 2012 March 7
is complicated by issues related to the high count rate, requiring a different analysis approach that will be discussed in a forthcoming paper.
Finally, the 2006 December 5 and 6 events in Table \ref{tab:used_events} were discarded since no data were collected by the PAMELA tracker due to a scheduled maintenance procedure.
Minor limitations affected the events on 2012 July 23 (only first $\sim$13 hours of data taking available) and 2014 February 25 (first $\sim$9 hours missing), for which partial results are provided.

Several
other
interesting events appear in Table \ref{tab:used_events}. The most energetic of them is the 2014 January 6 event, originating behind the western limb \citep{ref:THAKUR2014},
which triggered a sub-GLE with a $\sim$2.5\% increase in the SPNM count rate. 
A smaller sub-GLE
occurred on 2012 January 27 \citep{ref:BELOV2015}, with a $\lesssim$1.5\% SPNM increase.
Two long-duration ($>$6 days) SEP events were generated by the 2014 February 25 eastern limb and the 2014 September 1 back side events \citep{ref:LARIO2016,ref:PLOTNIKOV2017}.
Another far side eruption, the 2012 July 23 event, produced an extreme solar storm -- probably the most powerful recorded since the Carrington event in 1859 -- which only marginally affected the near-Earth environment due to the poor magnetic connection \citep{ref:RUSSEL2013,ref:LIU2014,ref:GOPALSWAMY2016,ref:RILEY2016}.
The 2012 January 23 and the 2013 May 22 events are remarkable since they involved interacting CMEs \citep{ref:JOSHI2013,ref:DING2014,ref:MAKELA2016}.
Finally, despite the far side source regions, the 2014 January 6 and September 1 events were found to be associated with significant $\gamma$-ray emission reported by the Fermi-LAT instrument \citep{ref:ACKERMANN2017}.

\subsection{Results}
\begin{figure}[!t]
\centering
\begin{tabular}{cc}
\includegraphics[width=3.4in]{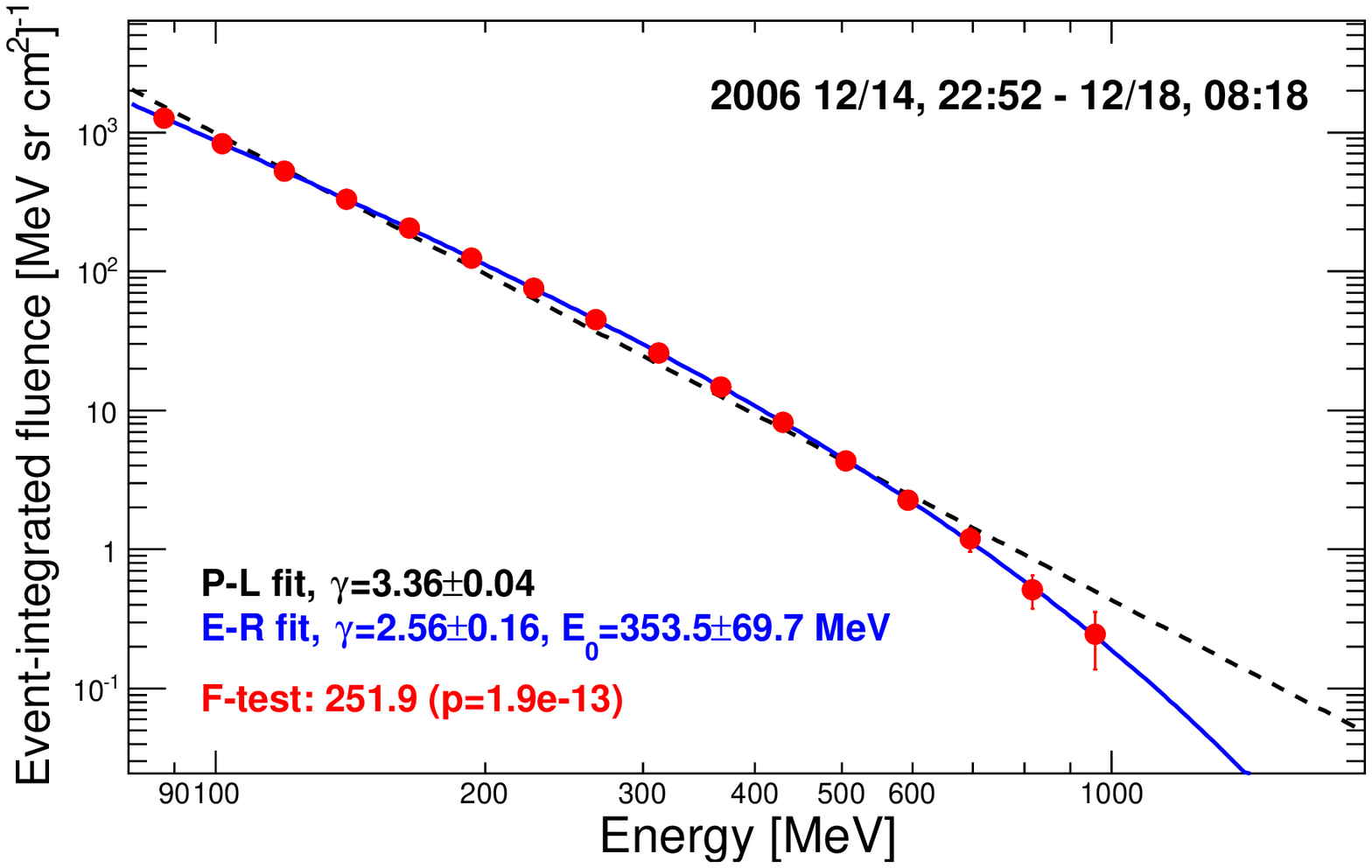} &
\includegraphics[width=3.4in]{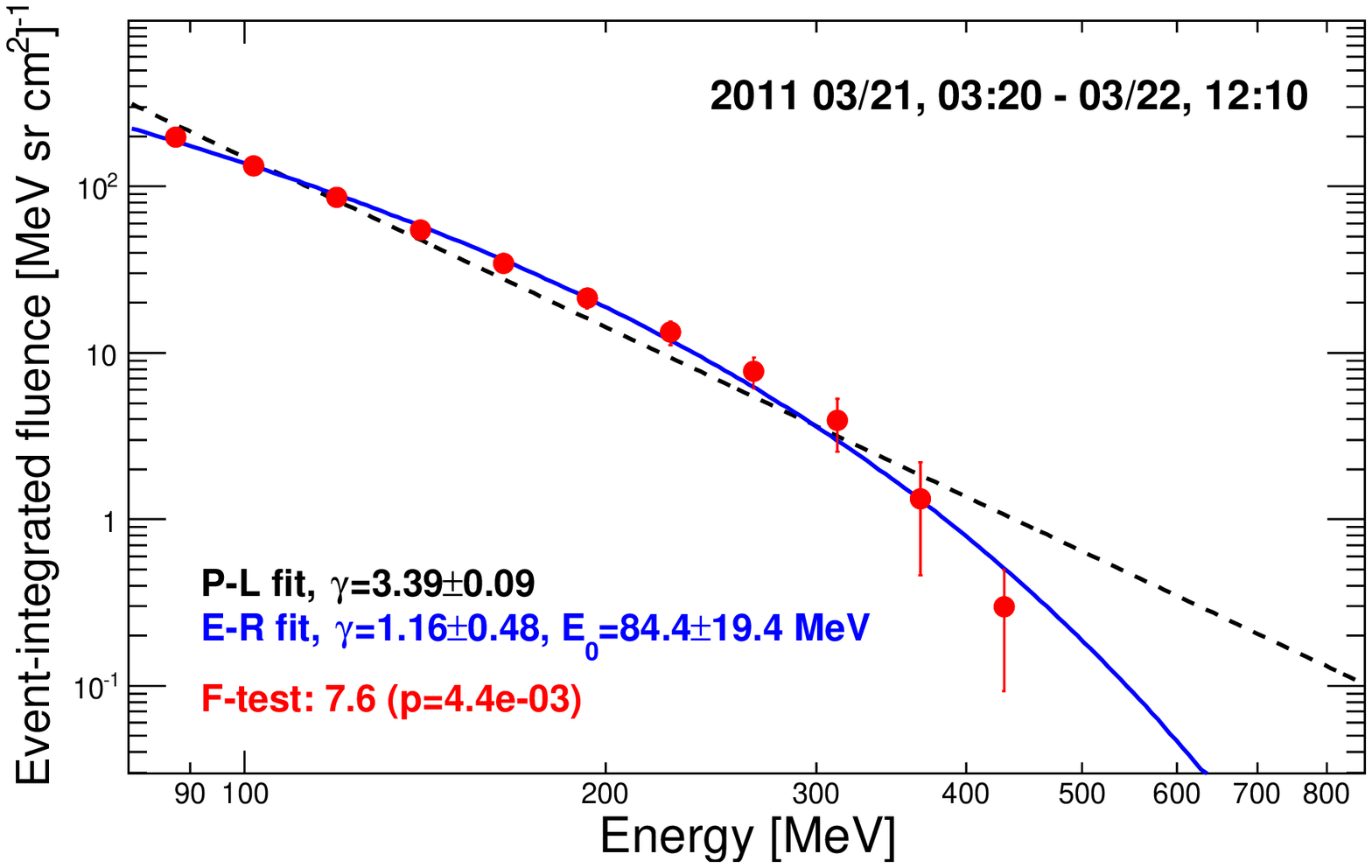} \\
\includegraphics[width=3.4in]{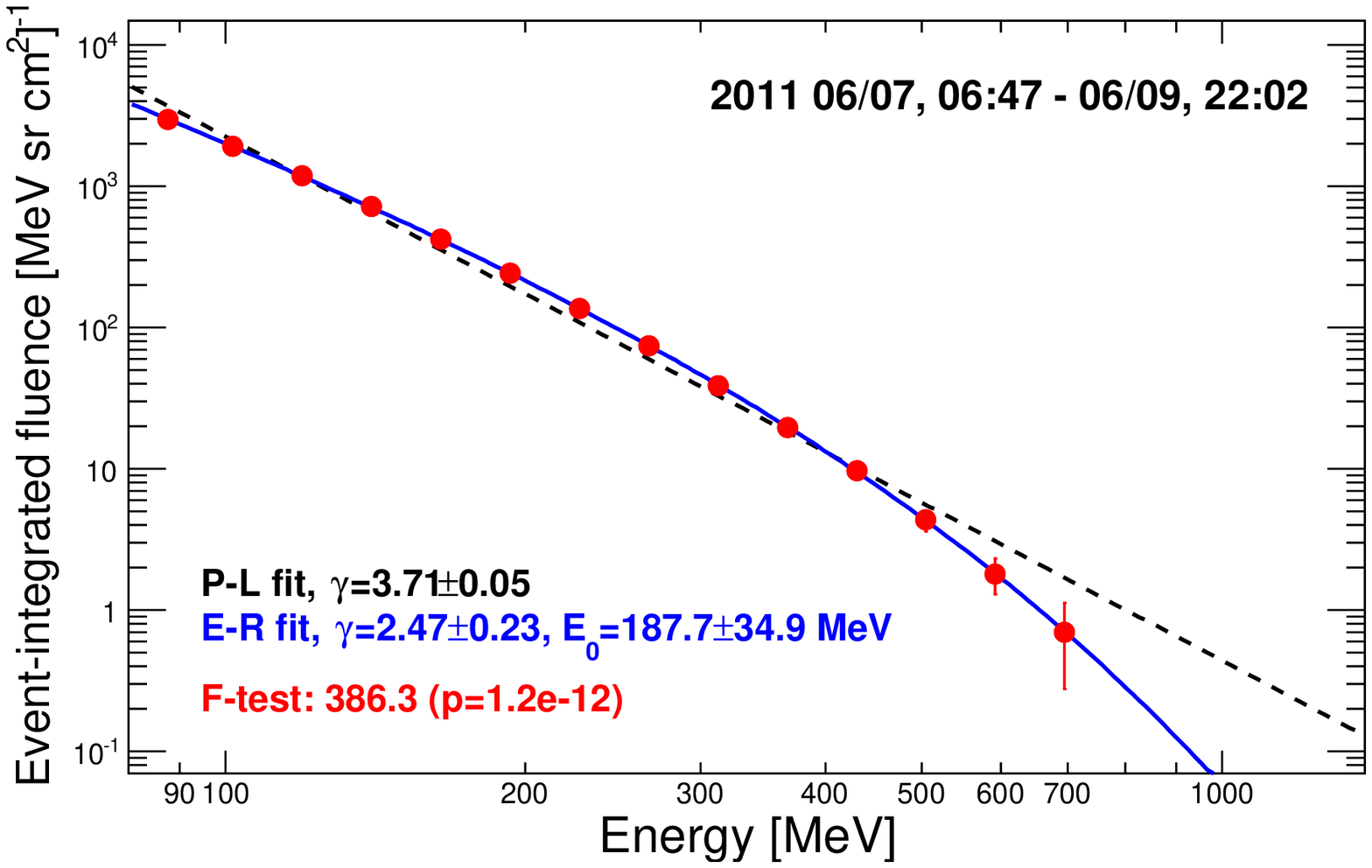} &
\includegraphics[width=3.4in]{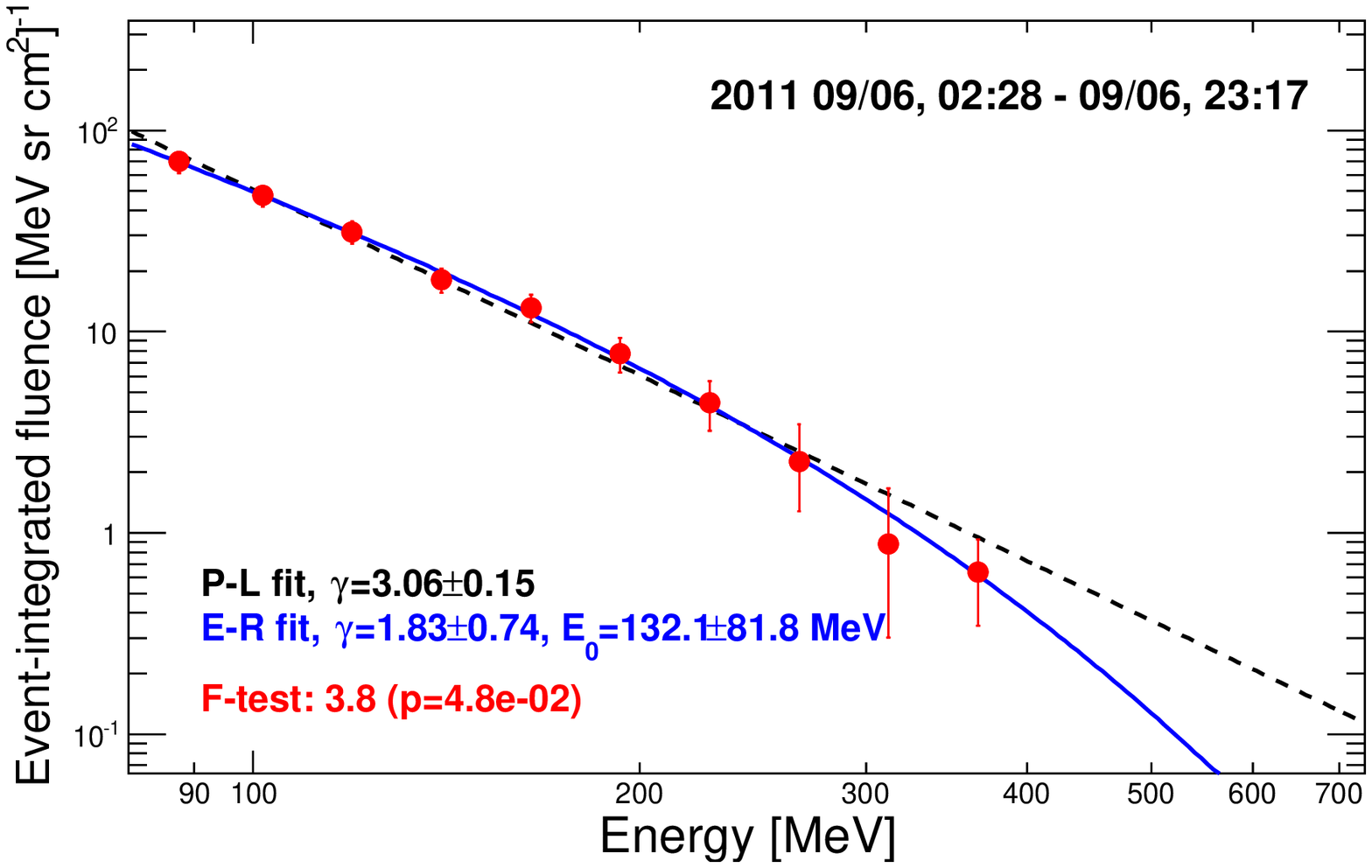} \\
\includegraphics[width=3.4in]{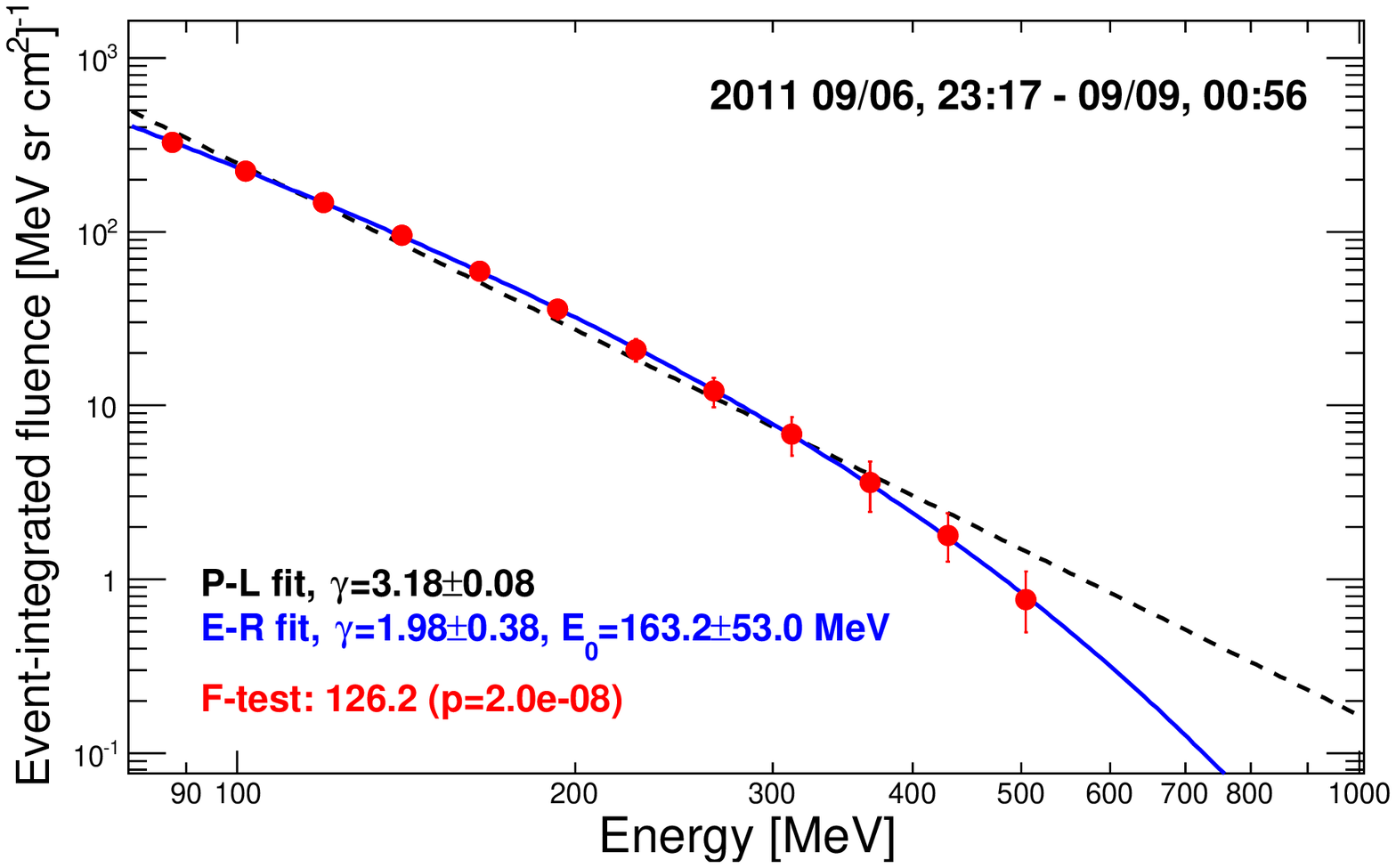} &
\includegraphics[width=3.4in]{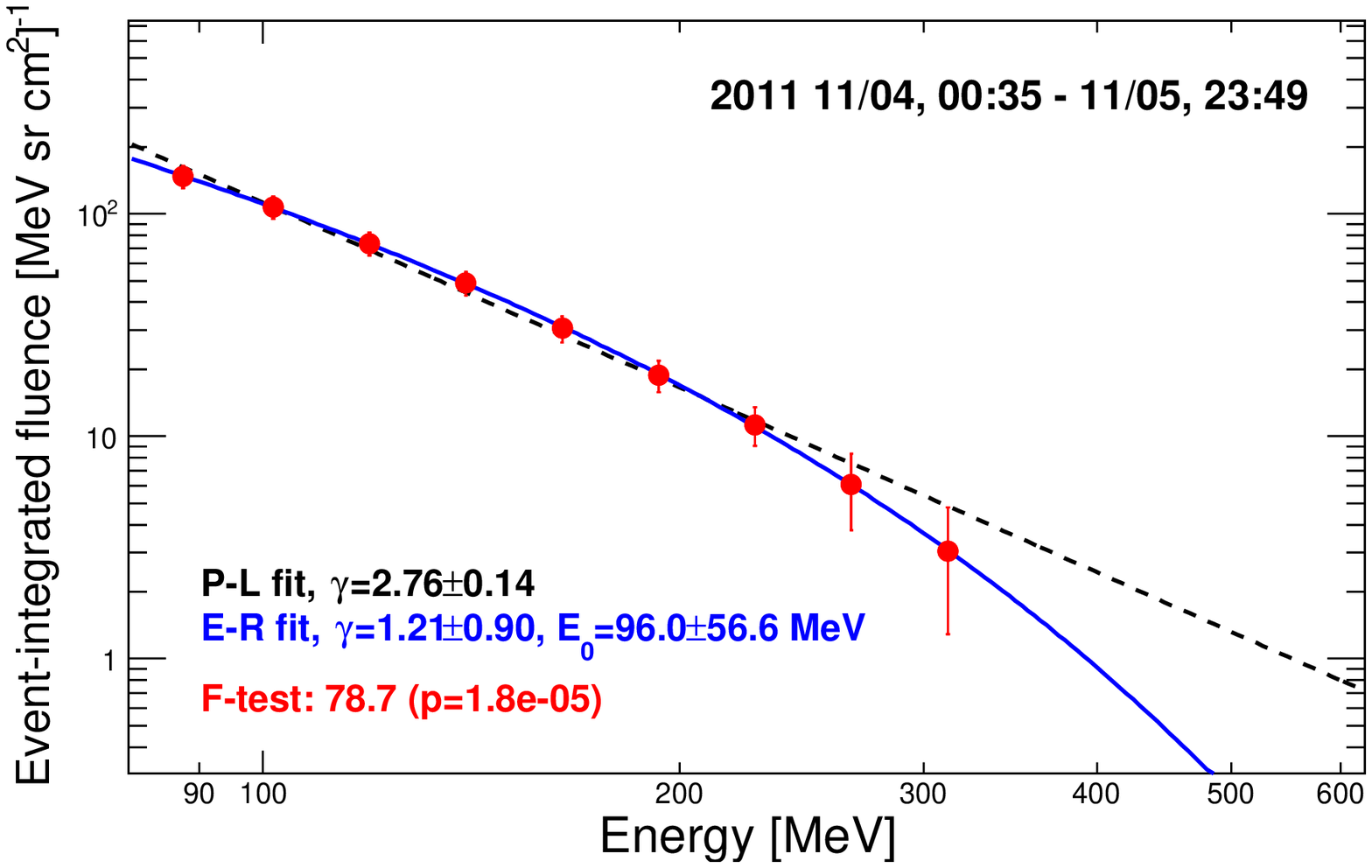} \\
\includegraphics[width=3.4in]{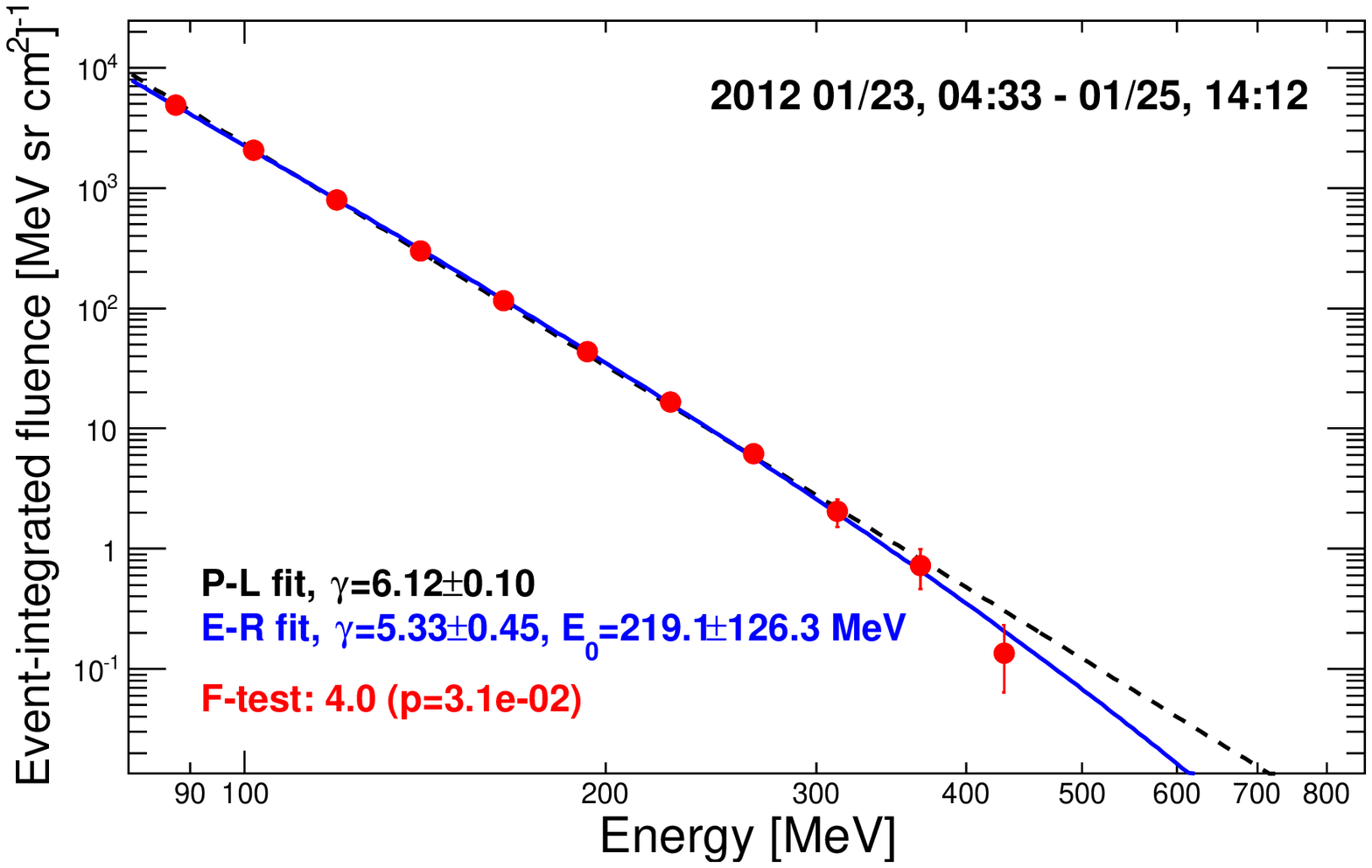} &
\includegraphics[width=3.4in]{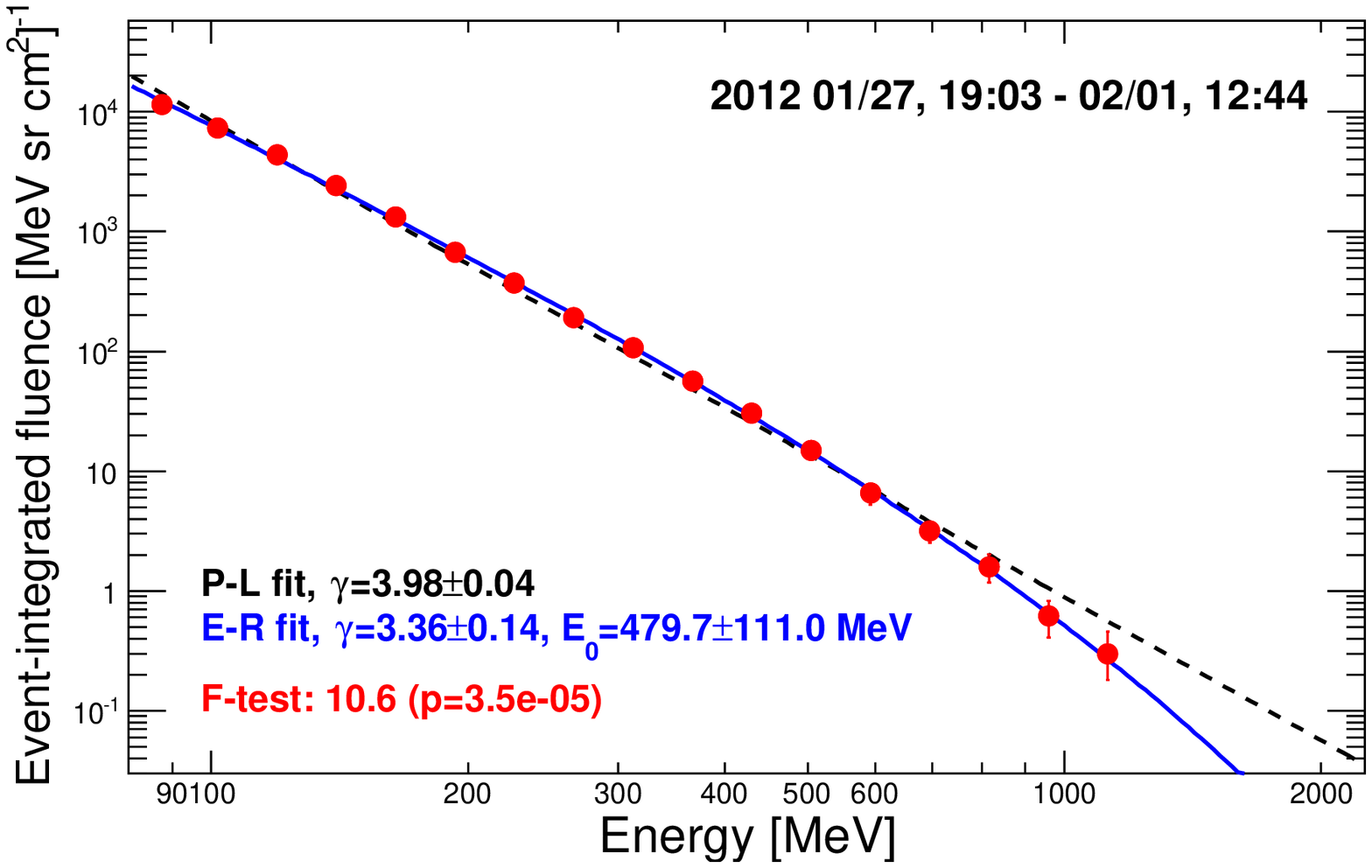} \\
\end{tabular}
\caption{SEP event-integrated fluence spectra measured by PAMELA. The error bars include both statistical and systematic uncertainties. For each event, the start/stop dates (UT) along with the fits with a simple power law (black dashed lines) and the E-R function (blue lines) are reported, including the fit parameters and the $F$-test results (with associated $p$-values).}
\label{fig:fluences_part1}
\end{figure}

\begin{figure}[!t]
\centering
\begin{tabular}{cc}
\includegraphics[width=3.4in]{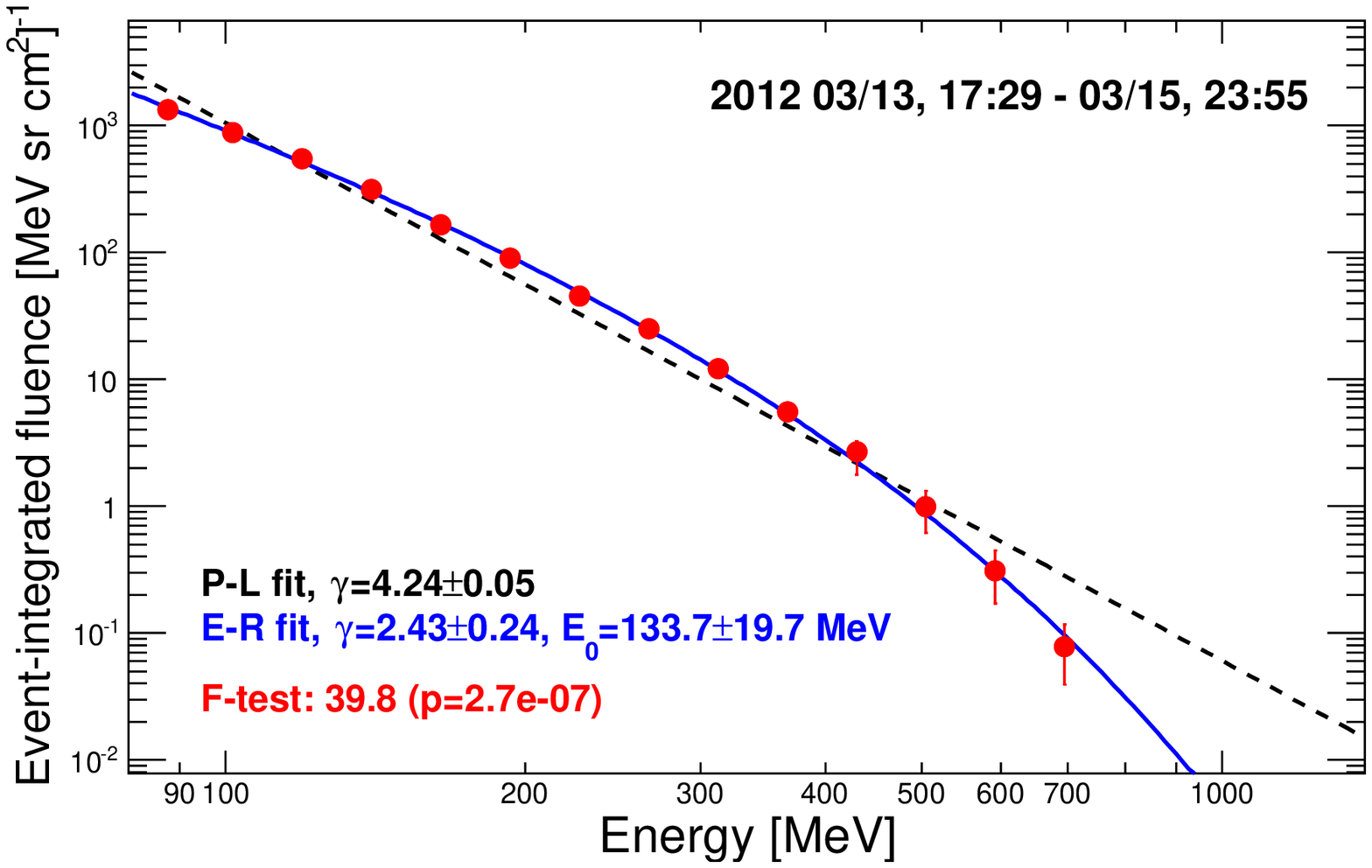} &
\includegraphics[width=3.4in]{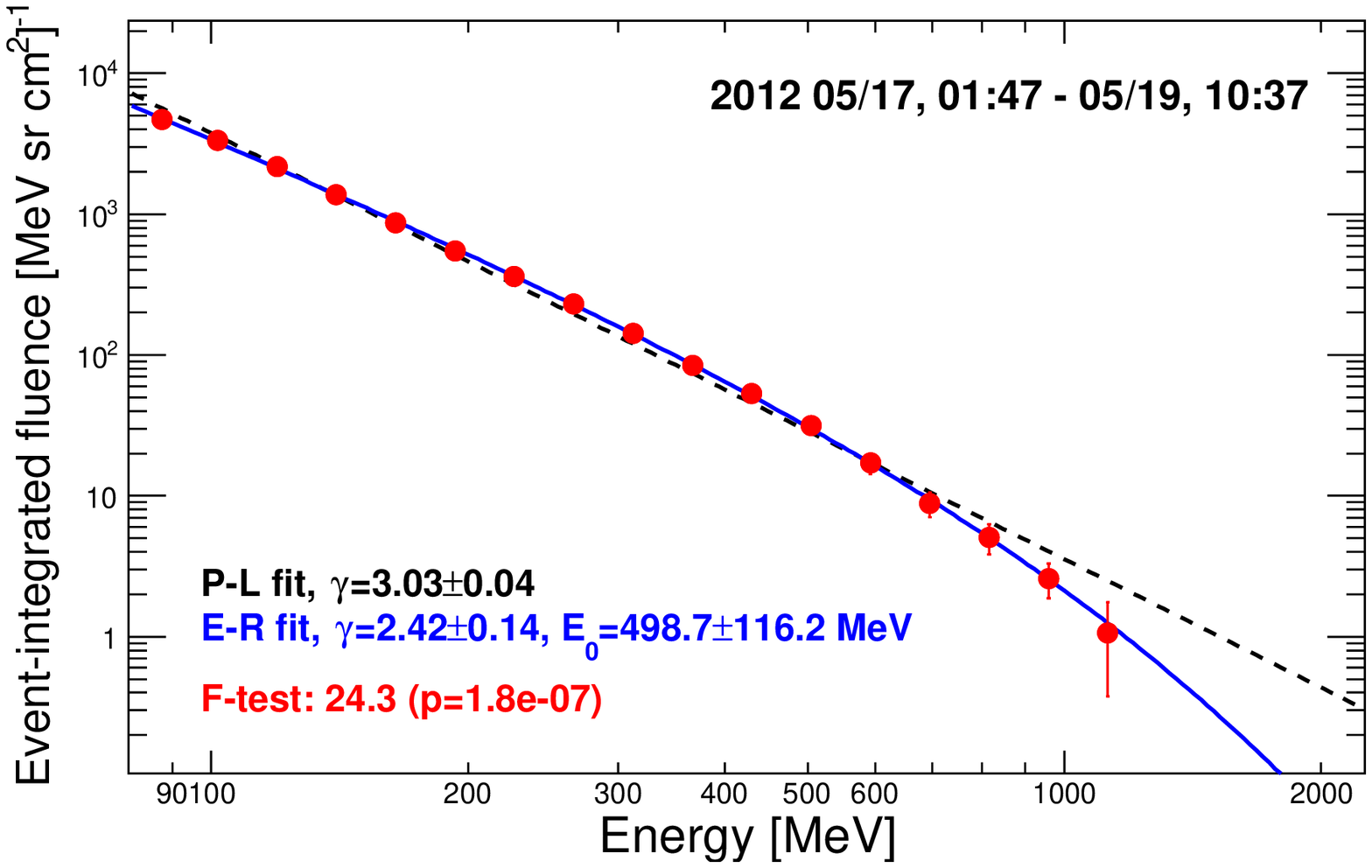} \\
\includegraphics[width=3.4in]{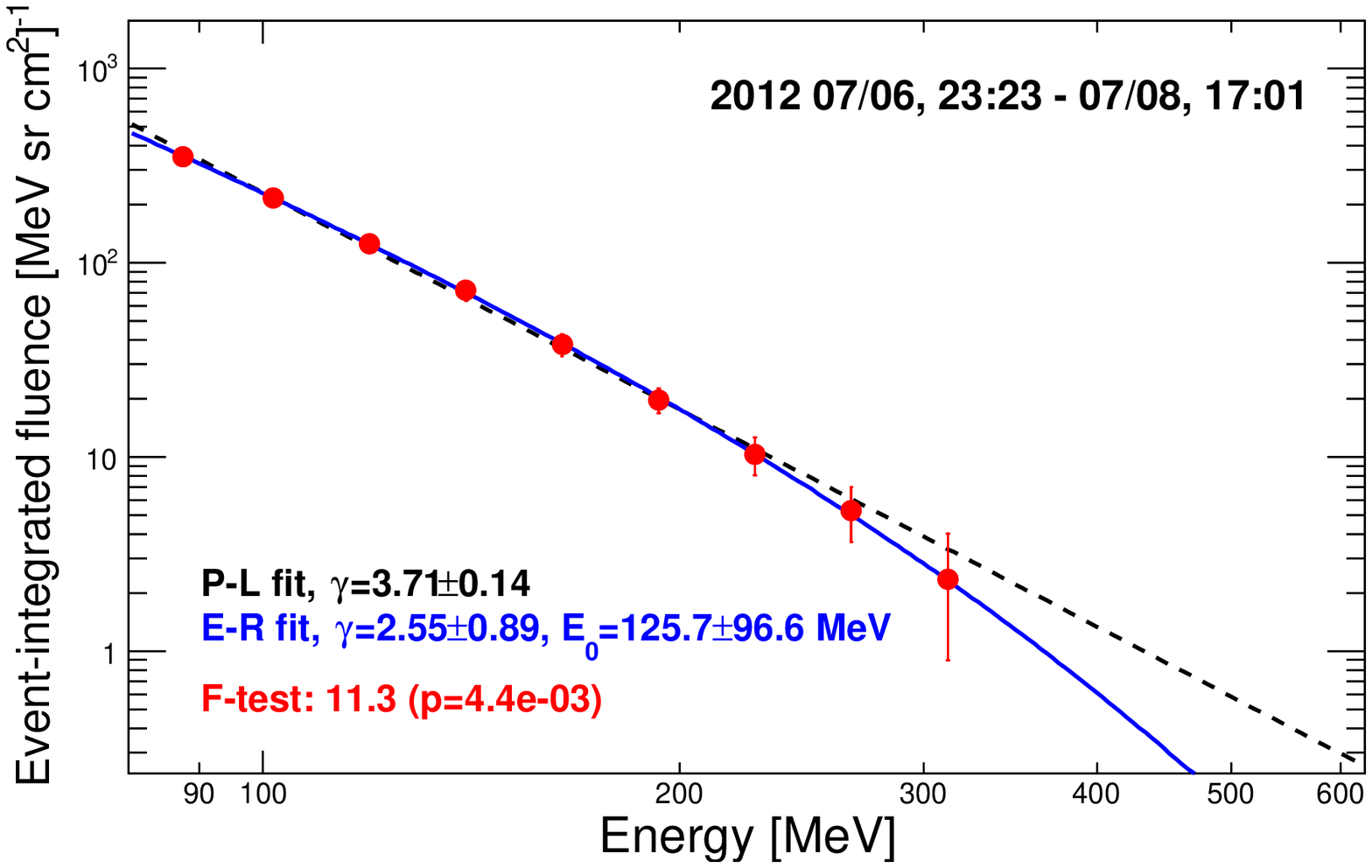} &
\includegraphics[width=3.4in]{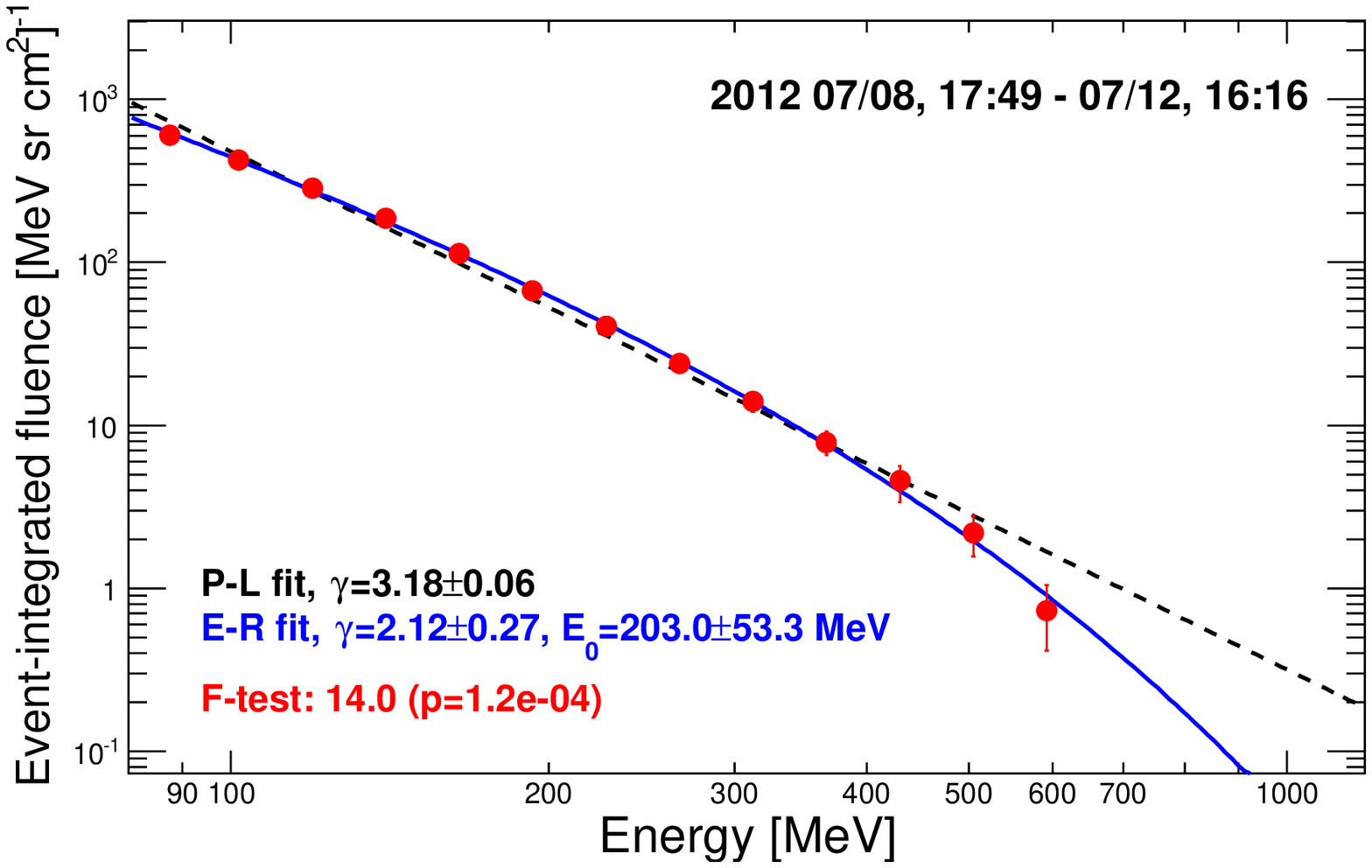} \\
\includegraphics[width=3.4in]{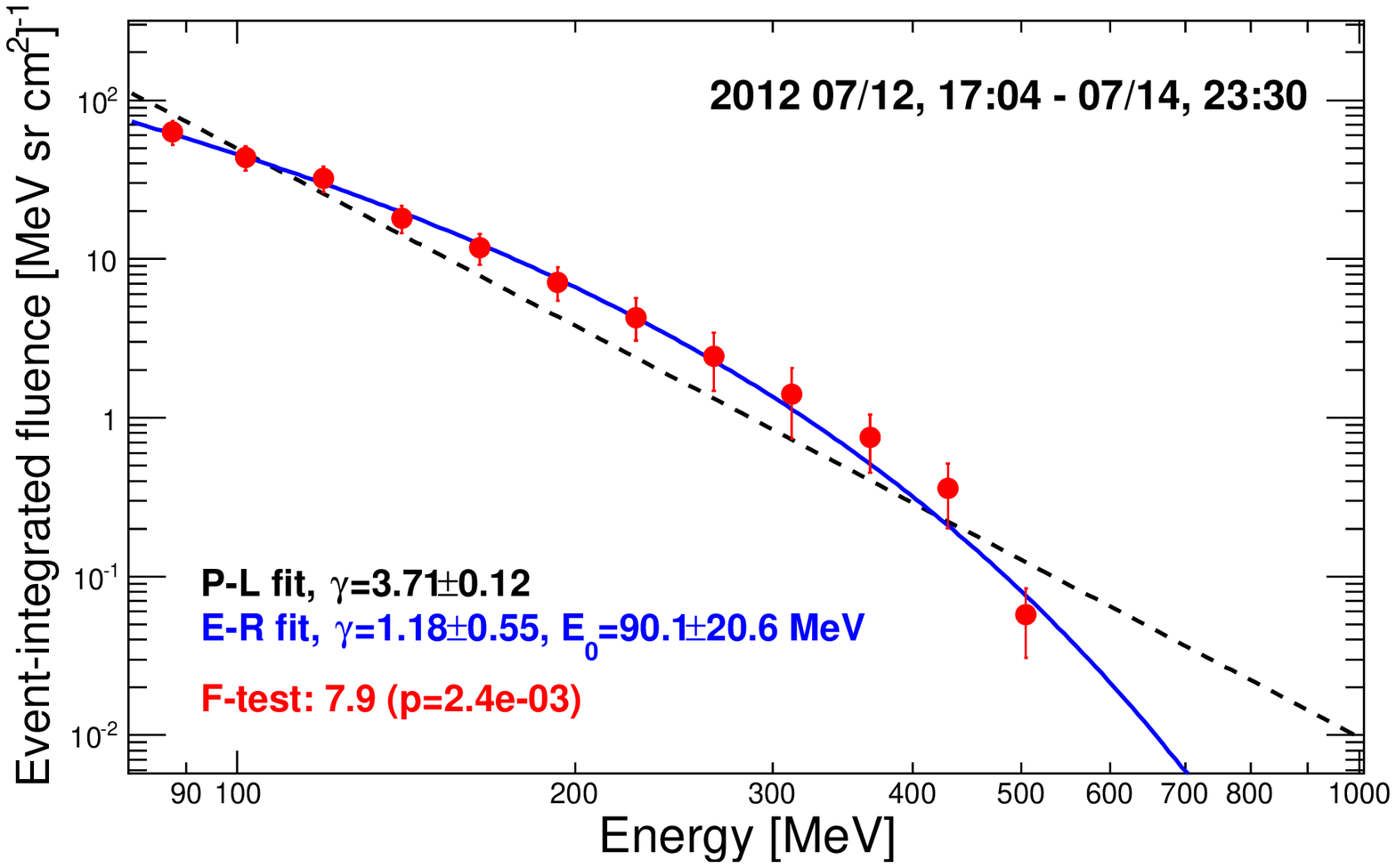} &
\includegraphics[width=3.4in]{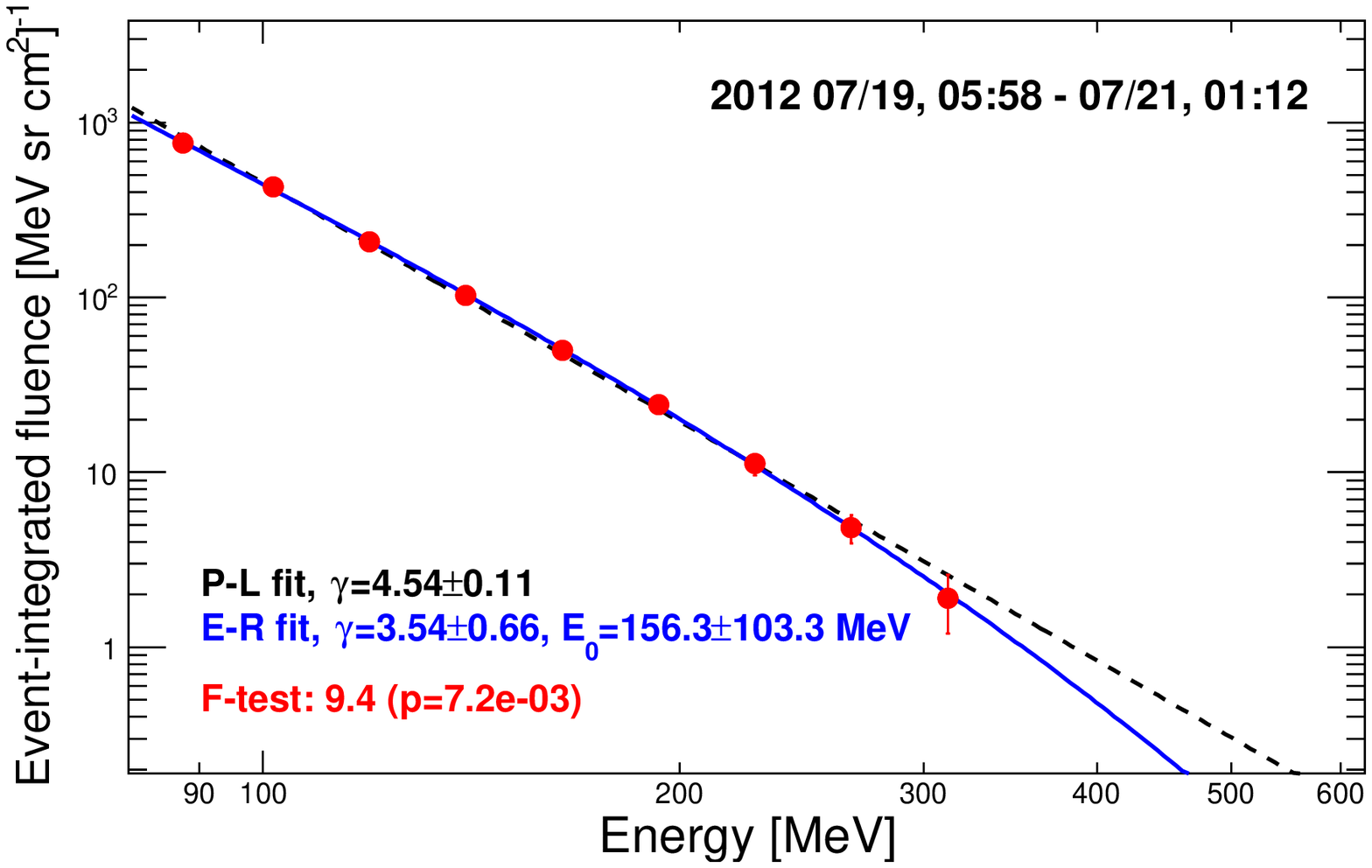} \\
\includegraphics[width=3.4in]{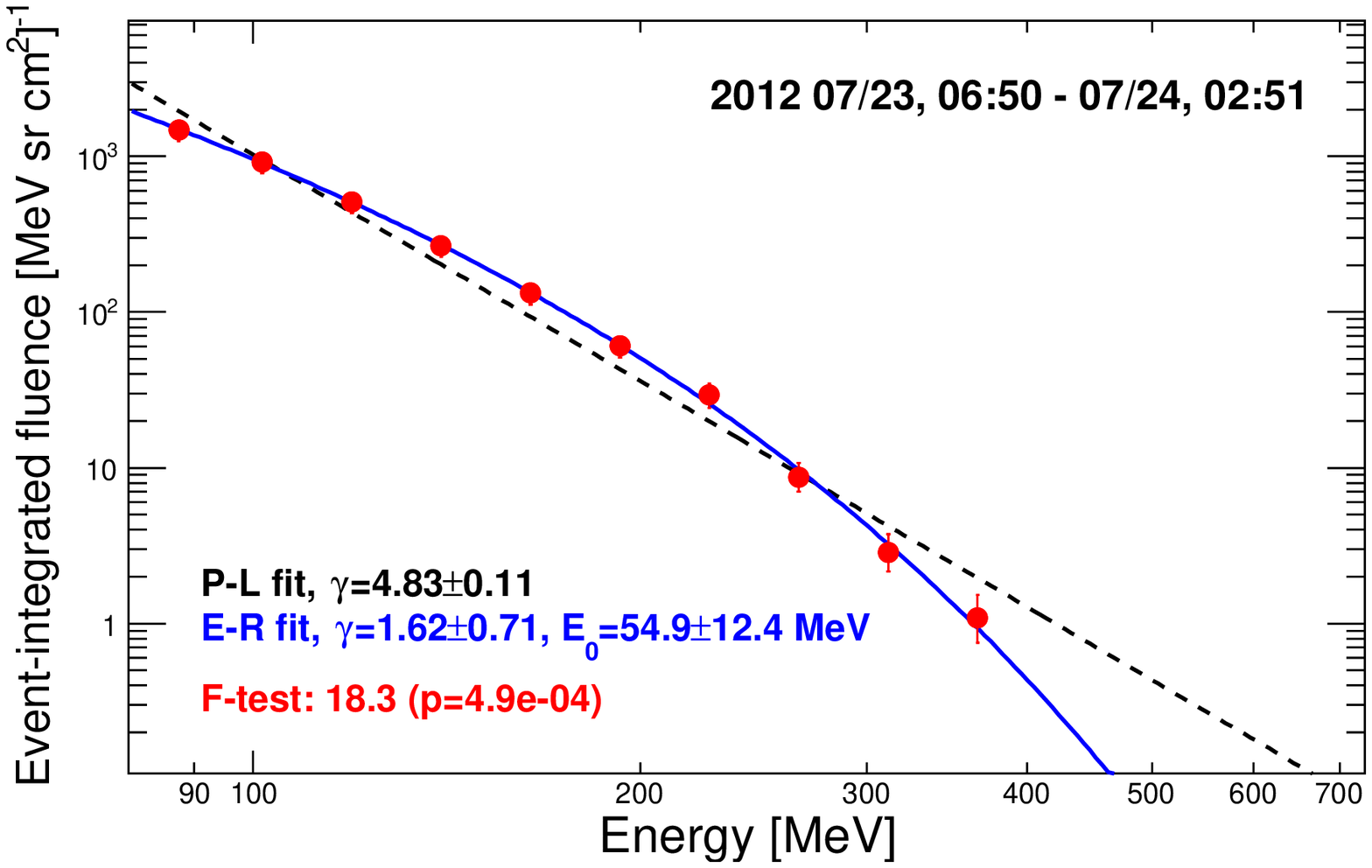} &
\includegraphics[width=3.4in]{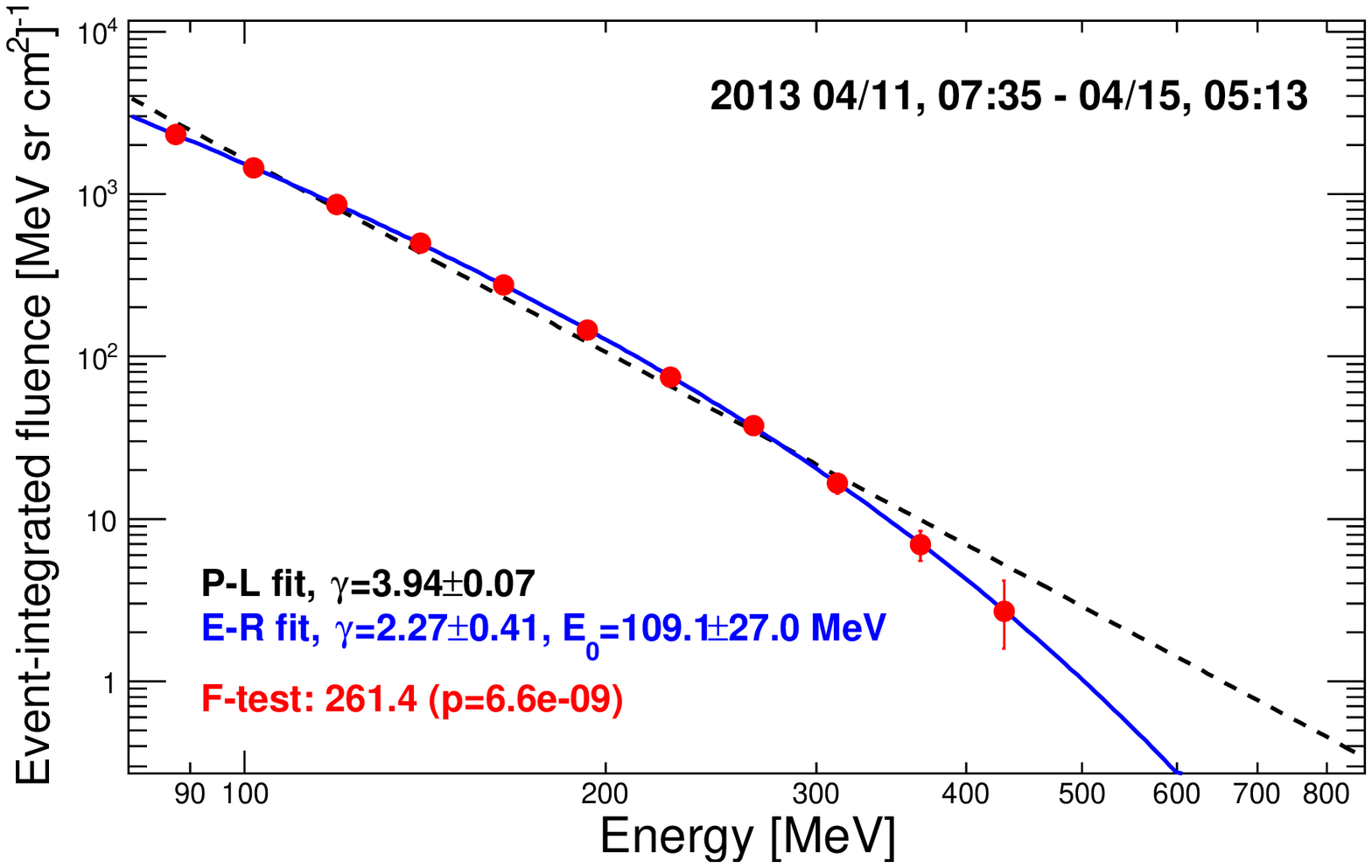} \\
\end{tabular}
\caption{SEP event-integrated fluence spectra measured by PAMELA in the same format as Figure \ref{fig:fluences_part1}.}
\label{fig:fluences_part2}
\end{figure}

\begin{figure}[!t]
\centering
\begin{tabular}{cc}
\includegraphics[width=3.4in]{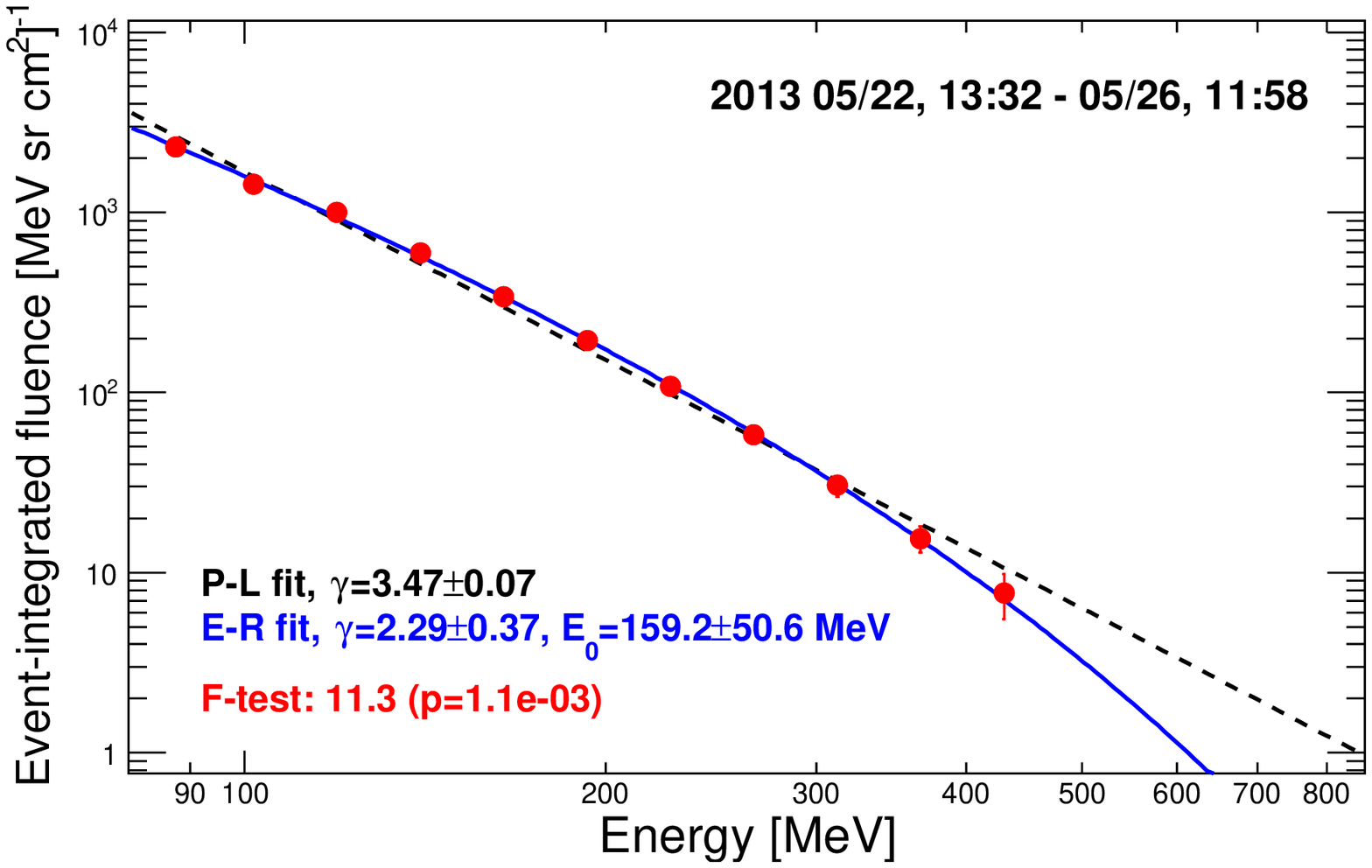} &
\includegraphics[width=3.4in]{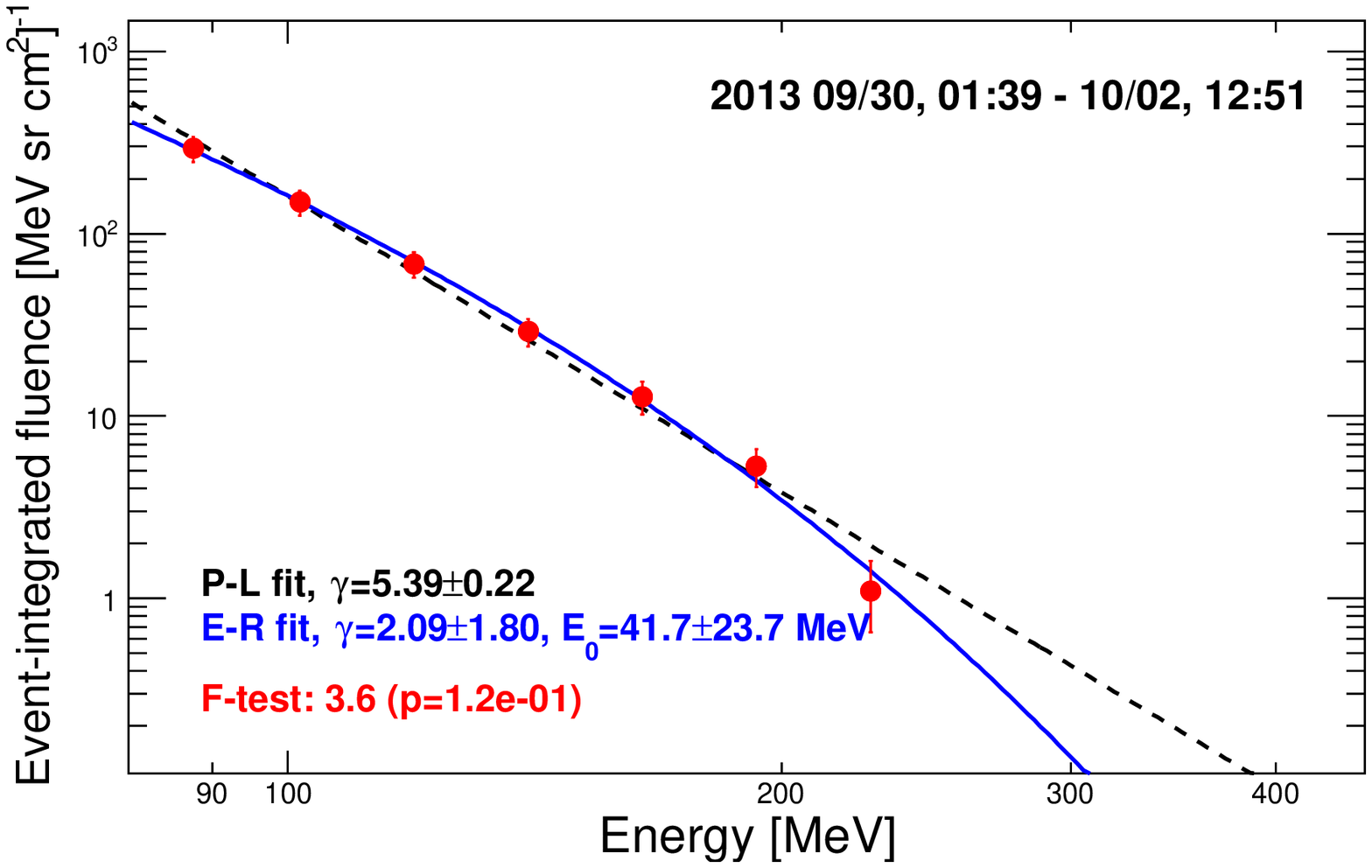} \\
\includegraphics[width=3.4in]{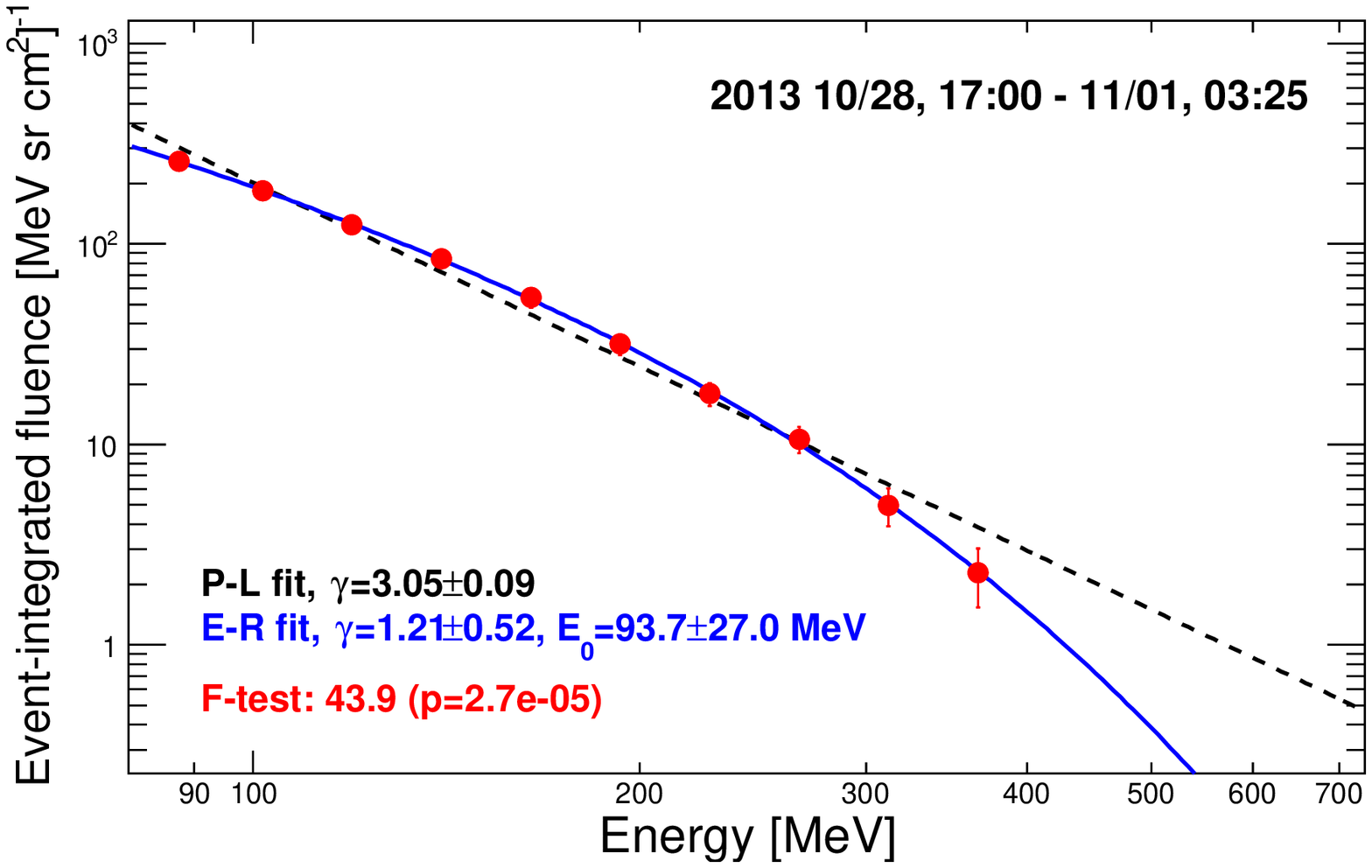} &
\includegraphics[width=3.4in]{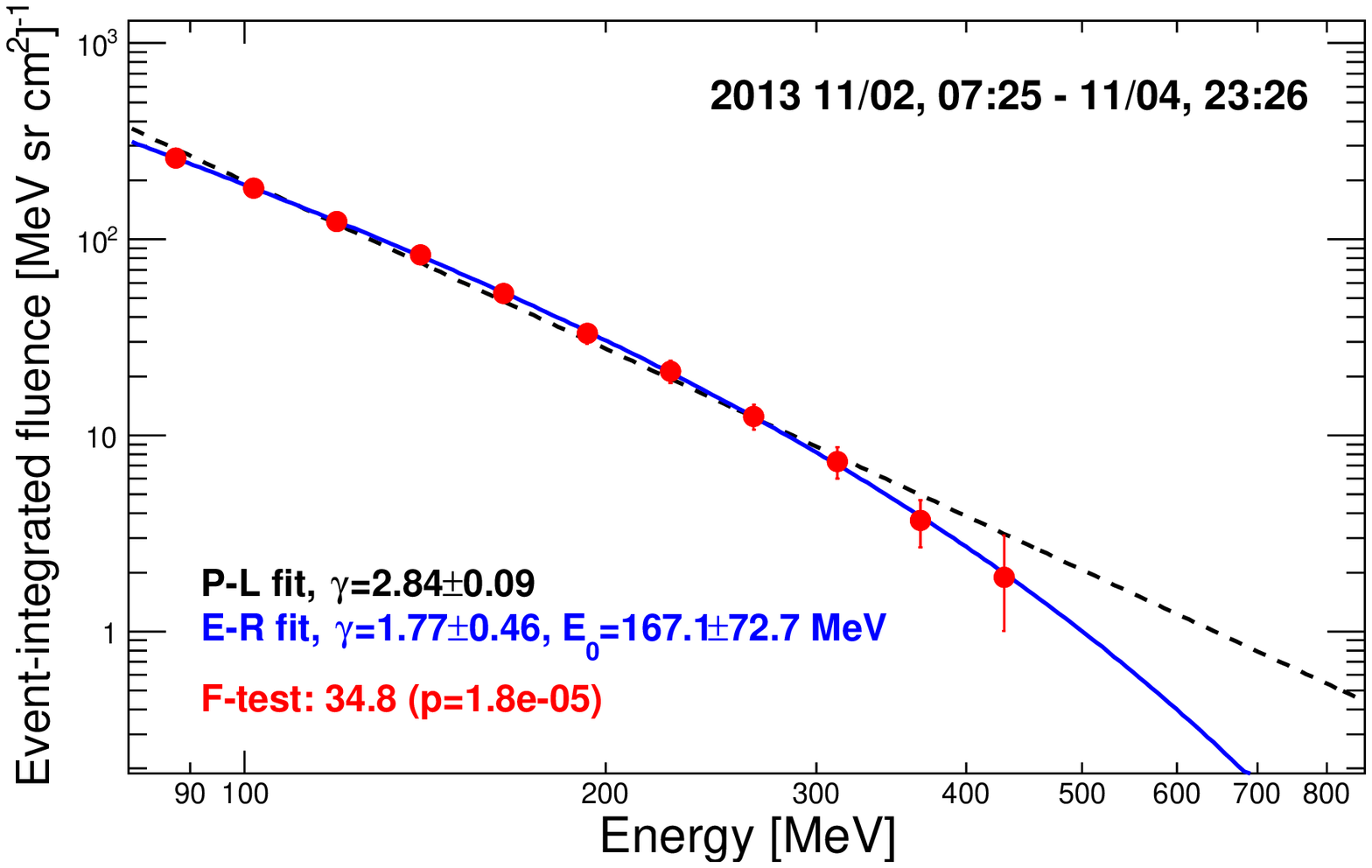} \\
\includegraphics[width=3.4in]{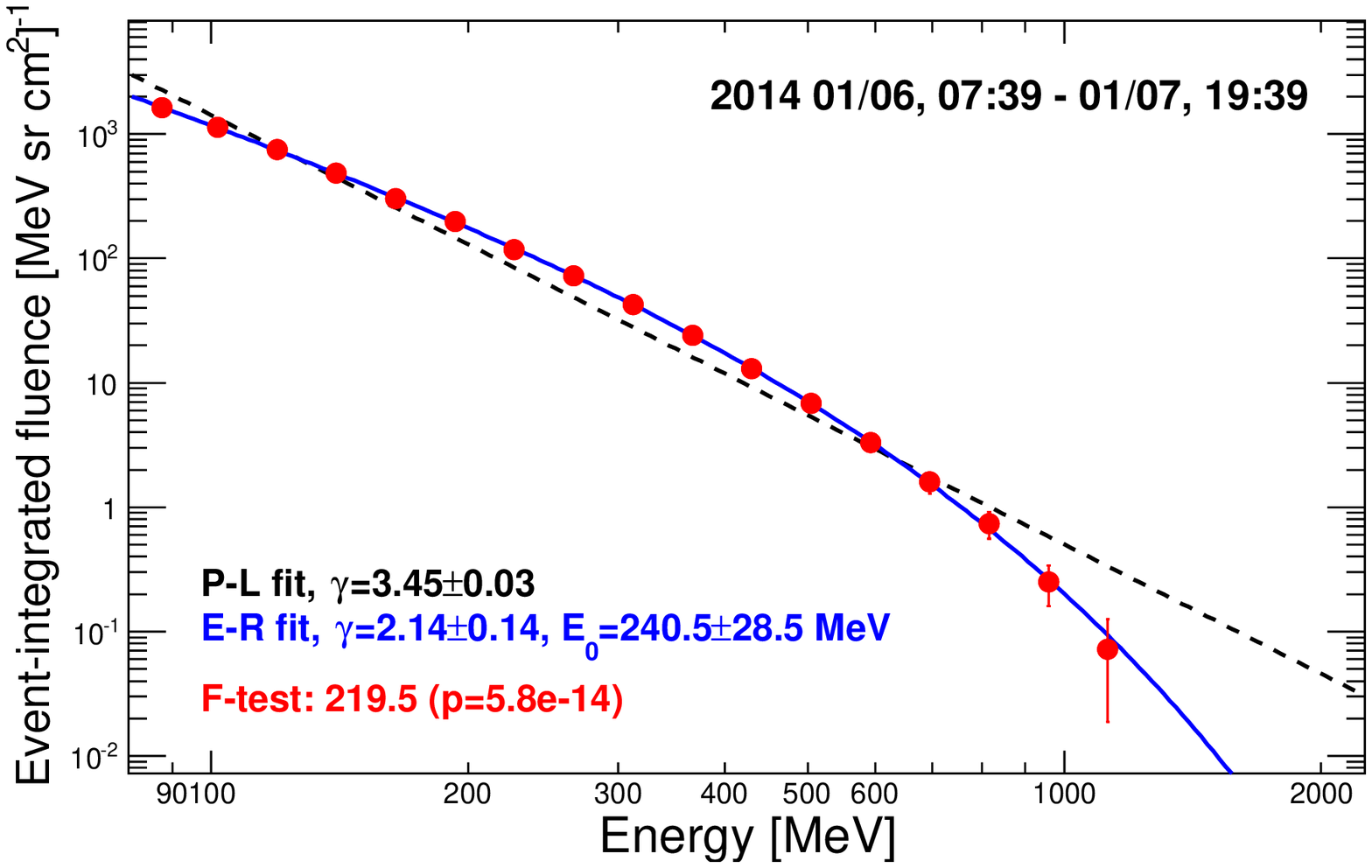} &
\includegraphics[width=3.4in]{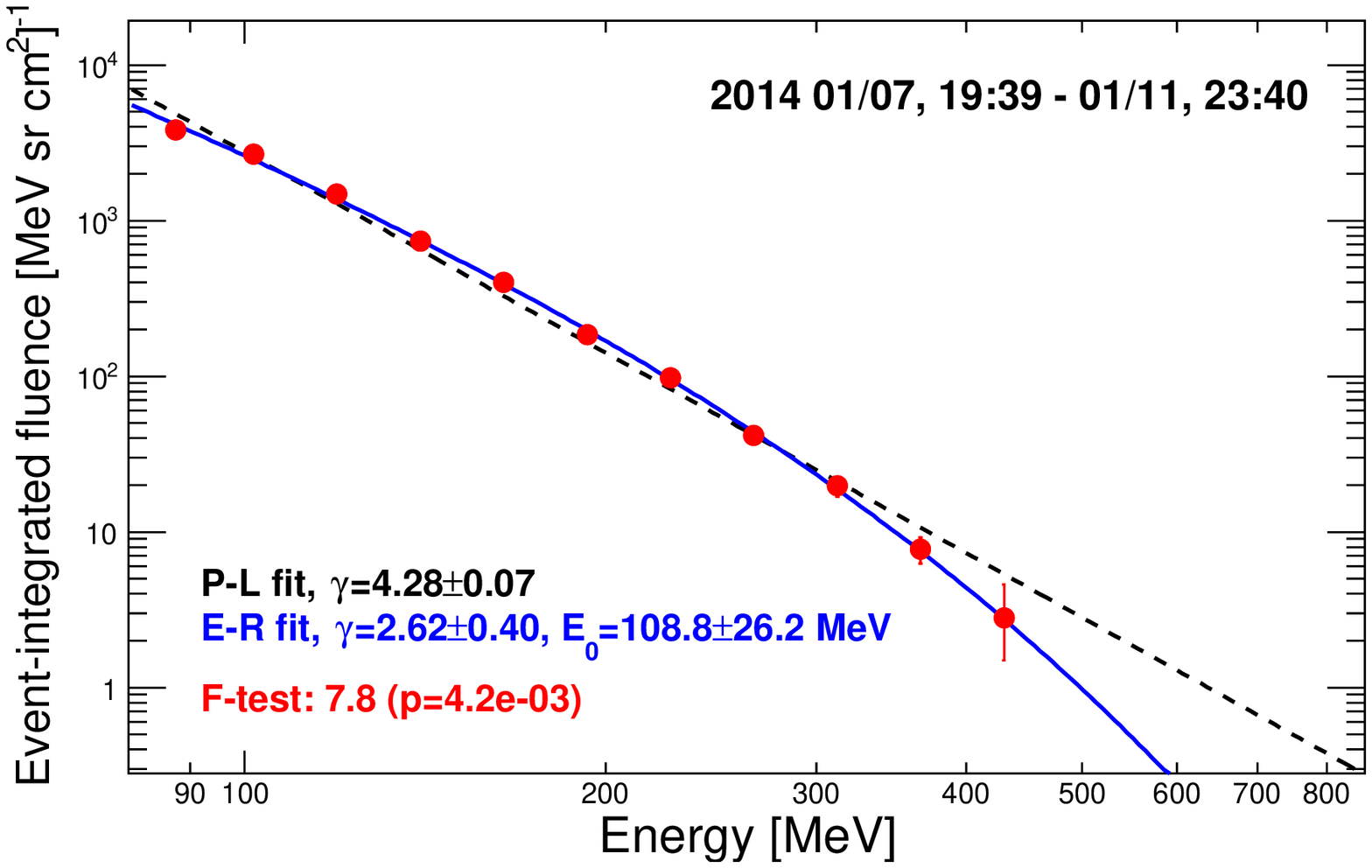} \\
\includegraphics[width=3.4in]{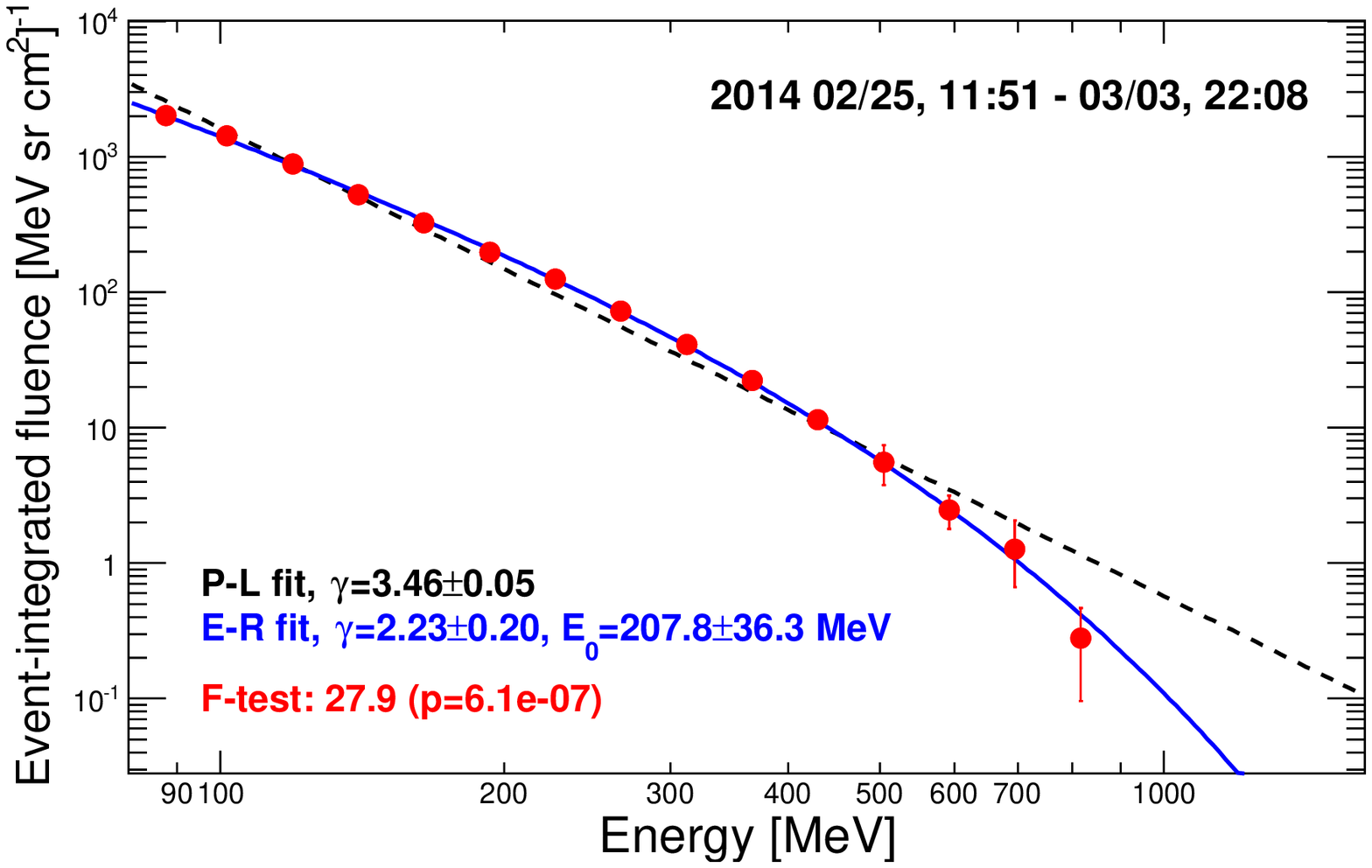} &
\includegraphics[width=3.4in]{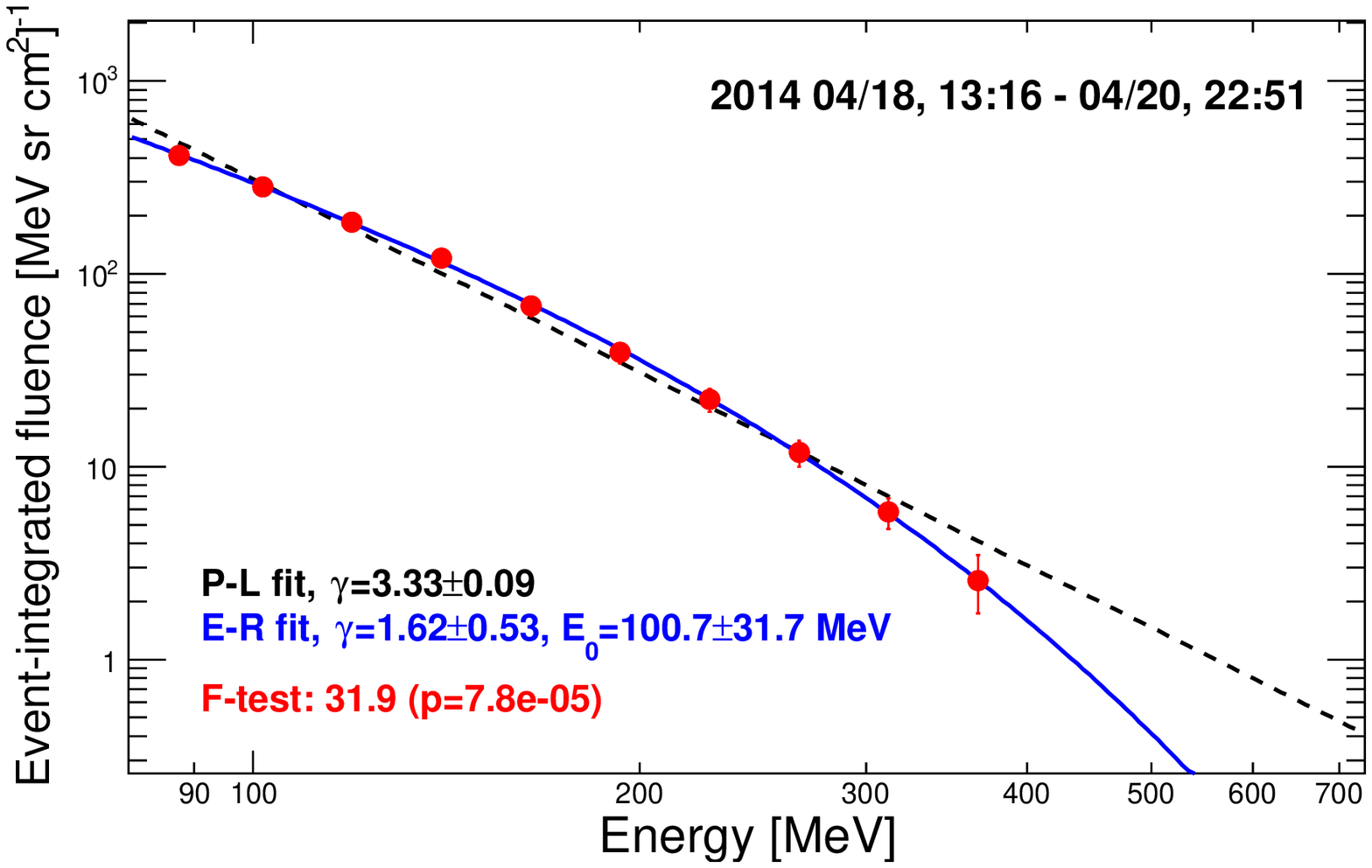} \\
\end{tabular}
\caption{SEP event-integrated fluence spectra measured by PAMELA in the same format as Figure \ref{fig:fluences_part1}.}
\label{fig:fluences_part3}
\end{figure}

\begin{figure}[!t]
\centering
\begin{tabular}{cc}
\includegraphics[width=3.4in]{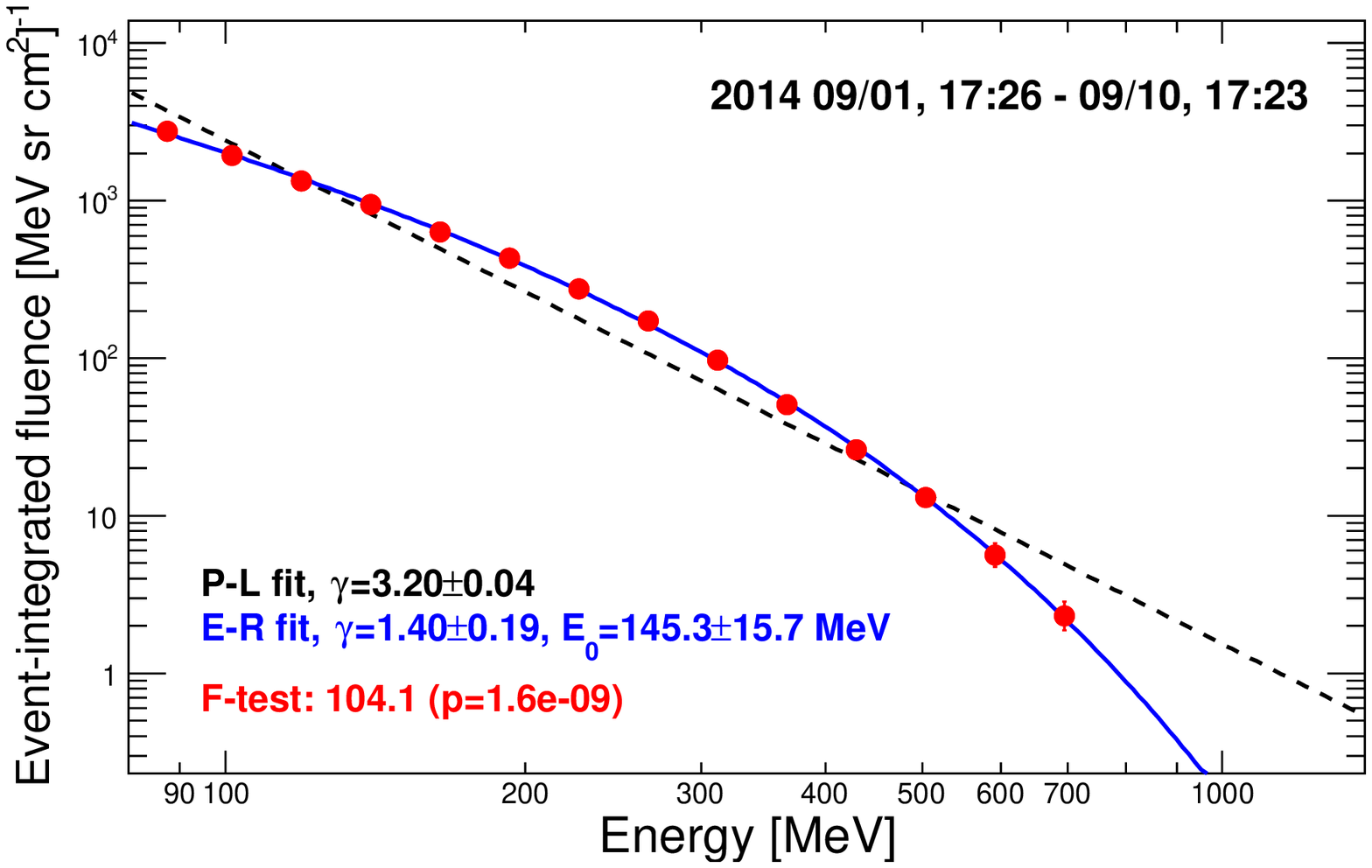} &
\includegraphics[width=3.4in]{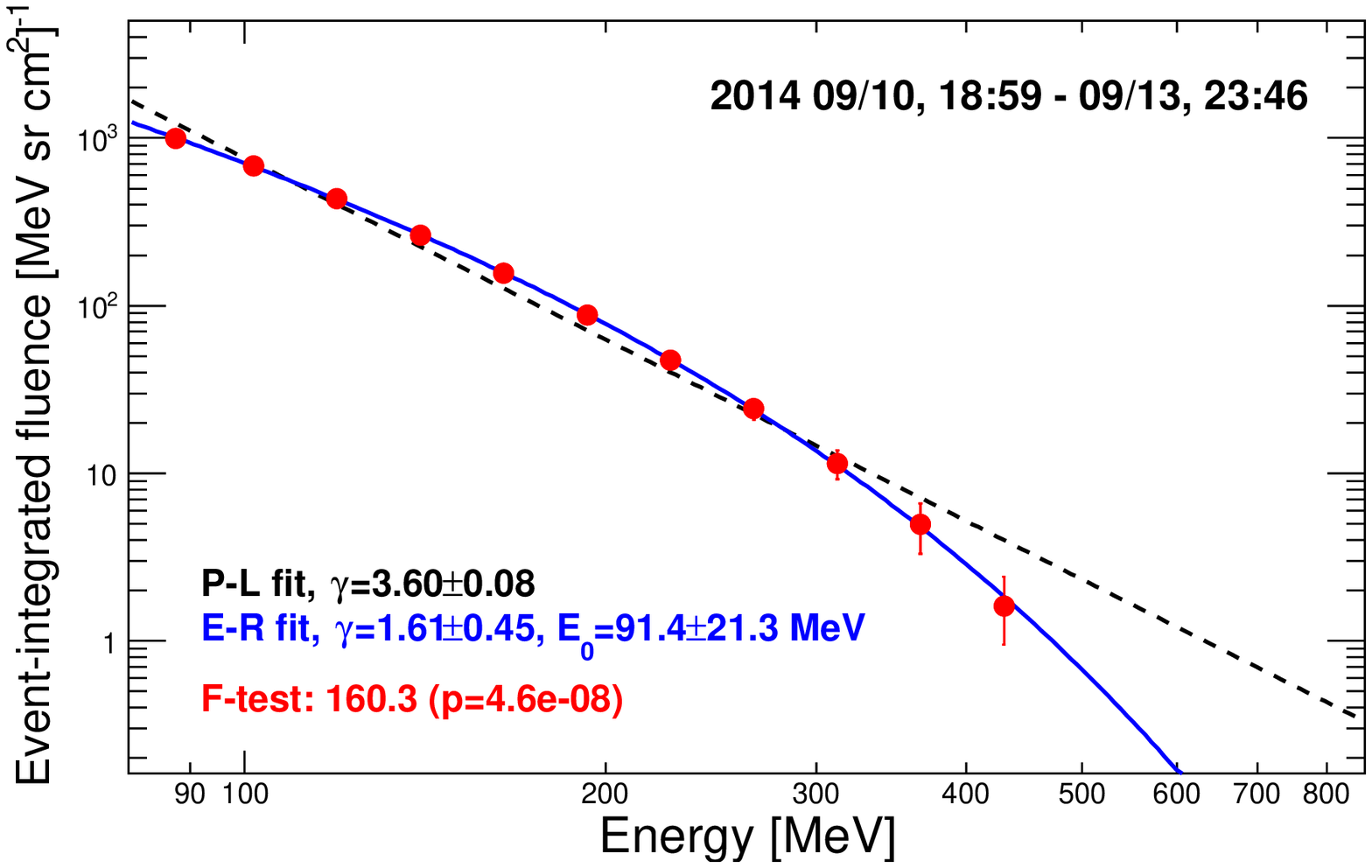} \\
\end{tabular}
\caption{SEP event-integrated fluence spectra measured by PAMELA in the same format as Figure \ref{fig:fluences_part1}.}
\label{fig:fluences_part4}
\end{figure}

Figures \ref{fig:fluences_part1}, \ref{fig:fluences_part2}, \ref{fig:fluences_part3} and \ref{fig:fluences_part4} display the event-integrated fluences estimated for the 26 selected SEP events, in chronological order.
The error bars include both statistical and systematic uncertainties. The start/stop dates are reported in each panel, along with the fitting results obtained by using a simple power law (black dashed lines) and the E-R function (blue solid lines).
It should be noted that the start dates do not coincide, in general, with the SEP event onset times, but they represent the
timestamp of the 48-minute temporal bin in which PAMELA started to
detect the SEPs.
The analyzed SEP sample includes five pairs of overlapping events: 
2006 December 13--14, 2011 September 6--7, 2012 January 23--27, 2012 July 7--8 and 2014 January 6--7.
Consequently, the integration interval used for the first event of each pair is limited by the onset of the subsequent event, and the spectrum of the latter comprises a contribution from the previous event.
As discussed in Section \ref{Data set}, the fluence values derived for the 2012 July 23 and the 2014 February 25 events represent lower limits.

The presence of the high-energy
rollover is evident for all the spectra measured with adequate statistical precision, and it is quantitatively supported by the $F$-test results (including $p$-values), also shown in the panels:
the $F$ $>$ 1 values demonstrate that all measured spectra are better reproduced by the E-R function, and
the power law hypothesis can be excluded at $<$1\% significance level for all the events except for 2011 September 6 and 2012 January 23,
for which it can be rejected at $<$5\% significance level: the former is a weak and short-duration
event, thus affected by statistical limitations and the latter is actually characterized by a very soft spectrum. 

\begin{sidewaystable}[!t]
\centering
\footnotesize
\setlength{\tabcolsep}{3pt}
\renewcommand{\arraystretch}{1.1}
\begin{tabularx}{\linewidth}{c|c|c|c|c|c|c|c}
\hline
 \multicolumn{1}{c|}{SEP event} & \multicolumn{3}{c|}{E-R fit parameters} & \multicolumn{2}{c|}{Event-integrated fluences [sr cm$^{2}$]$^{-1}$} & \multicolumn{2}{c}{Peak fluxes [sr s cm$^{2}$]$^{-1}$}\\
\hline
Start/stop date & A [MeV sr cm$^{2}$]$^{-1}$ & $\gamma$ & $E_{0}$ [MeV] & $>$80 MeV & $>$1000 MeV & $>$80 MeV & $>$1000 MeV\\
\hline
2006 12/14, 22:52 - 12/18, 08:18 & (2.04$\pm$0.11)$\times$10$^{3}$ & 2.56$\pm$0.16 & 353.5$\pm$69.7 & (6.7$\pm$0.3)$\times$10$^{4}$ & (3.8$\pm$1.1)$\times$10$^{1}$ & (2.8$\pm$0.1)$\times$10$^{0}$ & (3.6$\pm$1.6)$\times$10$^{-3}$ \\
2011 03/21, 03:20 - 03/22, 12:10 & (5.84$\pm$1.00)$\times$10$^{2}$ & 1.16$\pm$0.48 & 84.4$\pm$19.4 & (1.0$\pm$0.0)$\times$10$^{4}$ & \nodata & (1.7$\pm$0.1)$\times$10$^{-1}$ & \nodata \\
2011 06/07, 06:47 - 06/09, 22:02 & (5.95$\pm$0.40)$\times$10$^{3}$ & 2.47$\pm$0.23 & 187.7$\pm$34.9 & (1.4$\pm$0.1)$\times$10$^{5}$ & (7.6$\pm$4.0)$\times$10$^{0}$ & (2.5$\pm$0.1)$\times$10$^{0}$ & (4.9$\pm$3.9)$\times$10$^{-4}$ \\
2011 09/06, 02:28 - 09/06, 23:17 & (1.59$\pm$0.49)$\times$10$^{2}$ & 1.83$\pm$0.74 & 132.1$\pm$81.8 & (3.7$\pm$0.2)$\times$10$^{3}$ & \nodata & (1.2$\pm$0.2)$\times$10$^{-1}$ & \nodata \\
2011 09/06, 23:17 - 09/09, 00:56 & (6.72$\pm$0.82)$\times$10$^{2}$ & 1.98$\pm$0.38 & 163.2$\pm$53.0 & (1.8$\pm$0.1)$\times$10$^{4}$ & \nodata & (4.7$\pm$0.3)$\times$10$^{-1}$ & \nodata \\
2011 11/04, 00:35 - 11/05, 23:49 & (4.11$\pm$1.71)$\times$10$^{2}$ & 1.21$\pm$0.90 & 96.0$\pm$56.6 & (8.7$\pm$0.5)$\times$10$^{3}$ & \nodata & (1.0$\pm$0.1)$\times$10$^{-1}$ & \nodata \\
2012 01/23, 04:33 - 01/25, 14:12 & (1.17$\pm$0.19)$\times$10$^{4}$ & 5.33$\pm$0.45 & 219.1$\pm$126.3 & (1.4$\pm$0.1)$\times$10$^{5}$ & \nodata & (3.1$\pm$0.2)$\times$10$^{0}$ & \nodata \\
2012 01/27, 19:03 - 02/01, 12:44 & (1.99$\pm$0.12)$\times$10$^{4}$ & 3.36$\pm$0.14 & 479.7$\pm$111.0 & (5.2$\pm$0.2)$\times$10$^{5}$ & (1.0$\pm$0.3)$\times$10$^{2}$ & (1.2$\pm$0.1)$\times$10$^{1}$ & (7.2$\pm$5.2)$\times$10$^{-3}$ \\
2012 03/13, 17:29 - 03/15, 23:55 & (3.34$\pm$0.26)$\times$10$^{3}$ & 2.43$\pm$0.24 & 133.7$\pm$19.7 & (6.2$\pm$0.3)$\times$10$^{4}$ & \nodata & (3.3$\pm$0.2)$\times$10$^{0}$ & (5.6$\pm$5.1)$\times$10$^{-4}$ \\
2012 05/17, 01:47 - 05/19, 10:37 & (7.09$\pm$0.40)$\times$10$^{3}$ & 2.42$\pm$0.14 & 498.7$\pm$116.2 & (2.8$\pm$0.1)$\times$10$^{5}$ & (5.3$\pm$1.4)$\times$10$^{2}$ & (1.8$\pm$0.1)$\times$10$^{1}$ & (8.9$\pm$4.7)$\times$10$^{-2}$ \\
2012 07/06, 23:23 - 07/08, 17:01 & (8.94$\pm$3.73)$\times$10$^{2}$ & 2.55$\pm$0.89 & 125.7$\pm$96.6 & (1.5$\pm$0.1)$\times$10$^{4}$ & \nodata & (5.8$\pm$0.5)$\times$10$^{-1}$ & \nodata \\
2012 07/08, 17:49 - 07/12, 16:16 & (1.16$\pm$0.10)$\times$10$^{3}$ & 2.12$\pm$0.27 & 203.0$\pm$53.3 & (3.5$\pm$0.1)$\times$10$^{4}$ & \nodata & (5.7$\pm$0.4)$\times$10$^{-1}$ & \nodata \\
2012 07/12, 17:04 - 07/14, 23:30 & (1.80$\pm$0.28)$\times$10$^{2}$ & 1.18$\pm$0.55 & 90.1$\pm$20.6 & (3.5$\pm$0.3)$\times$10$^{3}$ & \nodata & (1.0$\pm$0.1)$\times$10$^{-1}$ & \nodata \\
2012 07/19, 05:58 - 07/21, 01:12 & (1.86$\pm$0.51)$\times$10$^{3}$ & 3.54$\pm$0.66 & 156.3$\pm$103.3 & (2.7$\pm$0.1)$\times$10$^{4}$ & \nodata & (1.1$\pm$0.1)$\times$10$^{0}$ & \nodata \\
2012 07/23, 06:50 - 07/24, 02:51\tablenotemark{a} & (8.50$\pm$2.13)$\times$10$^{3}$ & 1.62$\pm$0.71 & 54.9$\pm$12.4 & (5.9$\pm$0.4)$\times$10$^{4}$ & \nodata & (1.4$\pm$0.1)$\times$10$^{0}$ & \nodata \\
2013 04/11, 07:35 - 04/15, 05:13 & (6.36$\pm$0.86)$\times$10$^{3}$ & 2.27$\pm$0.41 & 109.1$\pm$27.0 & (1.0$\pm$0.0)$\times$10$^{5}$ & \nodata & (2.1$\pm$0.1)$\times$10$^{0}$ & \nodata \\
2013 05/22, 13:32 - 05/26, 11:58 & (4.96$\pm$0.59)$\times$10$^{3}$ & 2.29$\pm$0.37 & 159.2$\pm$50.6 & (1.1$\pm$0.0)$\times$10$^{5}$ & \nodata & (3.2$\pm$0.4)$\times$10$^{0}$ & \nodata \\
2013 09/30, 01:39 - 10/02, 12:51 & (2.82$\pm$2.69)$\times$10$^{3}$ & 2.09$\pm$1.80 & 41.7$\pm$23.7 & (9.3$\pm$0.8)$\times$10$^{3}$ & \nodata & (2.4$\pm$0.2)$\times$10$^{-1}$ & \nodata \\
2013 10/28, 17:00 - 11/01, 03:25 & (7.32$\pm$1.40)$\times$10$^{2}$ & 1.21$\pm$0.52 & 93.7$\pm$27.0 & (1.5$\pm$0.1)$\times$10$^{4}$ & \nodata & (1.4$\pm$0.1)$\times$10$^{-1}$ & \nodata \\
2013 11/02, 07:25 - 11/04, 23:26 & (5.13$\pm$0.81)$\times$10$^{2}$ & 1.77$\pm$0.46 & 167.1$\pm$72.7 & (1.5$\pm$0.1)$\times$10$^{4}$ & \nodata & (2.4$\pm$0.2)$\times$10$^{-1}$ & \nodata \\
2014 01/06, 07:39 - 01/07, 19:39 & (2.87$\pm$0.16)$\times$10$^{3}$ & 2.14$\pm$0.14 & 240.5$\pm$28.5 & (9.4$\pm$0.4)$\times$10$^{4}$ & (3.3$\pm$0.8)$\times$10$^{1}$ & (2.3$\pm$0.1)$\times$10$^{0}$ & (1.4$\pm$0.4)$\times$10$^{-2}$ \\
2014 01/07, 19:39 - 01/11, 23:40 & (1.17$\pm$0.16)$\times$10$^{4}$ & 2.62$\pm$0.40 & 108.8$\pm$26.2 & (1.7$\pm$0.1)$\times$10$^{5}$ & \nodata & (4.9$\pm$0.3)$\times$10$^{0}$ & \nodata \\
2014 02/25, 11:51 - 03/03, 22:08\tablenotemark{b} & (3.74$\pm$0.24)$\times$10$^{3}$ & 2.23$\pm$0.20 & 207.8$\pm$36.3 & (1.1$\pm$0.0)$\times$10$^{5}$ & (1.6$\pm$0.7)$\times$10$^{1}$ & (8.3$\pm$0.4)$\times$10$^{-1}$ & \nodata \\
2014 04/18, 13:16 - 04/20, 22:51 & (1.15$\pm$0.22)$\times$10$^{3}$ & 1.62$\pm$0.53 & 100.7$\pm$31.7 & (2.2$\pm$0.1)$\times$10$^{4}$ & \nodata & (6.9$\pm$0.4)$\times$10$^{-1}$ & \nodata \\
2014 09/01, 17:26 - 09/10, 17:23 & (5.51$\pm$0.32)$\times$10$^{3}$ & 1.40$\pm$0.19 & 145.3$\pm$15.7 & (1.8$\pm$0.1)$\times$10$^{5}$ & (2.0$\pm$0.7)$\times$10$^{1}$ & (5.9$\pm$0.3)$\times$10$^{-1}$ & \nodata \\
2014 09/10, 18:59 - 09/13, 23:46 & (3.02$\pm$0.48)$\times$10$^{3}$ & 1.61$\pm$0.45 & 91.4$\pm$21.3 & (5.0$\pm$0.2)$\times$10$^{4}$ & \nodata & (8.5$\pm$0.5)$\times$10$^{-1}$ & \nodata \\
\hline
\end{tabularx}
(\textit{a}) Only first $\sim$13 hours available. (\textit{b}) First $\sim$9 hours missing.
\caption{Results of the event-integrated fluence fits based on the E-R function (Equation \ref{eq:E-R_function}). The PAMELA integration intervals (UT) are listed in column 1. Columns 2, 3 and 4 give the fit parameters, while columns 5 and 6 report the values of the fluences above 80 and 1000 MeV, respectively. Finally, the peak fluxes above the same energy thresholds are listed in the last two columns.}
\label{tab:all_fit_results}
\end{sidewaystable}

The results of the event-integrated fluences based on the E-R function, including the integration intervals, the fit parameters and the event-integrated fluences above 80 and 1000 MeV are reported in Table \ref{tab:all_fit_results}. In addition, the last two columns list the peak flux measured above the same energy thresholds. The estimate of the energy-integrated fluences and peak fluxes is based on the extrapolation of the E-R fits to higher energies; associated uncertainties rely on the corresponding one-sigma error bands. Only values with $<$100\% uncertainties are reported.

\begin{figure}[!t]
\centering
\begin{tabular}{c}
\includegraphics[width=4.in]{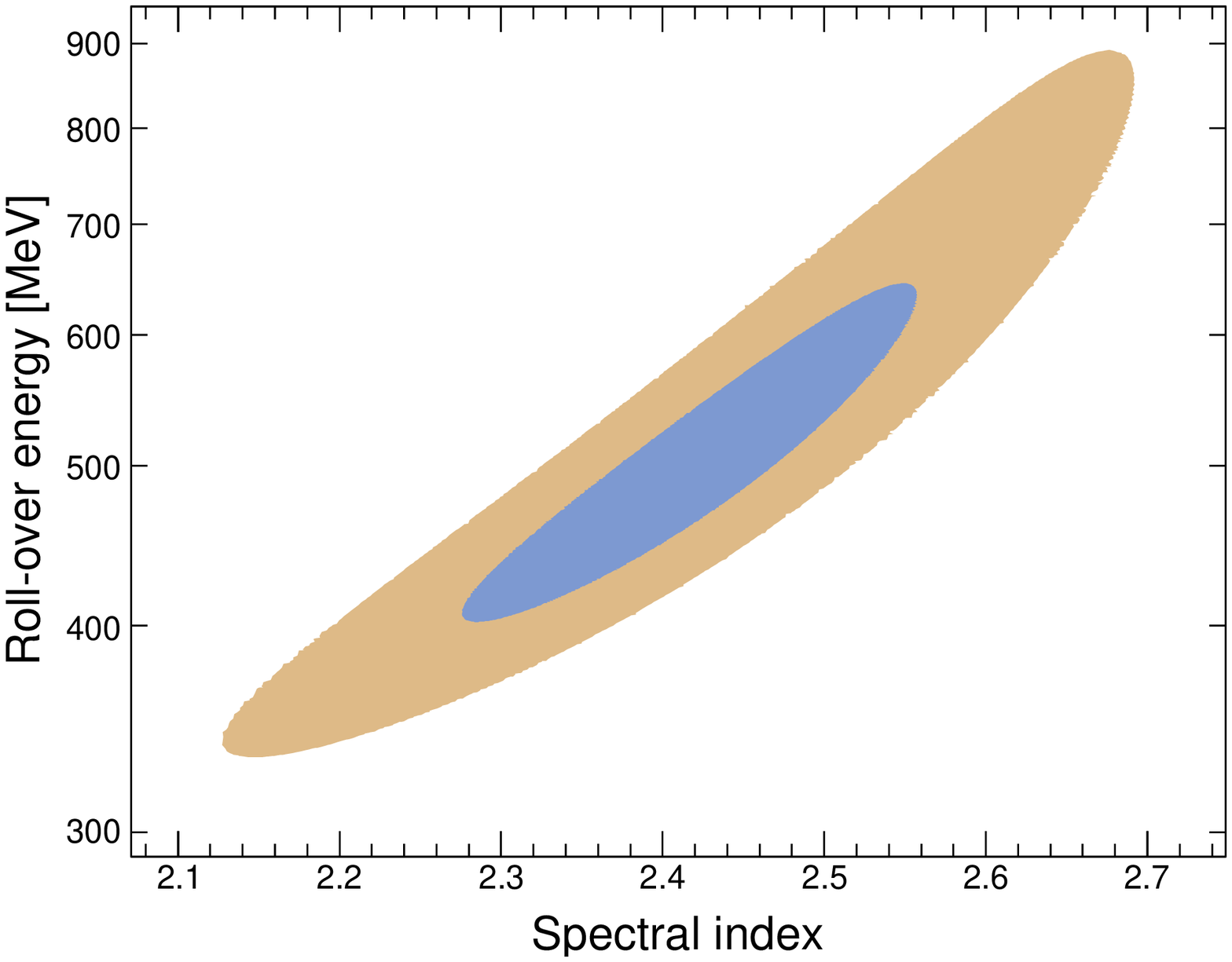}
\end{tabular}
\caption{1-sigma and 2-sigma covariance error ellipses between the rollover energy and the spectral index values from the E-R fit of the 2012 May 17 event-integrated fluence spectrum.}
\label{fig:sigma_contours}
\end{figure}

In fitting spectra, a cross correlation between the power law index and the rollover energy is unavoidable,
affecting parameter uncertainties. As an example, Figure \ref{fig:sigma_contours} shows the 1-sigma and the 2-sigma covariance error ellipses 
evaluated for the 2012 May 17 event; similar plots are obtained for all the investigated events. A significant correlation can be observed, with the rollover energy increasing with growing spectral index.
The reason for this is that when attempting an E-R fit to a given spectrum, if a steeper power law (larger spectral index) is chosen, the best fit is obtained when the greater falloff at higher energies in the power law is compensated by a larger rollover energy. 

\begin{figure}[!t]
\centering
\begin{tabular}{c}
\includegraphics[width=1.0\textwidth]{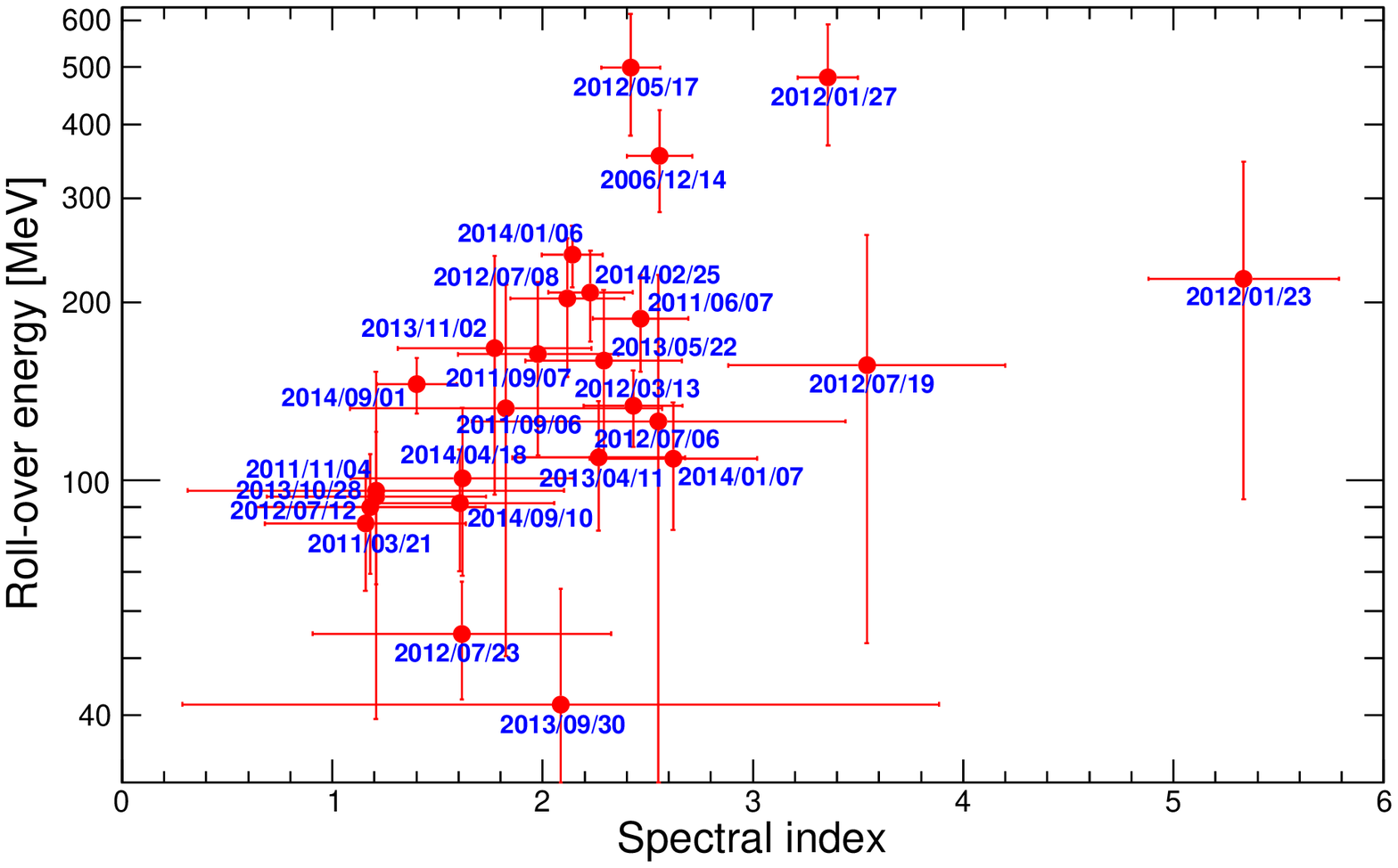}
\end{tabular}
\caption{Global distribution of the rollover energy versus spectral index values from the E-R fit of the event-integrated fluence spectra. Errors bars account for the parameter uncertainties.}
\label{fig:cRollover_vs_gamma}
\end{figure}
\begin{figure}[!t]
\centering
\begin{tabular}{cc}
\includegraphics[width=3.4in]{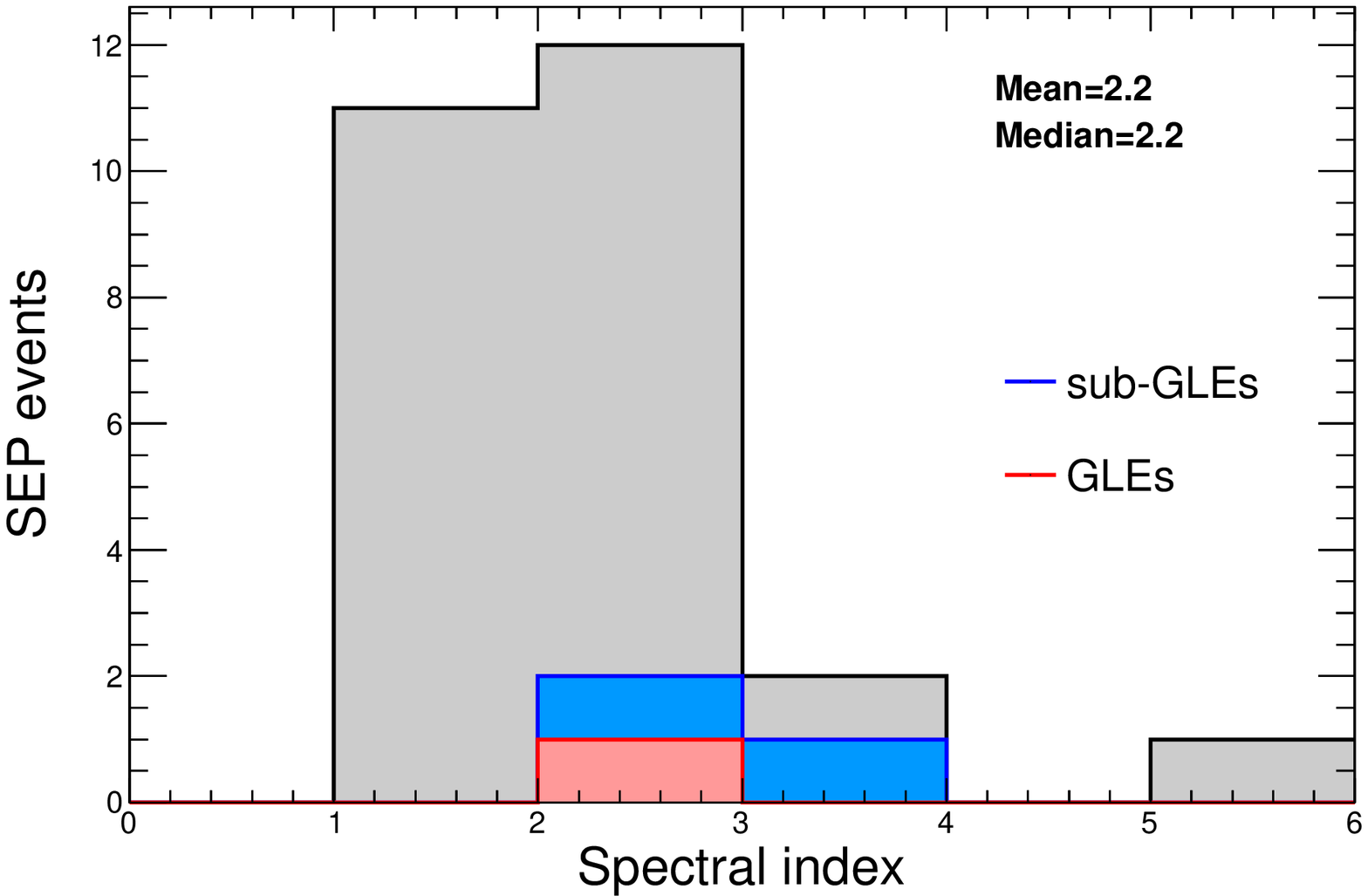} &
\includegraphics[width=3.4in]{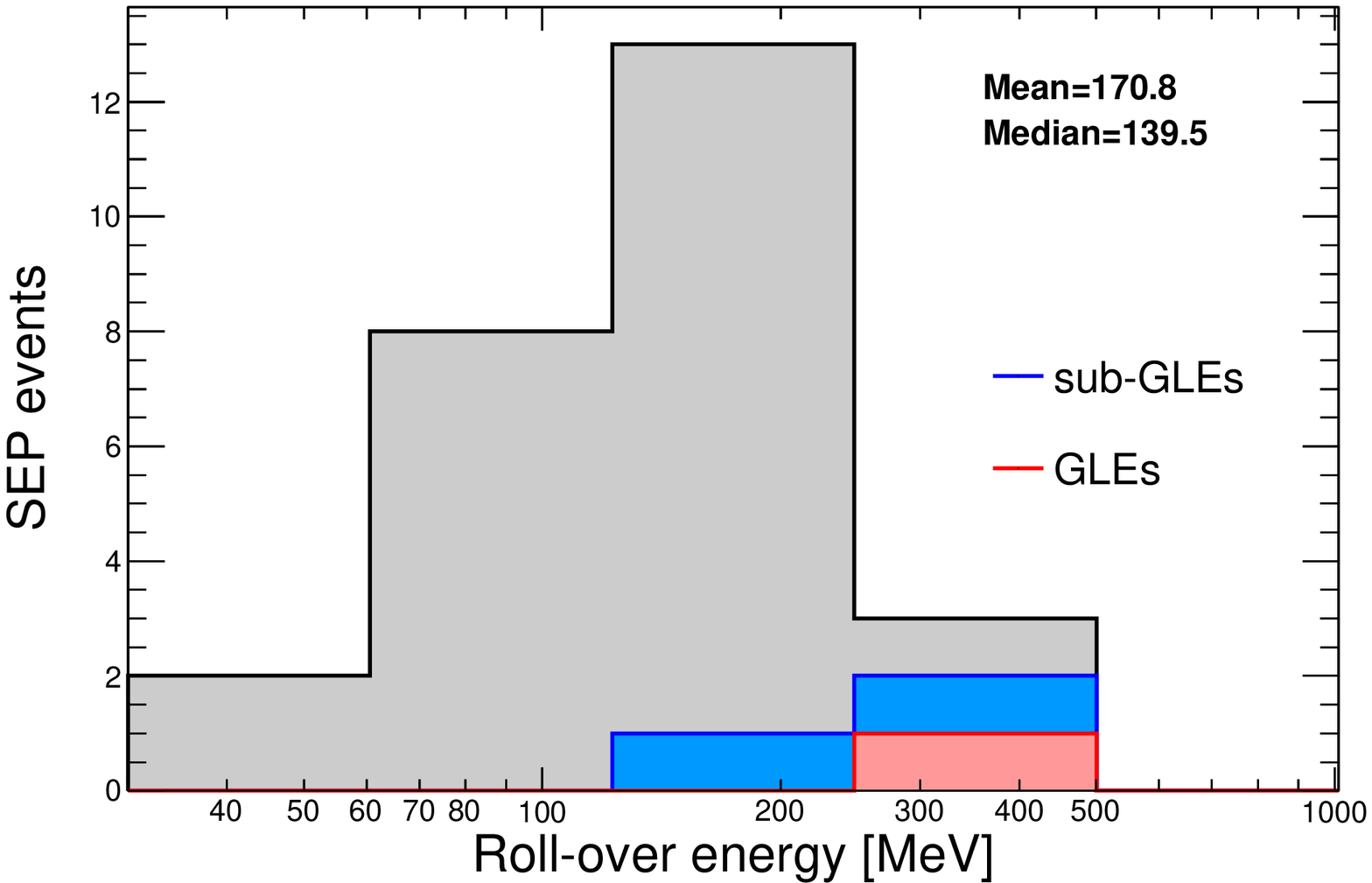} \\
\end{tabular}
\caption{Distributions of the spectral index (left) and rollover (right) parameters from the E-R fit of event-integrated fluence spectra. Mean and median values are reported in each panel.}
\label{fig:fit_par_distr}
\end{figure}

Figure \ref{fig:cRollover_vs_gamma} displays the rollover energy versus spectral index distribution from the E-R fits of all the fluence spectra.
The error bars indicate the related parameter uncertainties. An overall trend can be seen, with higher rollover energies associated with larger spectral indices;
the outlier point corresponds to
the 2012 January 23 event
that, as aforementioned, is peculiar because of the exceptionally soft spectrum ($\gamma$ $\sim$ 5.3).
The global positive correlation between rollover energy and power law index may be a manifestation of the effect illustrated in Figure \ref{fig:sigma_contours}; further statistical investigation is necessary to infer a more physical meaning to the trend.
The histograms of the spectral index and the rollover distributions are separately shown in the left and the right panels of Figure \ref{fig:fit_par_distr}. On average, the analyzed SEP sample is characterized by a spectral shape with $\gamma$=2.2 and $E_{0}$=170.8 MeV; the corresponding median values are 2.2 and 139.5 MeV, respectively. As discussed later, the 2012 May 17 GLE and the 2006 December 14, 2012 January 27 and 2014 January 6 sub-GLE events (also indicated in the plots) are properly incorporated in the SEP global distribution without exhibiting qualitative peculiarities, albeit the analyzed sample is statistically limited.

\begin{figure}[!t]
\centering
\begin{tabular}{cc}
\includegraphics[width=3.4in]{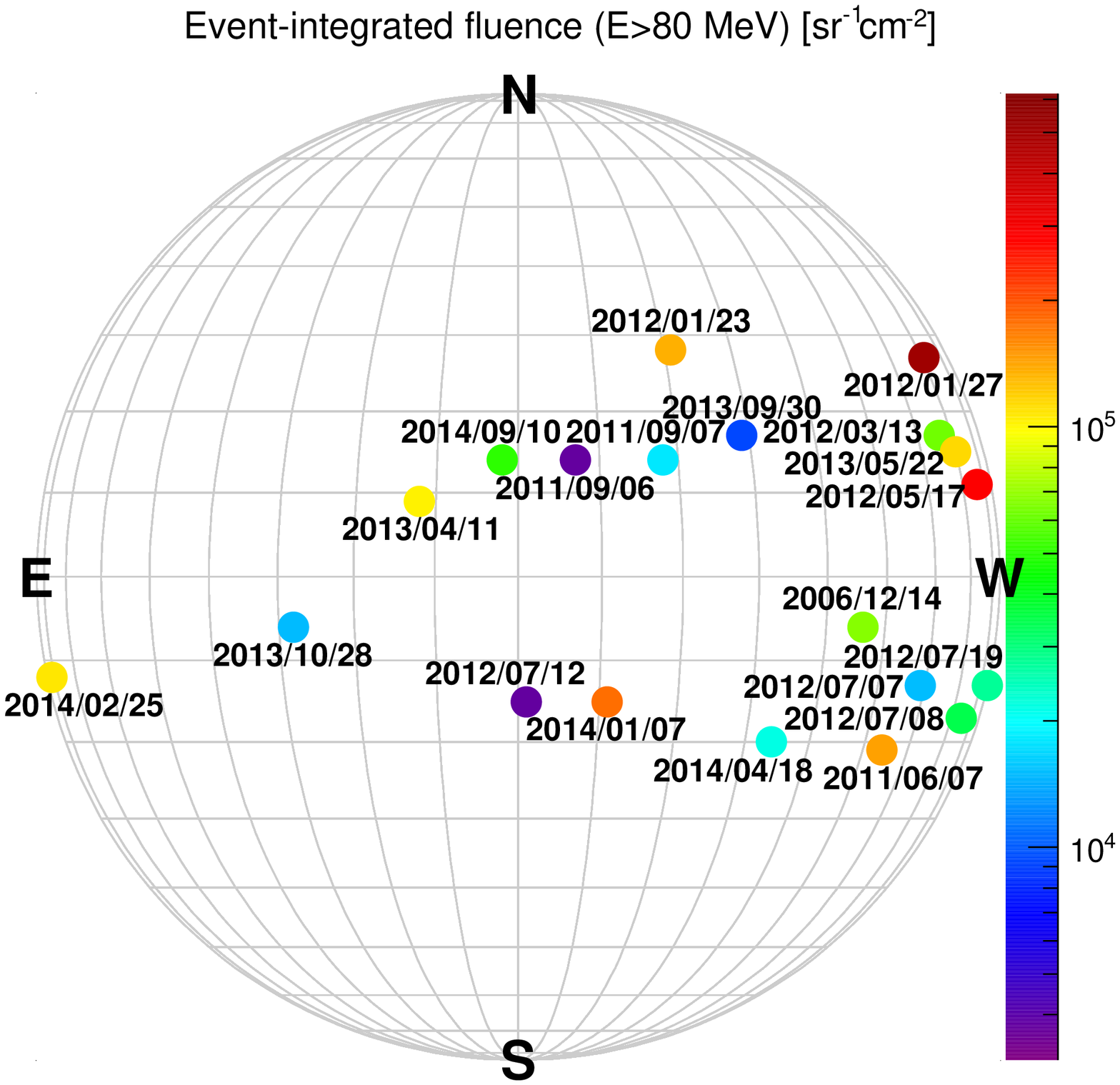} & \includegraphics[width=3.4in]{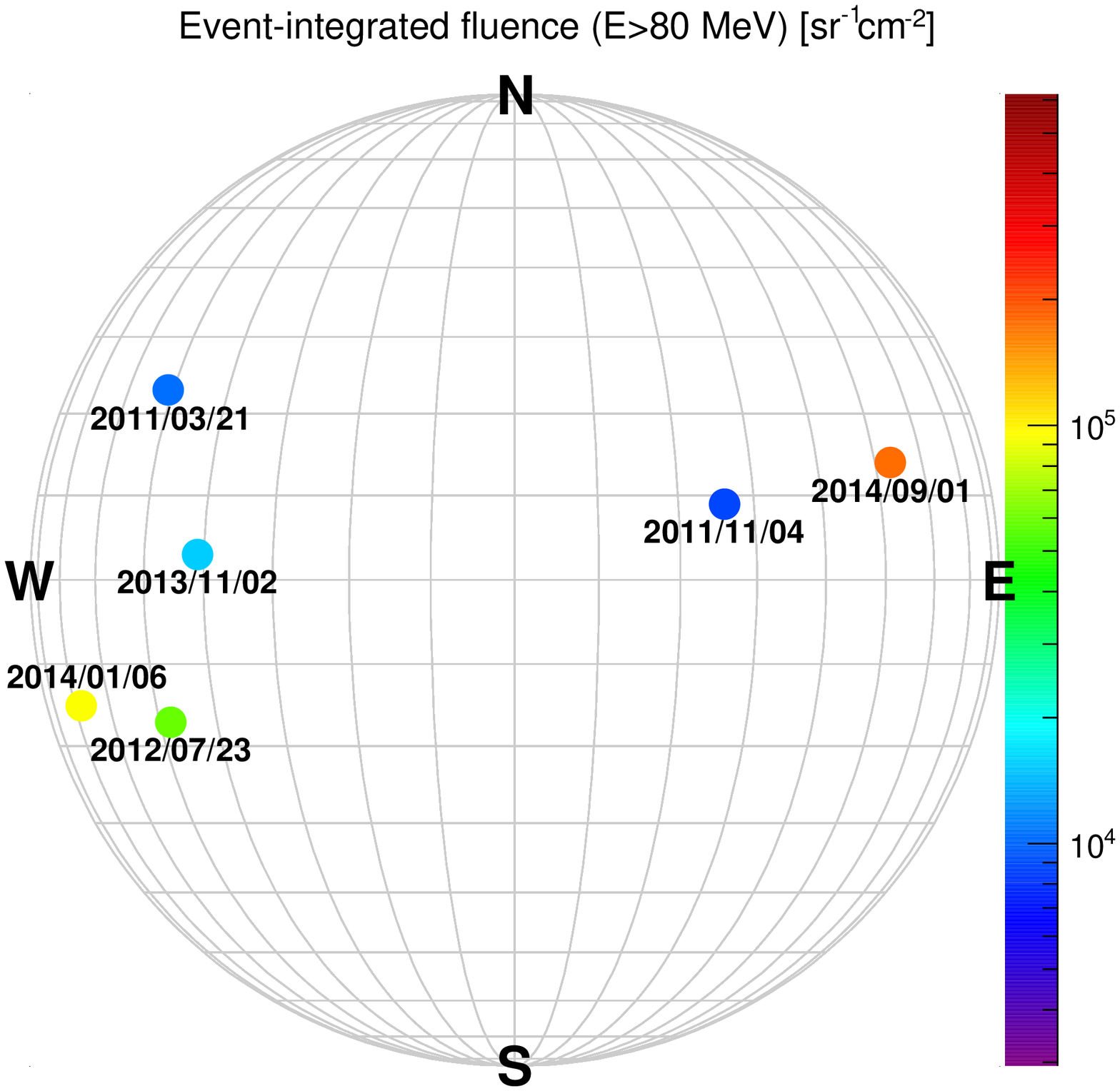} \\
\includegraphics[width=3.4in]{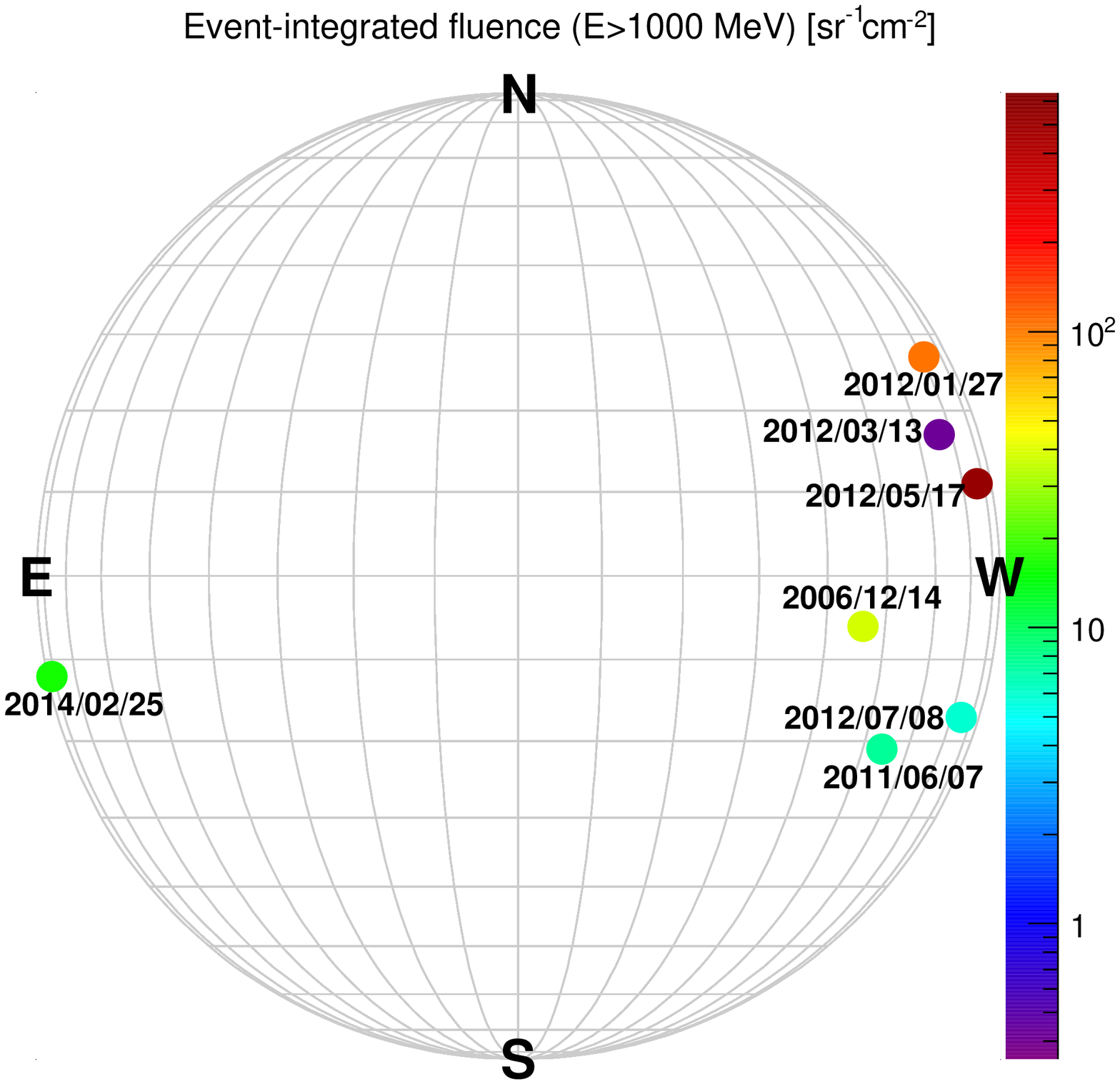} & \includegraphics[width=3.4in]{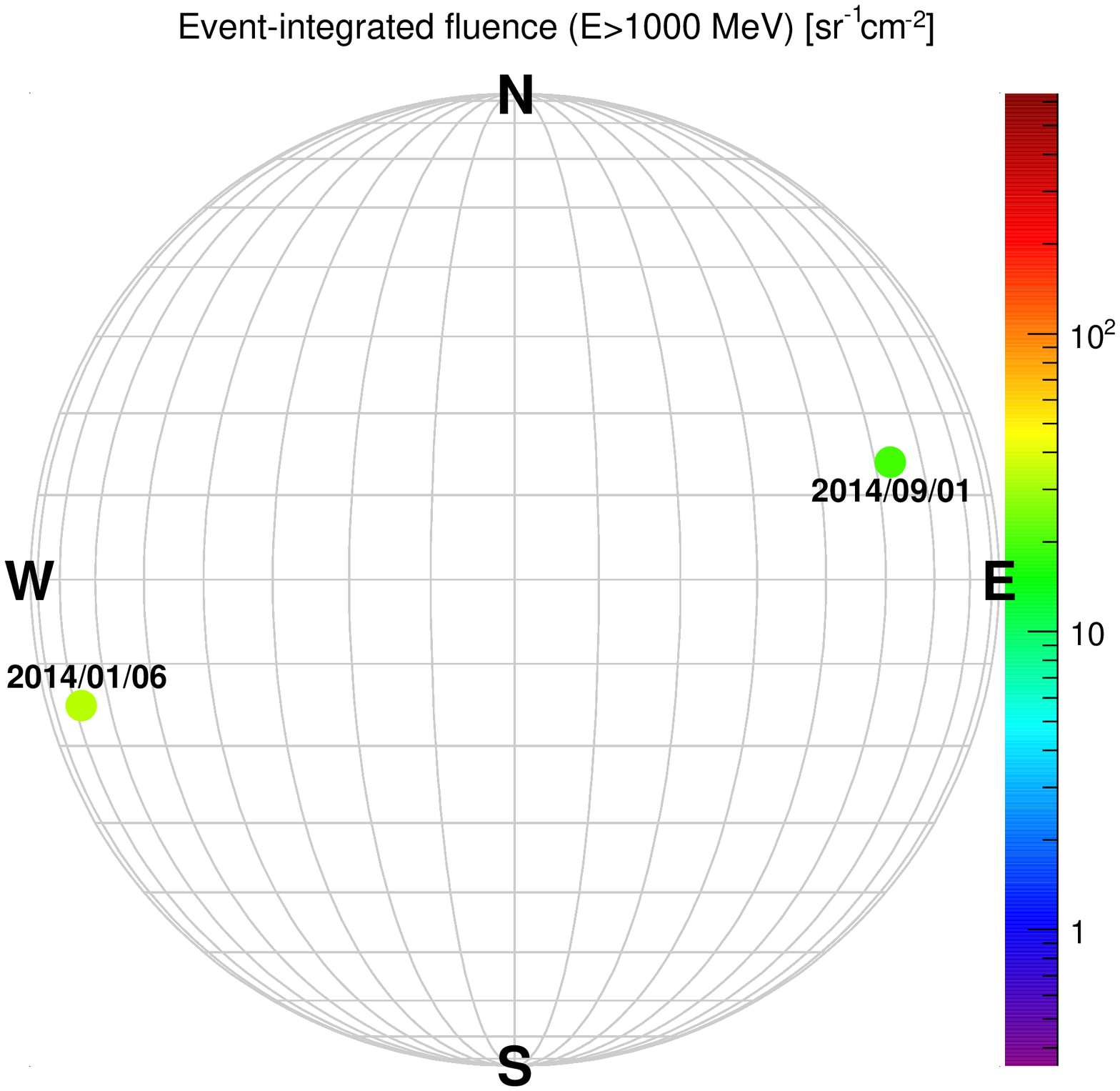} \\
\end{tabular}
\caption{Event-integrated fluences (color codes) as a function of source flare heliographic locations (see Table \ref{tab:used_events}), above two different energy thresholds: 80 and 1000 MeV (top and bottom panels, respectively). Front and back side events are reported in the left and the right panels, respectively.}
\label{fig:fluences_vs_location}
\end{figure}

Finally, Figure \ref{fig:fluences_vs_location} illustrates the event-integrated fluences (color codes) as a function of the source flare heliographic locations (see Table \ref{tab:used_events}), for two different energy thresholds (80 and 1000 MeV). Front and back side eruptions are displayed in the left and the right panels, respectively.
Only fluence values with $<$100\% uncertainties are shown. All the events with observable event-integrated fluences above 1000 MeV (bottom panels) were linked to flares occurring not far from the western limb (W46--W112)
with the exception of
the long-duration 2014 February 25 and the 2014 September 1 SEP events, which originated close to or behind the east limb (E82--E127).
However, the 2006 December 13 GLE and the 2012 March 7 sub-GLE events, not included in this work (see Section \ref{Data set}),
which also likely extended into this energy range, had parent flares with
longitudes closer to the central meridian (W23 and E27, respectively).
Thus, the PAMELA results demonstrate that poorly connected events can contribute significantly to the SEP fluence above 1000 MeV detected near the Earth.

\begin{figure}[!t]
\centering
\includegraphics[width=6.2in]{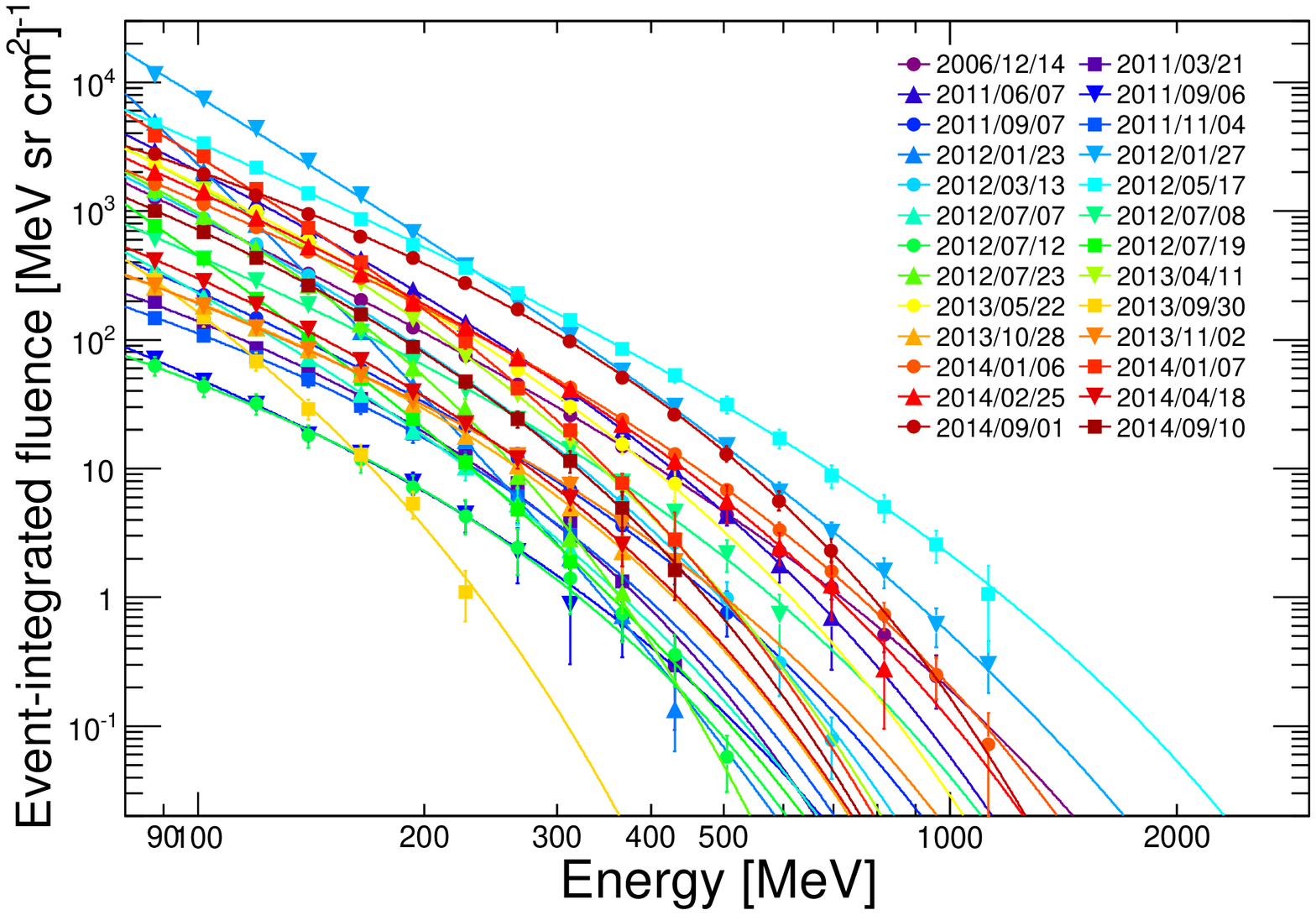}\\
\includegraphics[width=6.2in]{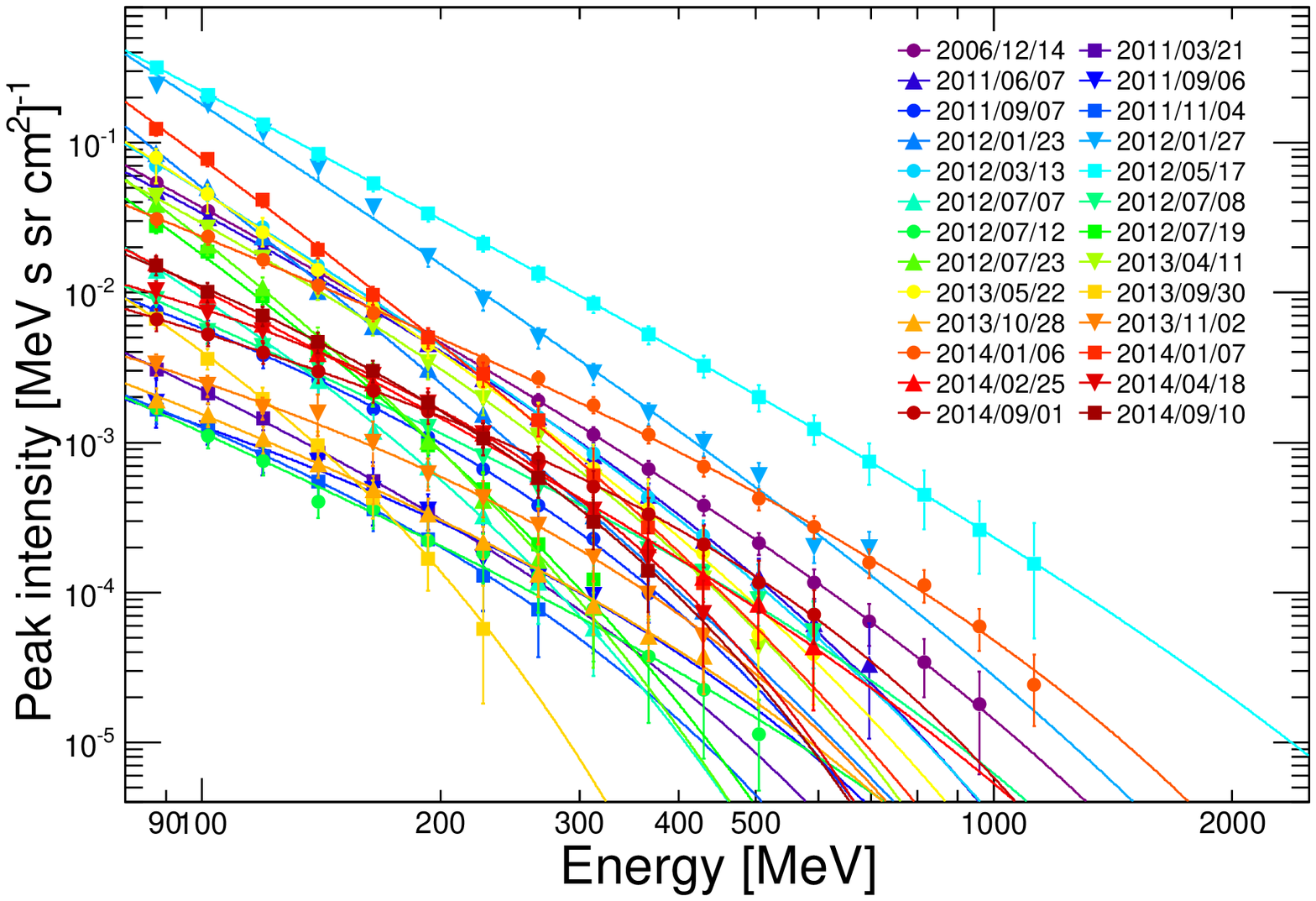}
\caption{Summary of PAMELA SEP measurements. Event-integrated fluences and peak differential spectra are shown in the top and the bottom panels, respectively. The error bars include both statistical and systematic uncertainties. The curves represent the fits performed with the E-R function.}
\label{fig:all_points}
\end{figure}

\section{Discussion}
The PAMELA mission provides accurate observations of SEP events over a wide energy range, from 
$\sim$80 MeV up to a few GeV, enabling, for the first time, a direct
investigation of spectral features at the highest SEP energies.
Results are summarized in Figure \ref{fig:all_points}, where the fluence and the peak spectra of all selected events are compared; the curves denote the corresponding E-R fits. 
PAMELA data as a whole span about five and four orders of magnitude in fluences and peak intensities, respectively.
While most of the spectra are of similar shape, there does appear to be 
a group of events with rapidly falling spectra but which are relatively intense at the lower energy range. One
is the 2013 September 30 event, which was associated with a small (C1.3) flare and with a huge filament eruption not located in active regions \citep{ref:GOPALSWAMY2015A,ref:HOLMANFOORD2015}. Remarkably, despite a relatively weak solar event, particles were accelerated up to a few hundred MeV.

\begin{figure}[!t]
\centering
\begin{tabular}{cc}
\includegraphics[width=3.4in]{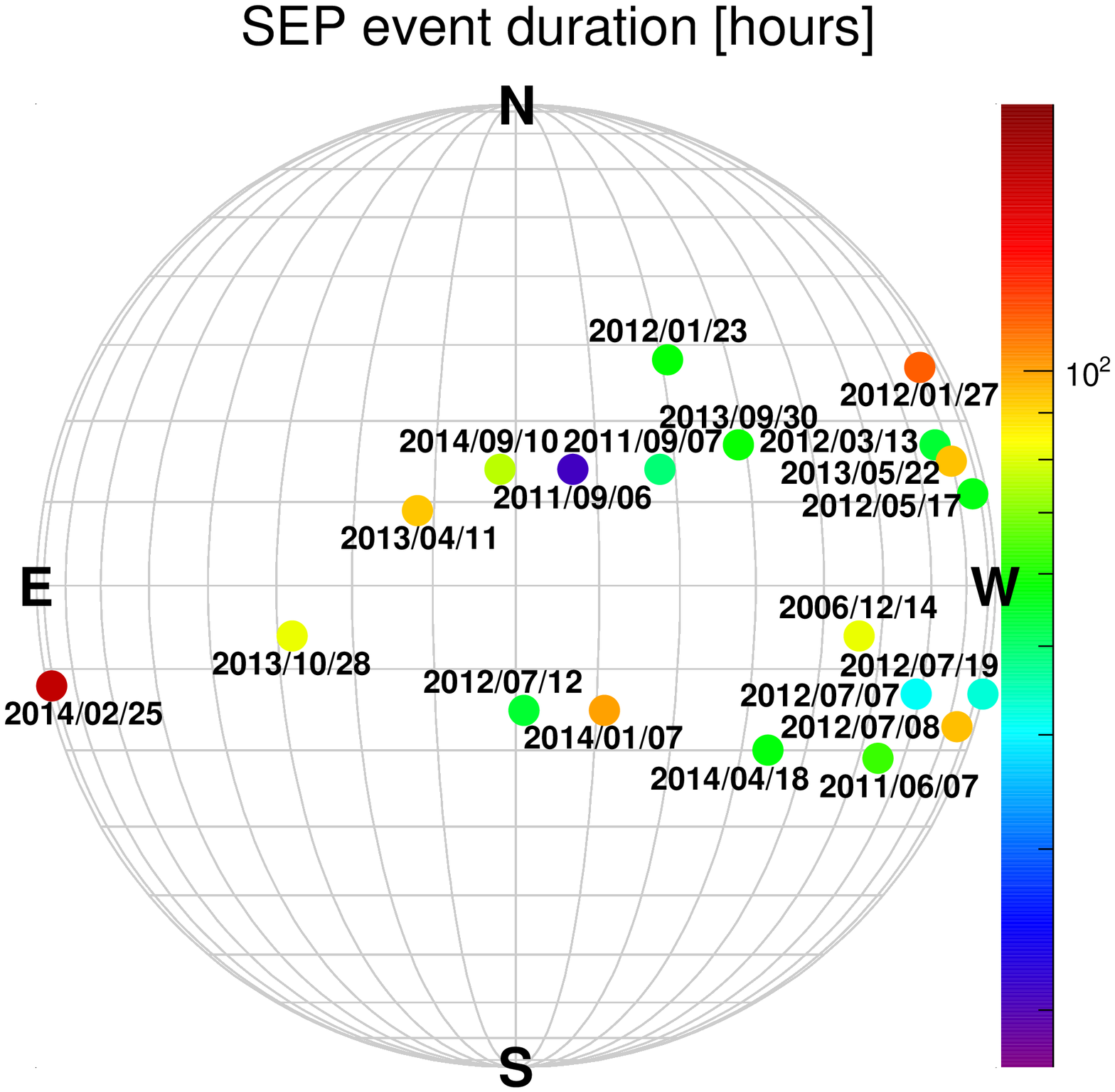} & \includegraphics[width=3.4in]{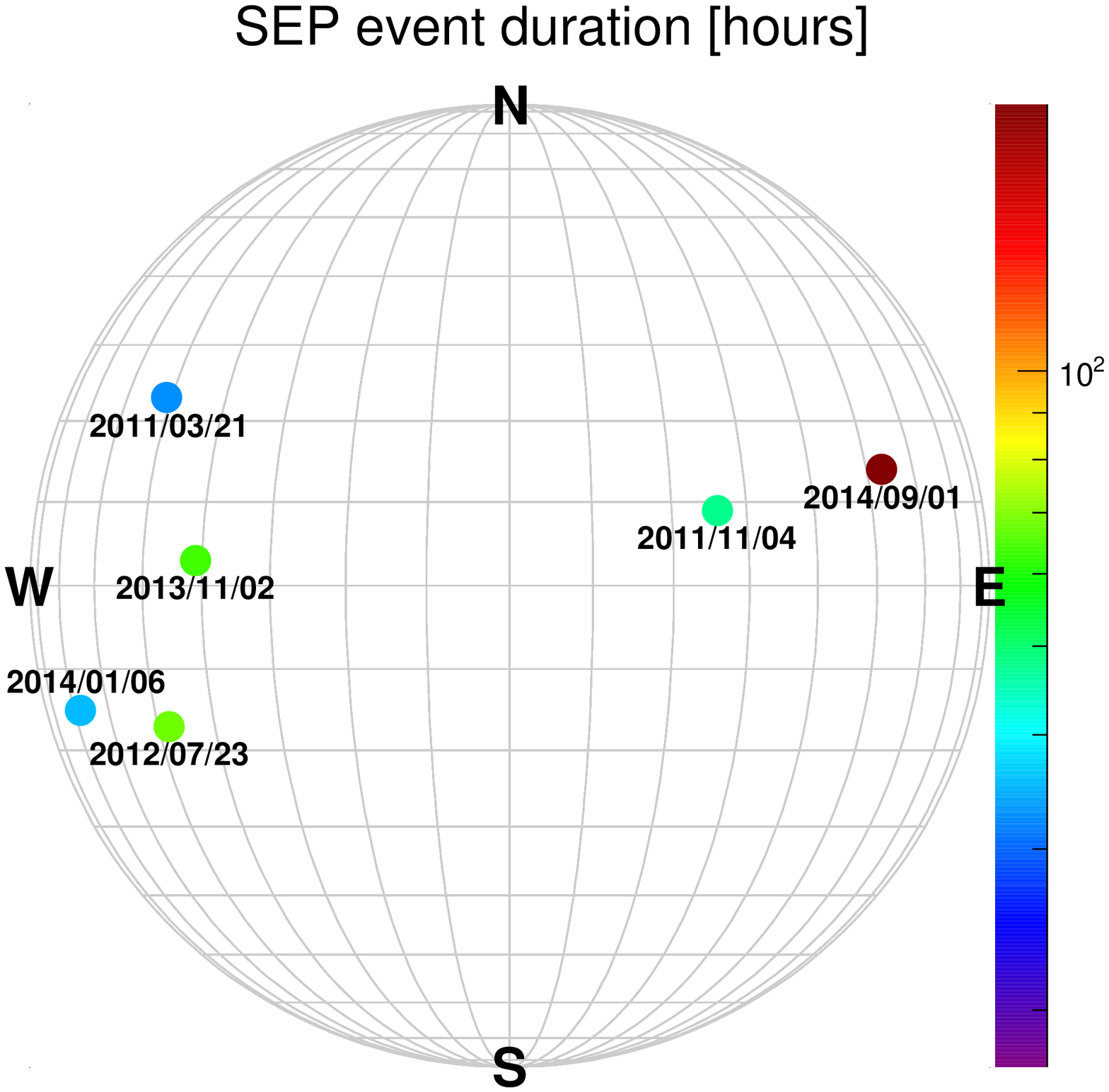} \\
\end{tabular}
\caption{SEP event duration (E$\gtrsim$80 MeV, color code) as a function of flare heliographic locations, for the SEP events registered by PAMELA 
(see Table \ref{tab:used_events}). 
Front and back side eruptions are reported in the left and the right panels, respectively.} 
\label{fig:flares_vs_duration}
\end{figure}

PAMELA measurements offer a unique opportunity to investigate the relationship between GLE and non-GLE events.
As reported in Table \ref{tab:all_fit_results}, the current SEP database comprises several events with
observable peak fluxes above 1000 MeV, including one GLE and three sub-GLE candidates. The main points can be summarized as follows.
\begin{itemize}
\item The most energetic event, on 2012 May 17, was observed as a GLE by NMs and exhibited the highest rollover energy value ($\sim$500 MeV), resulting in a harder spectrum at NM energies with respect to non-GLE events. Remarkably, it was associated with a modest (M5.1) flare and with a moderately fast CME (V$_{sky}$=1582 km s$^{-1}$). The detection of the highest-energy particles may have been due to the optimal magnetic connection between the solar event location and Earth \citep{ref:CARBONE2013,ref:MAY17PAPER,ref:ROUILLARD2016}. In particular, \citet{ref:GOPALSWAMY2013} have proposed that the CME non-radial motion combined with the favorable B0 angle (the inclination of the solar equator to the ecliptic) significantly improved the latitudinal connection of the shock nose, where particle acceleration to high energies is assumed to be most efficient. Furthermore, the onset phase of the SEP event was characterized by a pronounced pitch-angle anisotropy \citep{ref:MAY17PAPER,ref:BRUNO_JPCS}. 
\item The 2014 January 6 sub-GLE was generated by a behind western limb event associated with a $\lesssim$X3.5 flare and with a 1402 km s$^{-1}$ sky-plane speed CME \citep{ref:THAKUR2014}. Both STEREO spacecraft had a full view of the involved active region and detected a large filament
eruption at $\sim$07:50 UT \citep{ref:ACKERMANN2017}. As for the 2012 May 17 event, \citet{ref:GOPALSWAMY2015B} proposed that the non-radial motion of CME along with the favorable B0 angle rendered the shock nose latitudinally well-connected to Earth.
The SEP event duration ($\sim$35 hours) was limited by the onset of the January 7 event. It represents the most energetic sub-GLE in the PAMELA database.
With respect to the 2012 May GLE, the evaluated peak spectrum is harder below $\sim$500 MeV while it is slightly softer for energies $\gtrsim$1 GeV;
the corresponding flux intensity above 500 (1000 MeV) does not exceed $\sim$21\% ($\sim$15\%). Moreover, the cutoff energy of the fluence spectrum ($\sim$240 MeV) is relatively low compared to other GLE and sub-GLE events with a better longitudinal connection.
\item The 2012 January 27 sub-GLE originated from a X1.7 flare associated with a very fast CME (V$_{sky}$=2508 km s$^{-1}$).
It was characterized by a more gradual increase in intensity than might be expected for a well-connected western event. To account for this observation, \citet{ref:GOPALSWAMY2015B} proposed that there was poor latitudinal connection to the nose of the CME-driven shock (due to the unfavorable B0 angle) and the highest-energy particles did not reach Earth.
The event commenced during the decaying phase of the January 23 event. The arrival of two interplanetary shocks
on January 22 and 24 (\url{http://www.ssg.sr.unh.edu/mag/ace/ACElists/obs_list.html}) produced a small suppression of measured intensities \citep{ref:BELAKHOVSKY2017}, resulting in a relatively lower GCR background estimate (see Section \ref{Data analysis}). As displayed in the bottom panel of Figure \ref{fig:all_points}, the computed peak spectrum is significantly softer than the spectra of the other high-energy events; note that points above 700 MeV are not reported due to residual GCR contamination. The registered peak flux above 500 (1000 MeV) is $\sim$79\% ($\sim$50\%) with respect to that of the 2014 January 6 sub-GLE, but the much longer duration ($>$4 days) resulted in a higher event-integrated fluence (see Figure \ref{fig:flares_vs_duration}).
\item Finally, the well-connected 2006 December 14 event was associated with a X1.5 flare linked to a relatively slow CME (V$_{sky}$=1042 km s$^{-1}$). The SEP arrival occurred during the decaying phase of the 2006 December 13 GLE event, so measured intensities include a contribution from the previous event. However, they were significantly suppressed by Forbush decrease effects caused by the arrival of a fast interplanetary CME several hours prior to the SEP event onset, with remarkable space weather effects \citep{ref:VONROSENVINGE2009,ref:SEP2006,ref:GSTORM,ref:MUNINI2018}; consequently, the corresponding GCR background correction is found to be particularly small. On the whole, the resulting peak flux above 500 MeV (1000 MeV) is only $\sim$39\% ($\sim$26\%) with respect to the 2014 January 6 sub-GLE. Nevertheless, this event can be considered a potential ``failed'' sub-GLE.
\item For the rest of the sample the estimated energy-integrated peak fluxes above 1000 MeV are much less intense 
or characterized by large uncertainties due to the lack of high-energy data points.
This set also includes a few other events with statistically consistent $>$1000 MeV fluence values (see bottom panels in Figure \ref{fig:fluences_vs_location}). Among them, two events are peculiar because of the poor longitudinal connection and the long-duration (see Figure \ref{fig:flares_vs_duration}): the 2014 February 25 eastern limb and the 2014 September 1 back side eruptions \citep{ref:LARIO2016,ref:PLOTNIKOV2017}. Both events were associated with very intense (X4.9 and $\lesssim$X2.4, respectively) flares and fast CMEs (V$_{sky}$=2147 and 1901 km s$^{-1}$, respectively). Despite the magnetically unfavorable source region, relatively high-energy protons reached the Earth, giving rise to prolonged ($\sim$7 and $\sim$9 days, respectively) SEP events, characterized by similar time profiles exhibiting a slow increase of flux intensities as a consequence of the broad longitudinal spread of SEPs at 1 au \citep{ref:CANE1988,ref:REAMES1999}.
It can be speculated that these events would have triggered GLEs if located at well-connected heliographic longitudes.
\end{itemize}

\begin{figure}[!t]
\centering
\includegraphics[width=1.0\textwidth]{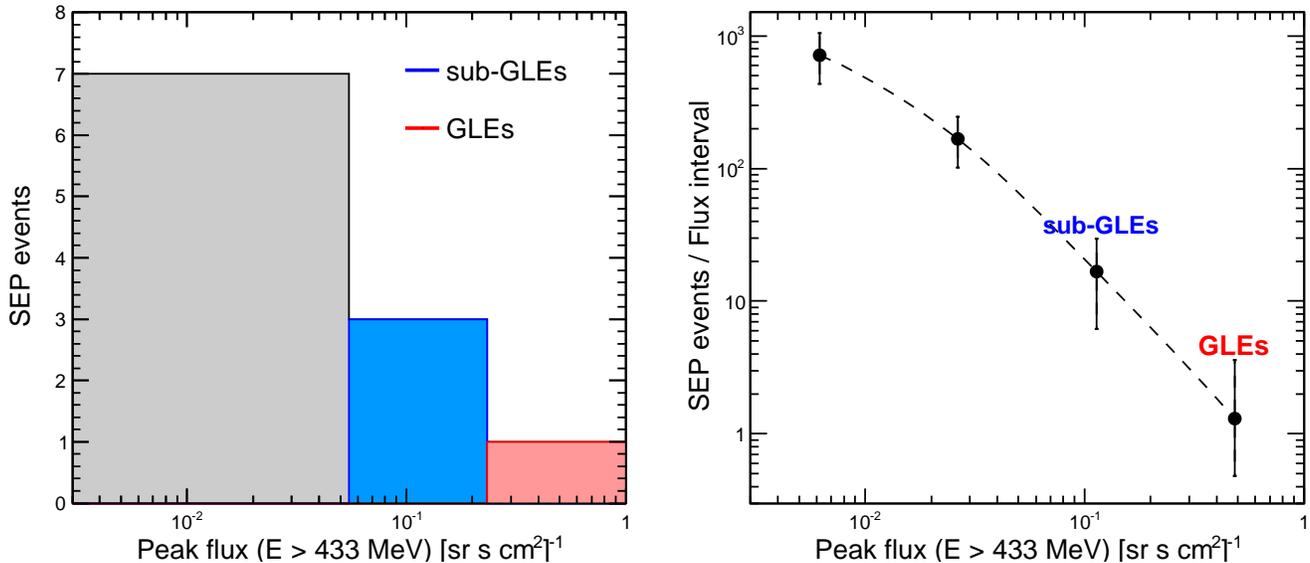}
\caption{Left: Number of SEP events as a function of the peak flux above 433 MeV (NM atmospheric threshold). Right: The resulting SEP size distribution, computed by dividing the number of events with given peak flux by the corresponding flux interval. The dashed curve is to guide the eye.}
\label{fig:cPeakDistr}
\end{figure}

As reported in the left panel of Figure \ref{fig:cPeakDistr}, PAMELA observations include several SEP events apart from the (sub-)GLEs with spectra extending well above the NM atmospheric threshold (1 GV or $\sim$433 MeV), but exhibiting relatively lower intensities. 
The occurrence rate of SEP events of a given peak flux 
is inversely related to the intensity itself.
This is demonstrated in the right panel of Figure \ref{fig:cPeakDistr}, where the SEP size distribution, given by the number of events in each log-equal peak flux bin divided by the corresponding bin width \citep{ref:BAZILEVSKAYA2005,ref:BELOV2005}, is displayed for energies above 433 MeV. The error bars include the statistical uncertainties.
Note that the first point is underestimated since the low intensities are close to the (time-dependent) threshold sensitivity and a fraction of events is missing; in addition, results do not account for peak flux uncertainties.
A power law fit of the last 3 points gives a 1.63$\pm$0.41 spectral index.
The highest-energy region of the distribution is populated by the sub-GLE and the GLE events;
based on the available sample, 
the potential peak flux thresholds (E$>$433 MeV) can be inferred to be $\sim$5--6$\times$10$^{-2}$ sr$^{-1}$s$^{-1}$cm$^{-2}$ and $\sim$2--3$\times$10$^{-1}$ sr$^{-1}$s$^{-1}$cm$^{-2}$, respectively.

Several concomitant factors may contribute to the SEP variability and to the rarity of the GLE events, such as the associated CME kinetic energy,
the shock morphology and evolution, the ambient conditions, the magnetic connection to Earth and the interplanetary transport. Hence, the absence of qualitative differences between the spectra of GLE, sub-GLE and non-GLE events is remarkable. These observations confirm that GLEs cannot be considered a separate category of SEP events, but they rather are the extreme end of a continuous spectral distribution.

\begin{figure}[!t]
\centering
\includegraphics[width=1.0\textwidth]{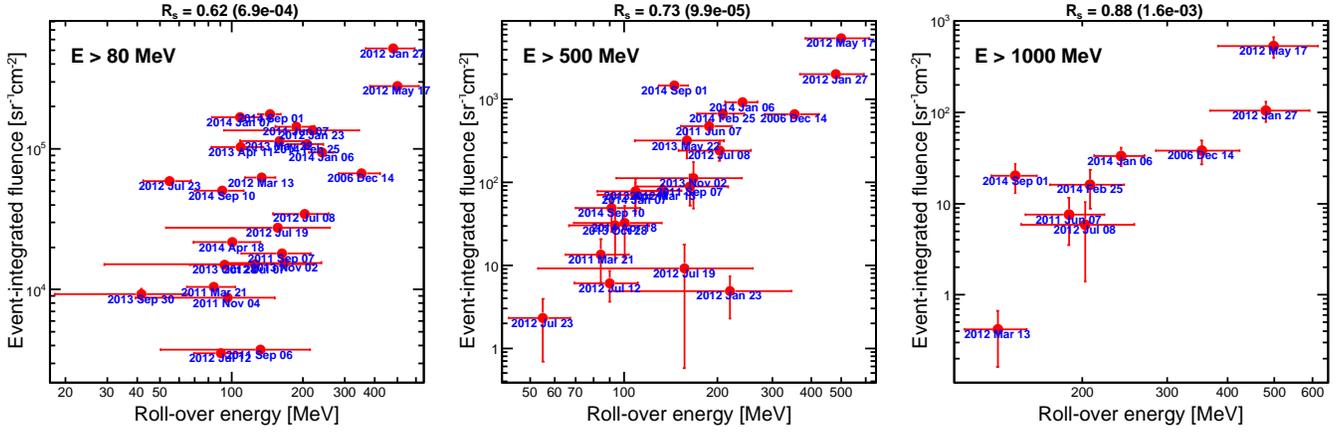}\\
\caption{Distributions of the event-integrated fluence for proton energies larger than 80 MeV (left), 500 MeV (center) and 1000 MeV (right) as a function of the rollover energy from the E-R fit of the event-integrated fluence spectra. The Spearman rank correlation coefficient $R_{s}$ is reported at the top of the panels, along with the corresponding $p$-value.}
\label{fig:cFluence_vs_Rollover}
\end{figure}

The spectra of SEPs, which may span more than 5 orders of magnitude in energy and more than 8 orders of magnitude in intensity, are determined by the processes by which the particles are accelerated and their subsequent transport. Moreover, spectral features measured at different energies may arise from particle acceleration in different locations (e.g., the flare region, corona or interplanetary space). Therefore, the spectral shape may exhibit the combined signatures of several dynamic processes that may be complex to disentangle. In addition, the spectrum will depend on the time at which it is measured, for example, at the peak of the event or integrated over the whole event to give a fluence spectrum. Previous studies have reported spectral breaks in fluence spectra occurring at tens of MeV (e.g. \citet{ref:LI2009,ref:MEWALDT2012,ref:MEWALDT2013,ref:DESAI2016}) that can be explained by
accounting for transport effects, especially diffusion \citep{ref:LILEE2015,ref:ZHAO2016}. The PAMELA mission is able to provide, for the first time, a direct measurement of SEP spectra at much higher energies -- above a few hundred MeV up to a few GeV.
high-energy events that are well-connected exhibit, in their first impulsive peak, a spectrum free of significant transport effects from a few solar radii to 1 au.   Because those particles are prompt and beam-like they have scattered very little over the great majority of their path, allowing us to investigate the acceleration at low altitudes.  For those events where we have no data from an impulsive peak (due to exposure effects) one could argue that the picture is less clear, where conceivably the transport effects from a few solar radii to 1 au could modify the spectrum.  However, at these energies, expect that any intervening scattering simply redirects particles into different areas of pitch-angle and real space with little effect on their energy.  Furthermore, one could also argue that for diffusive shock acceleration and transport are inseparable, but isolating the processes at low altitudes from those in interplanetary space is important for these high-energy events.

A picture of GLEs and high-energy SEP events emerges from the PAMELA data that appears to be consistent with traditional shock acceleration modeling and observations. Firstly, the ubiquity of power laws coupled to high-energy rollovers falls in line with a transient acceleration process with time- and space-limited operations, i.e. quasi-spherical CME shocks of limited extent that restrict the duration of acceleration at high-energy, where the limits arise from the limited time the shock is strong and the divergent geometry. 
Second, the more efficient the shock acceleration is, the greater the overall intensity of the particle event and the hardness of the spectrum. 
As reported in Figure \ref{fig:cFluence_vs_Rollover}, the measured fluences are well correlated with the rollover energies, with the most energetic events exhibiting the highest cutoffs; the correlation improves with increasing proton energy. Similar results are obtained for the peak fluxes (not shown).
Such intensities and spectra are typically only witnessed for well-connected events. When high-energy particles from poorly connected events are detected at Earth, they tend to be in long-duration, relatively weak, SEP events, where processes such as cross-field diffusion and co-rotation with the Sun delay their arrival and extend their duration at Earth.

Although the spectra are apparently consistent with shock acceleration, might a flare-accelerated component contribute? One expected signature might be a change in the spectral form such as, for example, a hardening, indicating the presence of an additional component, but there is no clear evidence of such in our events. A power law spectrum has been suggested for a flare-accelerated component \citep{ref:REAMES2015} but this appears to be inconsistent with the truncation of the observed spectra at high energies. On the other hand, the mechanisms that may lead to particle acceleration in flares are complex (e.g. \citet{ref:ZHARKOVA2011}) and hence a theoretical understanding of this process, and the spectra which result, is still incomplete. Thus, the presence of a spectral turndown may not necessarily be an argument against a flare-accelerated contribution. 
In general, while the smooth spectra characterized by E-R forms suggest that a single acceleration mechanism is occurring, contributions from other processes such as stochastic acceleration and magnetic field reconnection cannot be ruled out  \citep{ref:BAZILEVSKAYA2017}.
In addition, the spectra, including the rollover energy, must also be influenced by processes unrelated to particle acceleration including particle transport from the solar event to the spacecraft, and the spacecraft connection to the event. Thus, the next step will be to try to deduce, making an assumption of shock acceleration, how the power law index and spectral rollover relate to the shock parameters including the compression ratio, shock speed, Mach number, and diffusion coefficient at the time of release. Evidence for the influence of interplanetary conditions such as field-aligned and cross-field transport on the event to event spectral variability, will also be considered.
Further insight into the acceleration and transport mechanisms will also be provided by the PAMELA Helium observations, which will be elaborated and presented in a future work.

\section{Conclusions}
Thanks to its unique observational capabilities, the PAMELA satellite-borne experiment can investigate, for the
first time, SEP events over a wide energy region ($\gtrsim$80 MeV)
encompassing the low-energy observations by in situ spacecraft
and GLE observations made by the ground-based NM network.
In this work, 26 major SEP events, including one GLE and
potentially three sub-GLEs, registered by PAMELA between
2006 December and 2014 September, were studied. The
PAMELA results enable a more complete and clear picture of
SEPs that offers the possibility of constraining models of
particle acceleration to high energies. Consistent with the
diffusive shock acceleration theory, the measured SEP spectra
are well reproduced by a power-law modulated by an
exponential cutoff attributed to particles escaping the shock
region during acceleration. However, transport processes and
other effects such as the magnetic connection to Earth must
also contribute to the spectral variability, and further work is
required to explore the relative influences of acceleration and
transport processes on SEP spectra at the highest energies.
Furthermore, PAMELA observations shed new light on the
long-running question concerning the differences between GLE
and non-GLE events. Based on the intensity and the position of
the SEP sources, the event size distribution, and the spectral
shapes, we did not find any reason to consider the GLEs as a
separate class; rather, they represent the most energetic subset
of the SEP global distribution. Finally, potential limits for the
peak fluxes of the GLE and sub-GLE events, based on space
measurements, are provided for the first time.

\acknowledgments
The authors acknowledge partial financial support from The Italian Space Agency (ASI) under the program ``Programma PAMELA - attivit\`a scientifica di analisi dati in fase E''.
We also acknowledge support from a NASA Heliophysics and Solar Research grant, the National Science Foundation (SHINE), Deutsches Zentrum f\"{u}r Luftund Raumfahrt (DLR), the Swedish National Space Board, the Swedish Research Council, the Russian Space Agency (Roscosmos), and the Russian Ministry of Education and Science. 
A.~B. acknowledges support by an appointment to the NASA postdoctoral program at the NASA Goddard Space Flight Center administered by Universities Space Research Association under contract with NASA. I.~G.~R. acknowledges support
from NASA Living With a Star grant NNG06EO90A.

\end{document}